\documentclass[12pt]{article}

\usepackage[english]{babel}

\usepackage[a4paper,margin=2.5cm,marginparwidth=1.75cm]{geometry}

\usepackage{float}
\usepackage[caption = false]{subfig}
\usepackage[final]{graphicx}
\usepackage{amsmath}
\usepackage{amsthm}
\usepackage{amssymb}
\usepackage{graphicx}
\usepackage{multirow}
\usepackage{bbm}
\usepackage{enumerate}
\usepackage[colorlinks=true, allcolors=blue]{hyperref}
\usepackage[new]{old-arrows}
\usepackage{caption}
\usepackage{multirow}
\usepackage{cite}

\newtheorem{definition}{Definition}[section]

\newtheorem{proposition}[definition]{Proposition}
\newtheorem{lemma}[definition]{Lemma}
\newtheorem{corollary}[definition]{Corollary}

\usepackage{pst-node}
\usepackage{tikz-cd}
\numberwithin{equation}{section}
\usepackage[most]{tcolorbox}
\usepackage{caption}
\usepackage{subfig}

\renewcommand{\Re}{\operatorname{Re}}
\renewcommand{\Im}{\operatorname{Im}}

\DeclareMathOperator{\ind}{ind}

\newcommand{\LCS}{\mathrm{LCS}}
\newcommand{\LG}{\mathrm{LG}}
\newcommand{\C}{\mathrm{C}}
\newcommand{\gLCS}{\gamma_{\LCS}}
\newcommand{\gLG}{\gamma_{\LG}}
\newcommand{\gC}{\gamma_{\C}}

\usetikzlibrary{decorations.markings, arrows.meta}

\begin{document}

\begin{titlepage}

\setcounter{page}{1} \baselineskip=15.5pt \thispagestyle{empty}
\setcounter{tocdepth}{2}
	
\bigskip\

\vspace{.1cm}
\begin{center}
{\LARGE \bfseries Flux Vacua, Asymptotic Periods,\\[.5cm] and Relatively Anosov Representations}
\end{center}

\vspace{0.55cm}

\begin{center}
\scalebox{0.95}[0.95]{{\fontsize{14}{30}\selectfont 
Paul Balavoine, Severin Lüst}}

\vspace{0.5cm}

{\fontsize{11}{30}
\textsl{Sorbonne Universit\'e, CNRS, Laboratoire de Physique Théorique et Hautes Énergies, LPTHE, F-75005 Paris, France}
}
\vspace{0.75cm}

{\tt pbalavoine@lpthe.jussieu.fr, severin@lpthe.jussieu.fr} 

\vspace{1.25cm}
\normalsize{\bf Abstract} 
\end{center}

\noindent       
We develop a topological method to detect classical flux vacua of M-theory on Calabi-Yau fourfolds with a single complex structure modulus using only asymptotic period data and monodromy information.
To this end, we express the index of the F-term 
of the flux superpotential 
as a winding number along the boundary of the universal cover of the complex structure moduli space.
The index evaluates to a sum over zeros and poles of a suitably normalized boundary F-term, where local contributions  can be computed algebraically on a dense set of boundary points.
Assuming a conjectured bound on the remaining contributions, sufficiently many sign changes of the boundary F-term guarantee a flux vacuum in the interior.
We illustrate the method on the mirror Fermat sextic with explicit examples and a systematic flux scan.

A main technical ingredient in our construction is a continuity result for projective period maps: we prove that the non-holomorphic $(3,1)$-period map extends continuously to the boundary of the covering space if the monodromy representation is relatively Anosov. 
In general, this property does not hold for holomorphic $(4,0)$-periods.

\vfill

\noindent October 2026
\end{titlepage}

\tableofcontents

\section{Introduction}

Reliably identifying string theory vacua or backgrounds with no or minimal supersymmetry is a central problem of string phenomenology.
Quantum and stringy corrections beyond the leading supergravity approximation considerably complicate this task, often giving rise to Dine-Seiberg-like control issues \cite{Dine:1985he}.

Moreover, such corrections are notoriously difficult to compute and often only partially known.
These difficulties have led to extensive debates in the literature about the consistency and parametric control of constructions in asymptotic regions of moduli space, while a systematic analysis of generic vacua in its interior remains largely out of reach.%
\footnote{
At non-generic points in the interior of the moduli space, vacua have recently been constructed by either exploiting enhanced symmetries \cite{Grimm:2024fip, Chen:2025rkb,Mohseni:2025tig,Baykara:2026jzs} or using worldsheet methods \cite{Becker:2006ks,Bardzell:2022jfh, Becker:2022hse,Becker:2023rqi,Becker:2024ijy,Ishiguro:2024coq,Becker:2024ayh,Rajaguru:2024emw,Morros:2026yyi}.
However, such techniques are restricted to special loci and typically not available at arbitrary points in moduli space.
}

\medskip

In \cite{Lust:2024aeg}, it was proposed to use the behavior of an effective potential in controlled asymptotic regimes, where corrections are small or can even be ignored, to constrain the existence of critical points in the interior of the moduli space.
Concretely, a nonzero winding number or index of the gradient is a sufficient condition for the existence of at least one critical point.
This method, if applicable, is not limited to special points in moduli space.
However, this topological criterion does not determine the position or detailed properties of the critical points.

It was further proposed that this method is particularly suitable for detecting classical flux vacua of  type IIB string theory or M-theory compactified on Calabi-Yau three- or fourfolds, respectively.
Here, the index is computed for the F-terms of the GVW flux superpotential \cite{Gukov:1999ya}.
A nonzero result then guarantees the existence of at least one point in complex structure moduli space where the fluxes satisfy the appropriate classical self-duality condition.

In \cite{Lust:2024aeg}, the feasibility of this approach was demonstrated for the mirror Fermat sextic by explicitly evaluating the index as a contour integral along the boundaries of its one-dimensional complex-structure moduli space.
However, fixed fluxes generally break monodromy invariance and induce branch-cut discontinuities by making the superpotential multivalued.
To avoid these branch cuts, the integration contour must therefore enter the interior of moduli space, so the index can no longer be evaluated from asymptotic data alone.

\medskip

In this paper we demonstrate how to overcome this issue by working on the universal covering space of the complex structure moduli space.%
\footnote{The action of duality groups on moduli spaces and the properties of the relevant covering spaces and marked moduli spaces were recently discussed in the context of the Swampland program \cite{Raman:2024fcv,Delgado:2024skw}.}
On this covering space, the periods and hence the superpotential are single-valued and do not suffer from branch-cut discontinuities.
It is therefore possible to close the integration contour along its asymptotic boundary, allowing us to compute the index from boundary and monodromy information alone.
Concretely, we consider classical flux vacua of M-theory compactified on Calabi-Yau fourfolds with a single complex structure modulus.

Our approach and main results can be summarized as follows:
\begin{itemize}
    \item The universal covering space of the complex structure moduli space of a one-parameter Calabi-Yau manifold is given by the hyperbolic upper half-plane $\mathbb{H}$, isomorphic to the Poincar\'e disk.
    The moduli space is then recovered as the quotient $\mathbb{H}/\Gamma$ by a Fuchsian group $\Gamma$ that may be identified with the monodromy group.
    
    Period integrals naturally lift from the moduli space to $\Gamma$-equivariant functions on the interior of the covering space.
    We prove a sufficient condition for projective period maps to extend continuously to the boundary of the covering space.
    This condition requires the monodromy representation of $\Gamma$ to be relatively Anosov and imposes certain transversality and convergence conditions on the period map.
    
    For Calabi-Yau fourfolds with a single complex structure modulus, 
    the non-holo\-mor\-phic $(3,1)$-period maps satisfy these conditions and therefore extend continuously to the boundary of the universal covering space.
    The holomorphic $(4,0)$-periods fail the transversality requirement and, as we prove separately for certain fourfolds, do not admit a continuous boundary extension.
    
    \item The F-term $DW$ is a linear combination of the $(3,1)$-periods.
    Since the latter extend continuously to the boundary, the index of $DW$ on the covering space is well defined 
    and given by the sum formula
    \begin{equation}
        \ind_\mathbb{H}(DW) = \sum_{\substack{x\in\partial\mathbb{H}\\ f_g(x)\in\{0,\infty\}}} \ind_x(f_g) \,,
    \end{equation}
    where $f_g \propto DW$ denotes the normalized F-term for a given four-form flux $g$.
    The sum runs over all points $x$ on the boundary $\partial \mathbb{H}$ where the normalized F-term $f_g(x)$ either vanishes or diverges.
    
    For boundary points $x\in\Gamma\cdot\infty$, each local contribution $\ind_x(f_g)$ in the sum can be computed algebraically from just two inputs: the monodromy matrix relating $x$ to infinity and the asymptotic behavior of the periods at infinity.
    Such points are dense in $\partial\mathbb{H}$.
    
    A bound on the local indices $\ind_x(f_g)$ makes it possible to establish $\ind_\mathbb{H}(DW) \neq 0$ by counting sufficiently many sign-changing zeros of $f_g$ on the real axis. 
    A nonzero index then guarantees the existence of a flux vacuum in the interior of the covering space.
    This bound is rigorous for all points in the dense subset $\Gamma\cdot\infty \subset \partial \mathbb{H}$ but remains conjectural on its complement.
    
    \item We illustrate our findings for the mirror Fermat sextic and discuss a few explicit examples.
    We also present a computer-aided scan over a large number of fluxes, with an automated evaluation of bounds on the index.
\end{itemize}

The remainder of this paper is organized as follows.
In Section~\ref{sec:preliminaries} we review the relevant background material on classical flux vacua of M-theory, the nilpotent orbit expansion of Calabi-Yau periods, and the index construction of \cite{Lust:2024aeg}.
In Section~\ref{sec:coverinspace} we discuss period maps on the universal covering space of the complex structure moduli space and establish a connection between their continuous extension to its boundary and relatively Anosov period representations. 
In Section~\ref{sec:coveringspaceindex} we describe how to compute the index for flux vacua on the universal covering space of the moduli space and develop a sufficient condition for the existence of flux vacua in its interior.
In Section~\ref{sec:sextic} we apply our previous findings to the concrete example of the mirror Fermat sextic.
This section also contains explicit numerical examples and an automated scan over fluxes.
Section~\ref{sec:discussion} contains a brief concluding discussion of our results.

Detailed derivations and supplementary material can be found in the appendices.
In particular, Appendix~\ref{app:continuityproof} contains the proof of our main theorem, a sufficient condition for projective period maps to extend continuously to the boundary.

Two accompanying Mathematica notebooks implement the computation of boundary functions for Calabi-Yau fourfolds with triangle monodromy groups, recently classified in \cite{vandeHeisteeg:2024lsa}, and automated flux scans.

\medskip

Readers who are more interested in the mathematical aspects of our work may read Section~\ref{sec:coverinspace} together with Appendices~\ref{app:anosov} and \ref{app:asymptotichodge} independently from the rest of the paper.
Readers who are mostly interested in the physical aspects may skip Section~\ref{sec:coverinspace} on a first reading.
Its most relevant results for the subsequent discussion are briefly summarized at the beginning of Section~\ref{sec:coveringspaceindex}.

\section{Flux vacua and asymptotic period data}
\label{sec:preliminaries}

In this section we review the relevant background material on classical flux vacua of M-theory, Calabi-Yau fourfolds with a single complex structure modulus, and the nilpotent orbit expansion of their period maps.
We also briefly summarize the index construction of \cite{Lust:2024aeg} and discuss how the multivaluedness of the flux superpotential on the moduli space obstructs its evaluation from asymptotic data alone, motivating our subsequent formulation on the universal covering space.

\subsection{Flux compactifications of M-theory}

We begin by summarizing a few elementary facts on flux compactifications of M-theory to set the stage for our subsequent discussion.
We consider M-theory compactified on a Calabi-Yau four-fold $X_4$ with non-trivial four-form flux $G_4 \in H^4 (X_4)$.%
\footnote{The four-form flux is subject to a quantization condition and a tadpole cancellation condition 
\begin{equation}\label{eq:quantizationandtadpole}
    G_4 + \frac12 c_2(X_4) \in H^4(X_4, \mathbb{Z}) \,, \qquad\text{and}\qquad \frac12 \int G_4 \wedge G_4 + N_\mathrm{M2} = \frac{\chi(X_4)}{24} \,,
\end{equation}
with $c_2(X_4)$ the second Chern class of $X_4$, $\chi(X_4)$ its Euler characteristic, and $N_\mathrm{M2}$ the number of spacetime filling M2 branes in the compactification.
These conditions will not play an important role in our later discussion.
}

Classical flux vacua correspond to minima of the effective flux-induced potential
\begin{equation}\label{eq:Vflux}
    V_\mathrm{flux} \sim \frac{1}{\mathcal{V}^3} \int_{X_4} G_4 \wedge \star G_4 - G_4 \wedge G_4 \,,
\end{equation}
where $\mathcal{V}$ denotes the volume of $X_4$ and $\star$ the Hodge star operator associated with the Calabi-Yau metric on $X_4$.
This potential is positive semi-definite and vanishes precisely when $G_4$ is self-dual.
Classical Minkowski vacua are therefore characterized by the self-duality condition
\cite{Becker:1996gj, Dasgupta:1999ss}
\begin{equation}
    \label{eq:self-duality}
    G_4 = \star G_4 \,.
\end{equation}
In the following we assume that $G_4$ is primitive,
\begin{equation}
    J \wedge G_4 = 0 \,,
\end{equation}
or equivalently $G_4 \in H^4_\mathrm{prim}(X_4)$.
With this restriction, \eqref{eq:self-duality} becomes a condition only on the complex structure moduli of $X_4$.
Correspondingly, the dependence of the potential \eqref{eq:Vflux} on the Kähler moduli is only through the volume prefactor.

The primitive part of the middle cohomology admits the Hodge decomposition
\begin{equation}\label{eq:hodgedecomposition}
    H^4_\mathrm{prim}(X_4) = H^{4,0} \oplus H^{3,1} \oplus H^{2,2}_\mathrm{prim} \oplus H^{1,3} \oplus H^{0,4} \,.
\end{equation}
The Hodge star acts on a primitive $(4-p,p)$-form
with eigenvalue $(-1)^p$.
Consequently, the self-duality condition \eqref{eq:self-duality} implies that $G_4$ has no component in the $H^{3,1}$ or $H^{1,3}$ parts of the cohomology.

The vanishing of these components of the flux can be encoded in terms of an effective superpotential and K\"ahler potential \cite{Gukov:1999ya} (see also \cite{Haack:2001jz}),
\begin{equation}\label{eq:GVW}
    W(z^\alpha) = \int_{X_4} G_4 \wedge \Omega(z^\alpha) \,, \qquad K(z^\alpha, \bar z^{\bar\beta}) = - \log \int_{X_4} \Omega(z^\alpha) \wedge \overline{\Omega}(\bar z^{\bar\beta}) \,,
\end{equation}
via the F-term conditions
\begin{equation}\label{eq:fterm}
    D_\alpha W(z, \bar z) = \partial_\alpha W(z) + \partial_\alpha K(z, \bar z)\, W(z) = 0 \,,
\end{equation}
where $z = (z^\alpha)$, $\alpha = 1, \dots, h^{3,1}(X_4)$, denotes the complex structure moduli of $X_4$.
This shows that the self-duality condition gives $h^{3,1}$ complex equations for the $h^{3,1}$ complex-structure moduli.
Therefore, for a generic choice of flux, one expects these equations to fix the complex-structure moduli at isolated points in the corresponding moduli space.%
\footnote{For a large number of complex-structure moduli, this expectation is challenged by the tadpole conjecture \cite{Bena:2020xrh, Bena:2021wyr}, which states that stabilizing many moduli by fluxes requires a flux-induced tadpole that grows at least linearly with the number of stabilized moduli and may therefore exceed the available tadpole bound in \eqref{eq:quantizationandtadpole}.}

To solve the F-term equations \eqref{eq:fterm} explicitly, it is useful to express them in terms of periods.
The holomorphic four-form $\Omega$ and all its derivatives with respect to the complex structure moduli span the horizontal subspace 
\begin{equation}
    H^4_H(X_4) \subseteq H^4_\mathrm{prim}(X_4) \,.
\end{equation}
Its orthogonal complement in $H^4_\mathrm{prim}(X_4)$ is of type $(2,2)$ everywhere in the complex structure moduli space \cite{Braun:2014xka}.
Flux components along this complement are therefore automatically self-dual and do not enter the superpotential, so we can restrict our attention to $G_4 \in H^4_H(X_4)$.%
\footnote{Since the integral lattice $H^4(X_4,\mathbb{Z})$ is not always compatible with the decomposition into a horizontal part and its complement, the horizontal component of an integer quantized flux $G_4$ is in general only rational.}

We choose a basis $\alpha_I$, $I = 0, \dots, b_H -1$, with $b_H= \dim H^4_H(X_4)$, of the horizontal middle cohomology $H^4_H(X_4)$ and define the intersection form
\begin{equation}\label{eq:eta}
    \eta_{IJ} = \int_{X_4} \alpha_I \wedge \alpha_J \,.
\end{equation}
For future reference, we note that $\eta_{IJ}$ defines a Hermitian form that is definite on each Hodge component $H^{p,q}_H = H^4_H \cap H^{p,q} \subseteq H^{p,q}_\mathrm{prim}$.%
\footnote{This can be seen as follows: for every non-zero $\omega = \omega^I \alpha_I \in H^{4-p,p}_\mathrm{prim}$ one has
\begin{equation*}
    (-1)^p \, \omega^I \eta_{IJ} \overline{\omega}^J = (-1)^p \int_{X_4} \omega \wedge \overline{\omega} = \int_{X_4} \star \omega \wedge \overline{\omega} > 0 \,.
\end{equation*}}
This implies that $\eta_{IJ}$ has signature $(2+ h^{2, 2}_H, 2 h^{3,1})$.

In this basis, the periods $\Pi^I$ and the flux coefficients $g^I$ are defined by
\begin{equation}\label{eq:defper}
    \Pi^I(z) = \eta^{IJ} \int_{X_4} \Omega(z) \wedge \alpha_J \,,\qquad  g^I = \eta^{IJ} \int_{X_4} G_4 \wedge \alpha_J \,,
\end{equation}
where $\eta^{IJ}$ is the inverse of $\eta_{IJ}$.
The superpotential and K\"ahler potential \eqref{eq:GVW} become
\begin{equation} \label{eq:linearWK}
    W(z) = g^I \eta_{IJ} \Pi^J (z) \,, \qquad K(z, \bar z) = - \log \Pi^I(z) \eta_{IJ} \overline{\Pi}^J(\bar z) \,.
\end{equation}
The F-term conditions then read
\begin{equation}
    \label{eq:development}
    D_\alpha W(z, \bar z) = g^I \eta_{IJ} D_\alpha \Pi^J(z, \bar z) = 0 \,.
\end{equation}
For a single complex-structure modulus, i.e.\ $h^{3,1}(X_4) = 1$, we also sometimes write $DW$ instead of $D_\alpha W$.

Solving these equations is generally difficult since, at generic points in the moduli space, the periods $\Pi^I$ depend on the complex structure moduli in a complicated way.
As we will review next, they simplify close to the asymptotic boundaries of the moduli space.

\subsection{One-dimensional complex structure moduli spaces}
\label{sec:onedimmodulispace}

In the remainder of this paper, we restrict to Calabi-Yau fourfolds with a single complex structure modulus, i.e.
\begin{equation}
     h^{3,1}(X_4) = 1 \,.
\end{equation}
We also focus primarily on the case 
\begin{equation}
    h^{2,2}_{H}(X_4)=1 \,,
\end{equation}
for which the horizontal middle cohomology is five-dimensional, $b_H(X_4)=5$.
Parts of our general discussion, including most of the continuity results in Section~\ref{sec:coverinspace} and Appendix~\ref{app:anosov}, also apply to arbitrary $h^{2,2}_{H}(X_4)$, or equivalently to arbitrary $b_H(X_4) \geq 5$ in the one-modulus case.

We denote the corresponding one-dimensional complex structure moduli space by $\mathcal{M}_{X_4}$.
The periods $\Pi^I$ introduced in \eqref{eq:defper} are locally holomorphic functions on $\mathcal{M}_{X_4}$ away from a finite number of points $s_\nu$ and satisfy a Picard-Fuchs differential equation. 
The points $s_\nu$ are the singular points of this equation.

\subsubsection{Monodromy}

The periods are in general not globally single-valued but experience monodromies around the points $s_\nu$.
Analytic continuation along a loop $\gamma$ acts linearly on the period vector,
\begin{equation}\label{eq:monodromymap}
    \Pi^I \mapsto \left(M_\gamma{}\right)^I{}_J \, \Pi^J \,.
\end{equation}
In particular, each small loop $\gamma_\nu$ encircling one of the points $s_\nu$ gives a local monodromy matrix $M_\nu = M_{\gamma_\nu}$.

The intersection form $\eta_{IJ}$ is invariant under monodromy transformations,
\begin{equation}\label{eq:etamonodromy}
    \eta_{IJ} \left(M_\gamma{}\right)^I{}_K \left(M_\gamma{}\right)^J{}_L = \eta_{KL} \,.
\end{equation}
This ensures that the K\"ahler potential \eqref{eq:linearWK} is invariant under monodromy and that the corresponding K\"ahler metric, the Weil-Petersson metric, is well-defined on $\mathcal{M}_{X_4}$. 

By the monodromy theorem \cite{landman1967picard, landman1973picard}, $M_\nu$ is quasi-unipotent, i.e.\ its eigenvalues are roots of unity.
This means that there exists a positive integer $k_\nu > 0$ such that $M_\nu^{k_\nu}$ is unipotent.
Equivalently, after passing to the local $k_\nu$-fold cover around $s_\nu$, the local monodromy becomes unipotent and can be written as
\begin{equation}\label{eq:Ni}
    M_\nu^{k_\nu} = e^{N_\nu} \,,
\end{equation}
where $N_\nu$ is a nilpotent matrix.
For later reference, we denote the degree of nilpotency of $N_\nu$ by $0 \leq r_\nu \leq 4$, i.e.
\begin{equation}\label{eq:degreeofnilpotency}
    N_\nu^{r_\nu+1} = 0 \,, \qquad N_\nu^{r_\nu} \neq 0 \,.
\end{equation}
If $M_\nu$ has finite order, then $M_\nu^{k_\nu} = \mathbf{1}$, 
so that $N_\nu =0$ and $r_\nu = 0$.
On the other hand, if $M_\nu$ has infinite order, then the unipotent part is non-trivial and $N_\nu \neq 0$, i.e.\ $r_\nu \geq 1$.

\subsubsection{Nilpotent orbit expansion}

The local behavior of the periods is described by the nilpotent orbit theorem \cite{Schmid:1973cdo}.
Let $\psi$ be a local coordinate near a singular point $s_\nu$, with $s_\nu$ located at $\psi = 0$, and introduce
\begin{equation}
    \label{eq:deft}
    t = \frac{1}{2\pi i k_\nu }\log \psi 
\end{equation}
as a logarithmic coordinate on the local $k_\nu$-fold cover.
Then, the periods have the asymptotic form 
\begin{equation}\label{eq:nilpotent}
    \Pi (t) = e^{t N_\nu} A_\nu(e^{2 \pi i t}) \,,
\end{equation}
where $N_\nu$ is the nilpotent matrix introduced in \eqref{eq:Ni}.
The functions $A_\nu$ are holomorphic near the origin in the $q = \psi^{1/k_\nu} = e^{2 \pi i t}$ plane, and therefore admit a series expansion
\begin{equation}\label{eq:nilpotentexpansion}
    A_\nu(q) = \sum_{n=0}^\infty a_{\nu,n} \, q^n \,.
\end{equation}
This expansion converges for sufficiently small $|q|$.
Thus, the nilpotent orbit expansion \eqref{eq:nilpotent} applies for sufficiently large $\Im t$, whereas large $|t|$ alone does not guarantee its validity. 

For finite-order points one has $N_\nu = 0$, and the periods are holomorphic at $s_\nu$ on the finite cover.
Around infinite-order points one may separate the periods into a leading polynomial part and subleading exponential corrections,
\begin{equation}
    \Pi (t) = P_\nu(t) + e^{t N_\nu} \mathcal{O}(e^{2 \pi i t}) \,,
\end{equation}
where
\begin{equation}\label{eq:polynomialperiodexpansion}
    P_\nu(t) = e^{t N_\nu} A_\nu(0) =  e^{t N_\nu} a_{\nu,0} 
\end{equation}
is a polynomial in $t$ of degree at most the nilpotency degree $r_\nu$ of $N_\nu$, see \eqref{eq:degreeofnilpotency}.

The points $s_\nu$ lie at either finite or infinite distance with respect to the Weil-Petersson metric, i.e.\ the K\"ahler metric obtained from the K\"ahler potential \eqref{eq:linearWK}.
This classification is not the same as the distinction between finite-order and infinite-order monodromy.
Typical examples of points at finite distance include conifold points and Landau-Ginzburg (or Gepner) points.
For Calabi-Yau fourfolds both have finite-order monodromy.%
\footnote{In contrast, conifold points of Calabi-Yau threefolds have infinite order monodromy but are still at finite distance with respect to the Weil-Petetersson metric.}
An important class of infinite distance points is given by large-complex-structure (LCS) points.
These have maximally unipotent monodromy, i.e.\ $r_\nu = 4$, and are therefore also known as MUM points.%
\footnote{For a more exhaustive discussion, including examples of points where the classification by Weil-Petersson distance and the classification by monodromy order do not agree, see \cite{vandeHeisteeg:2024lsa}. For a discussion of all the general types of singularity occurring in fourfolds from the point of view of the limiting Hodge structure, see \cite{Grimm:2019ixq}.}

\subsubsection{Large complex structure}

Large complex structure points are of particular importance for our subsequent discussion.
Under mirror symmetry they get mapped to large-volume limits in the K\"ahler moduli space of the mirror-dual Calabi-Yau manifold.
In this description, the polynomial part $P_\mathrm{LCS}(t)$ of the periods corresponds to the classical and perturbative periods, while the exponential corrections in \eqref{eq:nilpotent} are interpreted as non-perturbative quantum corrections,
namely worldsheet instantons.

For $h^{2,2}_{H}(X_4)=1$,%
\footnote{For $h^{2,2}_{H}(X_4)>1$, the polynomial expansion at large complex structure still contains only five linearly independent periods.
In the interior of the moduli space, the additional periods become independent once exponentially suppressed corrections, mirror dual to quantum effects, are included.
Such ``quantum periods'' were discussed in more detail in \cite{Gerhardus:2016iot}.}
one can choose a suitable rational basis of $H^4_H(X_4, \mathbb{Q})$ and logarithmic coordinate $t$ in which the polynomial part of the periods at large complex structure takes the simple universal form \cite{Marchesano:2021gyv, vandeHeisteeg:2024lsa}
\begin{equation}\label{eq:genperiods}
    P_\mathrm{LCS}(t) = \begin{pmatrix}
        1\\
        t\\
        \frac{t^2}{2}\\
        \frac{t^3}{6} + i c \\
        \frac{t^4}{24} + i c \, t
    \end{pmatrix} \,,
\end{equation}
where $c$ is a model-dependent constant.%
\footnote{The constant $c$ is given in terms of topological data of the mirror fourfold $\tilde X_4$,
\begin{equation}
    \label{eq:defc3}
    c = \frac{\zeta(3)}{8 \kappa \pi^3}  \int_{\tilde X_4} c_3(\tilde X_4) \wedge J \,, \qquad\qquad  \kappa = \int_{\tilde X_4} J^4 \,,
\end{equation}
where $J$ is the primitive integral generator of the one-dimensional Kähler cone of $\tilde X_4$, $c_3(\tilde X_4)$ is its third Chern class, and $\zeta$ is the Riemann zeta function. The value of $c$ doesn't appear to be consistent with \eqref{eq:sextickappac}.}

From this expansion, one can straightforwardly determine the monodromy of the periods around the LCS point by sending $t \rightarrow t+1$.
The resulting monodromy matrix is displayed in Appendix~\ref{app:LCSrationalbasis}, together with an explicit expression for the vector $a_{\mathrm{LCS}, 0}$ that reproduces \eqref{eq:genperiods} from the leading-order term \eqref{eq:polynomialperiodexpansion} in the nilpotent orbit expansion.

We also note that the intersection form in this basis is given by
\begin{equation} \label{eq:def_eta}
    \eta_{IJ} = \kappa \begin{pmatrix}
        0 & 0 & 0 & 0 & 1\\
        0 & 0 & 0 & -1 & 0\\
        0 & 0 & 1 & 0 & 0\\
        0 & -1 & 0 & 0 & 0\\
        1 & 0 & 0 & 0 & 0
    \end{pmatrix} \,.
\end{equation}
In the examples we consider, one passes to an integral basis of $H^4_{H}(X_4,\mathbb{Z})$ by an appropriate rational change of basis.
In the resulting basis, additional lower-degree polynomial terms may appear in the asymptotic periods \eqref{eq:genperiods} \cite{Marchesano:2021gyv, vandeHeisteeg:2024lsa}.

With this information, one can directly compute the superpotential and K\"ahler potential \eqref{eq:linearWK} in the asymptotic limit at large complex structure, and hence also the K\"ahler covariant derivative $D_t W$.
Due to the polynomial form of the asymptotic periods, solving the resulting F-term condition \eqref{eq:development} for flux vacua is comparatively simple.
For explicit solutions, including the multi-moduli case, see, e.g.\ \cite{Demirtas:2019sip, Marchesano:2021gyv}.
Similar considerations hold at other asymptotic limits in the complex structure moduli space.

Away from any asymptotic limit, however, the problem becomes substantially harder.
In this regime, the exponential corrections to the periods in \eqref{eq:nilpotent} become important and must be taken into account.
Unless one can exploit an additional symmetry (see, e.g., \cite{Giryavets:2003vd, Denef:2004dm, DeWolfe:2004ns, DeWolfe:2005gy,Cicoli:2013cha, Braun:2020jrx, Lust:2022mhk, Candelas:2023yrg, Grimm:2024fip, Chen:2025rkb}), the F-term condition appears to be explicitly tractable only numerically in the interior of the moduli space.

In the following we explore whether one can extract topological information about the existence of flux vacua in the interior of the moduli space from the simple asymptotic form of the periods at large complex structure and near the other singular points $s_\nu$.%
\footnote{For a related but complementary idea, see, e.g., the moduli space holography proposal of  \cite{Grimm:2020cda, Grimm:2021ikg}, where bulk properties of the complex structure moduli space are encoded in asymptotic boundary data.}
Following an idea put forward in \cite{Lust:2024aeg}, we approach this question using the index of a vector field. 

\subsection{The index of a vector field}
\label{sec:index}

We briefly review the definition of the index of a vector field.
We consider a bounded open subset $U \subseteq \mathbb{R}^n$ with smooth boundary $\partial U$, a finite set of punctures $\{x_i\}_{i \in I} \subset U$, and a continuous vector field $X \colon \overline U \setminus \{x_i\}_{i \in I} \rightarrow \mathbb{R}^n$ that is non-zero everywhere on its domain.
In general, the $x_i$ correspond to zeros, poles, or other singular points of the vector field $X$.

The degree of a map is a higher-dimensional generalization of the two-dimensional winding number.
If the map
\begin{equation}
    f \colon \partial U \rightarrow S^{n-1} \,, \qquad x \mapsto \frac{X(x)}{\|X(x)\|} 
\end{equation}
is differentiable, its degree is given by
\begin{equation}
    \label{eq:degree}
    \deg f \int_{S^{n-1}} \omega = \int_{\partial U} f^* \omega\,,
\end{equation}
where $\omega$ is a top-form on the sphere $S^{n-1}$, and $f^* \omega$ denotes the pullback of $\omega$ by $f$.
The index of $X$ on $U$ is then defined as
\begin{equation}\label{eq:indexdegreedef}
    \operatorname{ind}_U(X) \equiv \deg f \in \mathbb{Z} \,.
\end{equation}
For continuous but non-differentiable $X$, the degree, and hence the index, is still well-defined by approximating $f$ with a homotopic differentiable map.

For any point $x \in U$ (including punctures), we define the local index 
\begin{equation}
    \ind_x(X) \equiv \ind_{B_x}(X) \,,
\end{equation}
where $B_x$ denotes a small ball centered at $x$ that contains no punctures other than possibly $x$ itself.
This definition is independent of the choice of $B_x$.
If $x$ is not a puncture, $X/\|X\|$ is well-defined and continuous at $x$, so shrinking the ball to a point yields $\operatorname{ind}_x(X) = 0$.
One can then show that the index is given by the sum of the local indices of the punctures,
\begin{equation}
    \operatorname{ind}_U(X) = \sum_{i \in I} \operatorname{ind}_{x_i}(X) \,.
\end{equation}
Consequently, $\operatorname{ind}_U(X) \neq 0$ implies that the set of punctures is not empty.

Choosing $\omega$ to be the standard volume form on the sphere,
\begin{equation}
    \omega = \epsilon_{i_1 \dots i_{n}} \frac{x^{i_1}}{(n - 1)!} d x^{i_2} \wedge \dots  \wedge d x^{i_{n}} \,,
\end{equation}
eqs.\ \eqref{eq:degree} and \eqref{eq:indexdegreedef} reduce to the explicit expression
\begin{equation}\label{eq:indexintegral}
    \mathrm{ind}_U (X) = \frac{1}{\mathrm{vol}(S^{n - 1})} \int_{\partial U} \epsilon_{i_1 \dots i_{n}} \frac{X^{i_1}}{(n - 1)! \|X\|^{n}} d X^{i_2} \wedge \dots  \wedge d X^{i_{n}}\,,
\end{equation}
with $\epsilon$ the Levi-Civita symbol, and $\|X\|$ the Euclidean norm of $X$.

In the two-dimensional case, which will be the main focus of this paper, the vector field $X$ can be encoded in a complex function
\begin{equation}
    g_X = X^1 + i X^2 \,.
\end{equation}
The index of $X$ on $U$ is then the winding number of $g_X$ along the boundary $\partial U$, corresponding to the total change of its argument or complex phase along $\partial U$ (divided by $2 \pi$).
This is well-defined for continuous $X$.
If $X$ is also differentiable, it can be written as
\begin{equation} \label{eq:winding_number}
    \operatorname{ind}_U(X) = \frac{1}{2 \pi i} \oint_{\partial U} d \log g_X \,,
\end{equation}
which agrees with \eqref{eq:indexintegral} for $n=2$.%
\footnote{
    Introducing the complex coordinate $t = x^1 + i x^2$, \eqref{eq:winding_number} can also be written as 
    \begin{equation*}
        \operatorname{ind}_U(X) = \frac{1}{2 \pi i} \oint_{\partial U} \frac{\partial_t g_X (t, \overline{t}) dt + \partial_{\overline{t}} g_X (t, \overline{t}) d \overline t}{g_X(t, \overline t)}\,. 
    \end{equation*}
    If $g_X$ is meromorphic, i.e.\ $\partial_{\bar t} g_X = 0$ away from its poles, Cauchy's argument principle is recovered.
}

If $X$ has no poles or other singular points in $U$, the index can therefore be used to detect zeros of $X$:
a non-zero winding number \eqref{eq:winding_number} guarantees that $X$ has at least one zero in $U$.
However, a vanishing winding number does not imply the absence of zeros, as individual zeros may have vanishing local index or the indices of several zeros may cancel.

\subsection{The multivaluedness problem}

We return to the problem of finding flux vacua in the interior of the complex structure moduli space.
For this purpose, we apply \eqref{eq:winding_number} to the K\"ahler-covariant derivative $DW$ of the flux superpotential \eqref{eq:linearWK}.
If we succeed in evaluating the integral in \eqref{eq:winding_number} along a suitably defined contour along the boundary $\partial \mathcal{M}_{X_4}$ of the complex structure moduli space, then
\begin{equation}
    \operatorname{ind}_{\mathcal{M}_{X_4}}(DW) \neq 0
    \label{eq:condition}
\end{equation}
implies the existence of at least one zero of $DW$ in the interior of $\mathcal{M}_{X_4}$, corresponding to a classical flux vacuum.

A naive choice for the integration contour  $\mathcal{C}$ around $\partial \mathcal{M}_{X_4}$ is given by
\begin{equation}\label{eq:naivecontour}
    \mathcal{C}=\sum_\nu \gamma_\nu \,,
\end{equation}
with $\gamma_\nu$ an infinitesimal loop around the singular point $s_\nu$.
Due to the simple asymptotic form \eqref{eq:nilpotent} of the periods near the points $s_\nu$, the evaluation of \eqref{eq:winding_number} is, in principle, explicitly possible.
Unfortunately, this naive approach fails because the periods $\Pi^I$ and hence also $DW$ are not single-valued on $\mathcal{M}_{X_4}$, as discussed in Section~\ref{sec:onedimmodulispace}.
The contour \eqref{eq:naivecontour} therefore does not define a closed contour for a single-valued integrand, and the corresponding index integral is not well-defined on $\mathcal{M}_{X_4}$.

One can imagine two possible ways to address this problem.
First, one may attempt to close the contour by encircling not only the singular points $s_\nu$ but also the branch cuts of $DW$ that are induced by the monodromies of the periods around them.
It was demonstrated in \cite{Lust:2024aeg} that this is possible in principle.
However, the resulting contour is not confined to the region asymptotically close to the singular points but probes the interior of the moduli space.
Therefore, the contour integral can, in general, not be evaluated anymore using only the simple asymptotic form \eqref{eq:nilpotent} of the periods.

A second possibility is to lift the problem to a suitable covering space of $\mathcal{M}_{X_4}$ on which the periods, and hence also $DW$, become single-valued.
In the remainder of this paper, we explore this second approach.

\section{The covering space and its boundary}
\label{sec:coverinspace}

In this section, we show how to extend the periods to the universal cover of the moduli space and discuss under which conditions they extend continuously to the boundary of the covering space.

In Section~\ref{sec:universalcover} we introduce the complex upper half-plane as the universal covering space. Section~\ref{sec:periodscover} then describes the periods as monodromy-equivariant functions on this cover, states their nilpotent orbit expansion there, and introduces the projective period map together with its normalized representatives.

In Section~\ref{sec:continuity} we discuss whether periods extend continuously to the boundary of the covering space and state our continuity criterion based on relatively Anosov period representations.
We test this criterion for the holomorphic $(4,0)$-periods and find that its conditions are not satisfied.
Indeed, we later show in Section~\ref{sec:sexticC1} for a concrete example that they cannot extend continuously to the boundary.

In Section~\ref{sec:31per} we repeat this analysis for the non-holomorphic $(3,1)$-periods and find that they do allow for a continuous boundary expansion, provided that the period representation is relatively Anosov.
We also show that they allow for a polynomial expansion that is valid for all points with sufficiently large absolute value in the upper half-plane, which will be used in our subsequent discussion in Section~\ref{sec:coveringspaceindex}.

Together with the relevant introductory material in Section~\ref{sec:onedimmodulispace} and Appendices~\ref{app:anosov} and \ref{app:asymptotichodge}, this section can be read independently from the rest of this paper.
We summarize its main results at the beginning of Section~\ref{sec:coveringspaceindex}.

\subsection{The universal covering space of the moduli space}
\label{sec:universalcover}

\subsubsection{Punctures and orbifold points}

In the one-dimensional case considered here,
$\mathcal{M}_{X_4}$ is a quasi-projective orbifold%
\footnote{This follows from Viehweg's quasi-projectivity theorem for moduli spaces of polarized varieties with semi-ample canonical bundle \cite{viehweg2012quasi}.}
and can therefore be written as
\begin{equation}\label{eq:quasiprojective}
    \mathcal{M}_{X_4} \simeq V \setminus \left\{ p_1, \dots, p_{n_p} \right\} \,,
\end{equation}
where $V  \simeq \overline{\mathcal{M}}_{X_4}$ is a compact Riemann orbifold and the points $p_i$ ($i=1, \dots, n_p$) are punctures.
We denote the interior orbifold points of $V$ by
\begin{equation}
    \left\{o_1, \dots, o_{n_o} \right\} \subset \mathcal{M}_{X_4} \,,
\end{equation}
and their orders by $k_a$ ($a=1, \dots, n_o$).
The underlying coarse space of $V$ is a compact Riemann surface.
In many standard examples, it is isomorphic to the Riemann sphere, $\left| V\right| \simeq \mathbb{CP}^1$.

The singular points $s_\nu$ ($\nu=1, \dots, n_p+n_o$) of the Picard-Fuchs equation introduced in Section~\ref{sec:onedimmodulispace} are precisely the punctures $p_i$ and the orbifold points $o_a$.
In our conventions, the orbifold points $o_a$ lie in the interior of the moduli space $\mathcal{M}_{X_4}$, and only the punctures $p_i$
constitute the compactification boundary of the moduli space, 
\begin{equation}
    \partial \mathcal{M}_{X_4} \equiv \overline{\mathcal{M}}_{X_4} \setminus \mathcal{M}_{X_4} = \{ p_1,\ldots,p_{n_p} \} \,.
\end{equation}
They correspond to infinite-order monodromy singularities.
On the other hand, the orbifold points correspond to points with finite-order monodromy.
Their orders $k_a$ agree with the order of the corresponding local monodromy, as introduced in \eqref{eq:Ni}.
Note that, with these conventions, degeneration points of the Calabi-Yau manifold $X_4$ may still lie in the interior of $\mathcal{M}_{X_4}$.
This is, for example, the case for a conifold point, which for Calabi-Yau fourfolds is an orbifold point of order two.

\subsubsection{Uniformization and hyperbolic covering space}

Since $\mathcal{M}_{X_4}$ is a one-dimensional complex orbifold, the uniformization theorem implies that its orbifold universal cover is either the Riemann sphere $\widehat{\mathbb{C}}$, the complex plane $\mathbb{C}$ or the upper half-plane $\mathbb{H}$, equivalently the disk $\mathbb{D}$.
The orbifold Euler characteristic is given by
\begin{equation}\label{eq:orbifoldeuler}
    \chi(\mathcal{M}_{X_4}) = 2 - 2g - n_p - \sum_{a=1}^{n_o} \left(1 - \frac{1}{k_a}\right) \,,
\end{equation}
where $g$ is the genus, $n_p$ the number of punctures, and $k_a$ the order of the orbifold point $o_a$.
The moduli spaces considered here contain at least one puncture, namely an LCS point, as expected from mirror symmetry.
Moreover, one-parameter Calabi--Yau moduli spaces with only one or two singular points $s_r$ have constant periods \cite{Griffiths1970,vandeHeisteeg:2024lsa}.
It follows that $\chi(\mathcal{M}_{X_4}) < 0$ unless $g=0$ and there are precisely one puncture and two orbifold points of order two.%
\footnote{The Riemann orbifold of signature $(0;2,2,\infty)$ has $\chi = 0$ and universal cover $\mathbb{C}$.}
However, this case is ruled out by the classification in~\cite{vandeHeisteeg:2024lsa}.
Therefore, the orbifold universal cover of $\mathcal{M}_{X_4}$ is $\mathbb{H}$.
Unless stated otherwise, we use $z = x + i y$ with $y > 0$ as a holomorphic coordinate on $\mathbb{H}$.

To make this explicit, we identify a discrete and faithful representation
\begin{equation}
    \iota: \pi^\mathrm{orb}_1(\mathcal{M}_{X_4}) \,\to\, \mathrm{PSL}(2, \mathbb{R}) \,,
\end{equation}
of the orbifold fundamental group $\pi^\mathrm{orb}_1(\mathcal{M}_{X_4})$.
The existence of such a representation is guaranteed by the uniformization theorem.
Its image 
\begin{equation}
    \Gamma_\mathrm{F} = \iota\left( \pi^\mathrm{orb}_1(\mathcal{M}_{X_4})\right) \,,
\end{equation}
is a Fuchsian group acting properly discontinuously on the upper half-plane by M\"obius transformations,
\begin{equation}\label{eq:moebius}
    \gamma \cdot z = \frac{a z + b}{c z + d} \,, \qquad
    \gamma = \begin{pmatrix} a & b \\ c & d \end{pmatrix} \in \Gamma_\mathrm{F} \subset \mathrm{PSL}(2, \mathbb{R}) \,,
\end{equation}
for $z \in \mathbb{H}$.
Note that since $\iota$ is faithful, it provides an isomorphism between $\pi_1^\mathrm{orb}(\mathcal M_{X_4})$ and $\Gamma_{\mathrm F}$.
We therefore identify both groups and use the same symbol $\gamma$ for an element of the orbifold fundamental group and its image $\iota(\gamma)\in\Gamma_{\mathrm F}$.

The moduli space is then isomorphic to the quotient
\begin{equation}
    \mathcal{M}_{X_4} \simeq \mathbb{H} / \Gamma_F \,, \label{eq:defM} 
\end{equation}
and the associated orbifold covering map is
\begin{equation}\begin{aligned}
    \label{eq:cover}
    \phi \colon \mathbb{H} &\rightarrow \mathcal{M}_{X_4} \\
    z &\mapsto [z] \,,
\end{aligned}\end{equation}
where $[z] = \Gamma_{\mathrm F}\cdot z$ denotes the $\Gamma_F$-orbit of $z \in \mathbb{H}$.

This construction is closely related to the marked moduli space, which was introduced in \cite{Raman:2024fcv} in the context of the Swampland program and later used in \cite{Delgado:2024skw} to argue for the emergence of dualities.
There, the points of the marked moduli space correspond to points in moduli space together with a choice of basis of observables, in our setting given by the integral middle cohomology.
The physical moduli space is then recovered by quotienting by the duality group.

\subsubsection{Fuchsian group and boundary of $\mathbb{H}$}

Elements of a Fuchsian group are classified by their fixed points in $\overline{\mathbb{H}} = \mathbb{H} \cup \partial \mathbb{H}$.
Elements with a unique fixed point in the interior of $\mathbb{H}$ are called elliptic and have finite order.
On the other hand, elements with a unique fixed point on the boundary $\partial \mathbb{H}$ are called parabolic and have infinite order.
Consequently, the orbifold points $o_a$ are the images of fixed points of elliptic elements of $\Gamma_\mathrm{F}$, while the punctures $p_i$ correspond to fixed points of parabolic elements (see Figure~\ref{fig:fundz} in Section~\ref{sec:sextic} for an illustration in the example of the mirror sextic).
Finally, non-identity elements of $\Gamma_\mathrm{F}$ that are neither elliptic nor parabolic are called hyperbolic and have two distinct fixed points on $\partial \mathbb{H}$.

The standard hyperbolic metric on $\mathbb{H}$ is given by
\begin{equation}\label{eq:hyperbolicmetric}
    \mathrm{d}s_{\mathbb{H}}^2 = \frac{\mathrm{d} z\,\mathrm{d}\bar z} {(\Im z)^2} = \frac{\mathrm{d}x^2+\mathrm{d}y^2}{y^2} \,,\qquad \text{for} \quad z = x + i y \in \mathbb{H} \,.
\end{equation}
It is invariant under $\mathrm{PSL}(2,\mathbb{R})$ and therefore defines a complete hyperbolic metric on $\mathcal{M}_{X_4}$.%
\footnote{This metric is in general distinct from the previously introduced Weil-Petersson metric on $\mathcal{M}_{X_4}$.}
With respect to this metric, an orbifold point $o_a$ of order $k_a$ gives a conical point with angle $2 \pi / k_a$, while each puncture $p_i$ corresponds to a cusp.
Both are finite volume singularities, i.e.\ they have neighborhoods of finite hyperbolic volume.
Since there are only finitely many such points and the remaining part of $\mathcal{M}_{X_4}$ is compact, $\mathcal{M}_{X_4}$ has finite hyperbolic volume.

Since $\Gamma_\mathrm{F}$ acts properly discontinuously on $\mathbb{H}$, the orbit $\Gamma_\mathrm{F} \cdot z$ of any $z \in \mathbb{H}$ has no accumulation points in the interior of $\mathbb{H}$.
On the other hand, its accumulation points on the boundary form the limit set $\Lambda(\Gamma_\mathrm{F})$,
\begin{equation}
    \Lambda(\Gamma_\mathrm{F}) \equiv \overline{\Gamma_\mathrm{F} \cdot z} \cap \partial \mathbb{H} \,.
\end{equation}
Equivalently, $x \in \partial \mathbb{H}$ belongs to $\Lambda(\Gamma_\mathrm{F})$ if there is a sequence $\gamma_n \in \Gamma_\mathrm{F}$ such that
\begin{equation}\label{eq:arbitraryboundarylimit}
    \gamma_n \cdot z \rightarrow x \,.
\end{equation}
Importantly, this definition is independent of the choice of $z$.

The Fuchsian group $\Gamma_\mathrm{F}$ is said to be of the first kind if
\begin{equation}\label{eq:fuchsianfirstkind}
    \Lambda(\Gamma_\mathrm{F}) = \partial \mathbb{H} \,,
\end{equation}
i.e., if its limit set is the entire boundary of $\mathbb{H}$.
Since $\mathcal{M}_{X_4}$ has finite hyperbolic volume, $\Gamma_\mathrm{F}$ is of the first kind \cite{katok1992fuchsian}.
In other words, one can reach every point $x \in \partial \mathbb{H}$ by choosing an arbitrary starting point $z \in \mathbb{H}$ and then approaching $x$ along a suitably chosen sequence $(\gamma_n \cdot z)$ as in \eqref{eq:arbitraryboundarylimit}.
This property will be important in our later discussion of the continuity of the periods at the boundary in Section~\ref{sec:continuity}.

The boundary of the upper half plane consists of the real line and the point at infinity,
\begin{equation}
    \partial \mathbb{H} = \mathbb{R} \cup \{\infty\} \,.
\end{equation}
For a Fuchsian group of the first kind, the orbit $\Gamma_\mathrm{F} \cdot \infty$ is dense in $\partial \mathbb{H}$ \cite{Beardon1983}.

After conjugating $\Gamma_\mathrm{F}$ by an element of
$\mathrm{PSL}(2,\mathbb{R})$, the point at infinity can be chosen to represent any of the punctures $p_i$.
Therefore, we may assume it to correspond to a large complex structure puncture, i.e., by a slight abuse of notation,
\begin{equation}\label{eq:LCSinfinity}
    \phi(\infty) = p_\mathrm{LCS} \,.
\end{equation}
Unless stated otherwise, we adopt this convention for the remainder of this paper.
We further refer to the points in $(\Gamma_\mathrm{F}\cdot\infty)\cap\mathbb{R}$ as \emph{$\infty$-equivalent} and to those in $\mathbb{R}\setminus(\Gamma_\mathrm{F}\cdot\infty)$ as \emph{$\infty$-inequivalent}.

\subsection{The periods on the covering space}
\label{sec:periodscover}

The linear transformation law \eqref{eq:monodromymap} of the periods under analytic continuation along loops $\gamma \in \pi^\mathrm{orb}_1(\mathcal{M}_{X_4})$ defines the monodromy (or period) representation
of the orbifold fundamental group,\footnote{Using the identification $\pi^\mathrm{orb}_1(\mathcal{M}_{X_4}) \simeq \Gamma_{\mathrm F}$, we also regard $\rho$ as a representation of $\Gamma_{\mathrm F}$.}
\begin{equation}\begin{aligned}
    \rho : \pi^\mathrm{orb}_1(\mathcal{M}_{X_4}) \,&\to\, \mathrm{GL}(b_H(X_4), \mathbb{R})  \\
    \gamma \,&\mapsto\, M_\gamma \,.
\end{aligned}\end{equation}
Its image is called the monodromy group $\Gamma$,
\begin{equation}
\Gamma =
\rho\bigl(\pi^\mathrm{orb}_1(\mathcal{M}_{X_4})\bigr) \,.
\end{equation}
If $\rho$ is faithful, we thus have
\begin{equation}
    \Gamma \simeq \Gamma_\mathrm{F} \simeq \pi^\mathrm{orb}_1(\mathcal{M}_{X_4}) \,.
\end{equation}
This is the case for the examples considered here.%
\footnote{In this case, the monodromy group plays the role of the duality group of \cite{Raman:2024fcv, Delgado:2024skw}, and the upper half-plane $\mathbb{H}$ coincides with their marked moduli space. For Calabi-Yau threefold examples, see e.g.\ \cite{Doran:2005gu, Brav_2014}.}
By \eqref{eq:etamonodromy}, the monodromy group preserves the intersection
form $\eta$, and hence $\Gamma \subset \mathrm{O}(b_H(X_4)-2,2)$.

Let $\mathcal{F}\subset\mathbb{H}$ be a fundamental domain for the action of $\Gamma_\mathrm{F}$ (see Figure~\ref{fig:fundz} in Section~\ref{sec:sextic} for an illustration).
The covering map identifies $\mathcal{F}$ with the moduli space $\mathcal{M}_{X_4}$.
Therefore, choosing a branch of the locally defined periods on $\mathcal{M}_{X_4}$ and pulling it back with $\phi$ defines the period vector $\Pi$ on $\mathcal{F}$.
The transformation of $\Pi$ under the monodromy group then extends it to the whole covering space,
\begin{equation}\label{eq:equivariance}
    \Pi(\gamma\cdot z) = \rho(\gamma)\,\Pi(z)\,,
\end{equation}
for $\gamma \in \Gamma_\mathrm{F}$ and $z \in \mathbb{H}$.
The period vector therefore defines a $\Gamma_\mathrm{F}$-equivariant holomorphic map
\begin{equation}
    \label{eq:generalperiods}
    \Pi:\mathbb{H} \rightarrow\mathbb{C}^{b_H(X_4)} \,.
\end{equation}

\subsubsection{Nilpotent orbit approximation}
\label{sec:nilpotentcovering}

We now translate the nilpotent orbit approximation \eqref{eq:nilpotent} to the upper-half-plane coordinate $z$.
The point $z = \infty$ corresponds to a puncture $p$ and its stabilizer in $\Gamma_\mathrm{F}$ is generated by a parabolic deck transformation
\begin{equation}
    \label{eq:zmonod}
    \gamma_\infty \cdot z = z + T \,,
\end{equation}
where $T > 0$.
Note that we may always choose $T = 1$ by rescaling $z$, but decide to keep it arbitrary for later convenience.
For simplicity, we assume that the local monodromy around $p$ is unipotent, i.e.\ $k = 1$ in \eqref{eq:Ni}.
This is, for example, the case for an LCS point.
The general case follows straightforwardly.

The invariant coordinate
\begin{equation}
    q = e^{2 \pi  i z /T} \,,
\end{equation}
is a local coordinate at this puncture.
The corresponding logarithmic coordinate, as defined in \eqref{eq:deft} reads%
\footnote{\label{note:tshift}
For another local coordinate $\tilde q$ at $p$ the corresponding logarithmic coordinate $\tilde t$ satisfies
\begin{equation*}
    \tilde t = t + t_0 + \mathcal{O}(q) = \frac{z}{T} + t_0 + \mathcal{O}(q) \,.
\end{equation*}
The shift amounts to a redefinition $\tilde A(\tilde q) = e^{(t-\tilde t) N} A(q)$ in the nilpotent orbit expansion, which otherwise retains its form.
}
\begin{equation}\label{eq:LCSt}
    t = \frac{1}{2 \pi i} \log q = \frac{z}{T} \,.
\end{equation}
Substituting \eqref{eq:LCSt} into the nilpotent orbit expansion \eqref{eq:nilpotent}, we obtain
\begin{equation}\label{eq:nilpotentcovering}
    \Pi(z) = e^{z N/T} A(q) 
\end{equation}
for sufficiently large $y = \Im z$.
Here $A(q)$ is the respective holomorphic function in \eqref{eq:nilpotent} with expansion \eqref{eq:nilpotentexpansion}, and $N$ is the nilpotent matrix satisfying
\begin{equation}\label{eq:NLCS}
    e^{N} = M_\infty = \rho(\gamma_\infty) \,.
\end{equation}

In particular, if we take the point at infinity to correspond to an LCS puncture, as in \eqref{eq:LCSinfinity}, then in the basis \eqref{eq:genperiods}, and for a suitable choice of local coordinate (see footnote~\ref{note:tshift}), the periods admit the polynomial approximation
\begin{equation}\label{eq:polynomialperiodsH}
    \Pi(z) = P_\mathrm{LCS}(z / T) + e^{zN/T} \mathcal{O}(q) \,,
\end{equation}
with $P_\mathrm{LCS}(t) = e^{tN} A(0)$ given in \eqref{eq:genperiods}.
It is important to note that this approximation is in general valid only as $y \to \infty$, but not in an arbitrary $|z| \rightarrow \infty$ limit.

\subsubsection{Projective and normalized periods}
\label{sec:projectiveperiods}

The nilpotent orbit approximation \eqref{eq:polynomialperiodsH} gives a simple expression for the periods near the boundary point at $z = \infty$.
We also wish to understand their behavior on the rest of the boundary $\partial \mathbb{H}$, namely the real line $\mathbb{R}$.

For an $\infty$-equivalent point $x \in \mathbb{R}$, i.e.\ $x \in \Gamma_\mathrm{F} \cdot \infty$, there exists $\gamma_x \in \Gamma_\mathrm{F}$ such that
\begin{equation}
    \gamma_x \cdot \infty = x \,.
\end{equation}
The equivariance \eqref{eq:equivariance} of the periods suggests computing a boundary value $\Pi(x)$ by taking the limit
\begin{equation}\label{eq:tentativeboundarylimit}
    \lim_{\Im z \rightarrow \infty} \Pi(\gamma_x \cdot z) = \rho (\gamma_x) \lim_{\Im z \rightarrow \infty} \Pi (z) \,.
\end{equation}
However, it follows from  their polynomial expansion \eqref{eq:polynomialperiodsH} that the periods diverge in this limit.
Consequently, $\Pi(x)$ is not well-defined as a finite boundary value.

We may attempt to remedy this behavior by introducing the projective period map
\begin{equation}\label{eq:projectiveperiods}
    \mathcal{P} : \mathbb{H} \rightarrow \mathbb{CP}^{b_H(X_4)-1} \,, \qquad \mathcal{P}(z) = \left[\Pi(z) \right] \,,
\end{equation}
with $\left[v \right] \in \mathbb{CP}^{b_H(X_4)-1}$ denoting the point represented by the homogeneous coordinates $v \in \mathbb{C}^{b_H(X_4)} \setminus \{0\}$.

This is a sensible definition because the holomorphic $(4,0)$-form $\Omega$, and hence the period vector $\Pi(z)$, are defined only up to multiplication by a non-vanishing holomorphic function, corresponding to a K\"ahler transformation.

Although the projective period map is insensitive to the overall growth of the periods, it is often useful to represent it by a vector that remains finite near the boundary.
To this end, we introduce the normalized period map
\begin{equation}\label{eq:normalizedperiods}
    \widehat \Pi^I_\ell(z) = \frac{\Pi^I(z)}{\ell_J \Pi^J(z)} \,,
\end{equation}
for a constant vector $\ell \in \mathbb{C}^{b_H(X_4)} \setminus \{0\}$.
It satisfies $\ell_I \widehat \Pi^I_\ell(z) = 1$ and is invariant under K\"ahler transformations.
Wherever $\ell_J \Pi^J(z) \neq 0$, it represents the same point in projective space, i.e.\ $\mathcal{P}(z) = [\Pi(z)] = [\widehat \Pi(z)]$.
As we will see below, there need not exist a single choice of $\ell_I$ for which $\ell_J \Pi^J$ is nonzero on the entire upper half-plane $\mathbb{H}$.
Consequently, the image of $\mathcal{P}$ need not be contained in a single affine patch of $\mathbb{CP}^{b_H(X_4)-1}$.

For the case of interest here, where the point at infinity corresponds to an LCS singularity, it is natural to choose $\ell$ such that
\begin{equation}
    \lim_{\Im z \rightarrow \infty} \ell_I \Pi^I(z) = 1 \,,
\end{equation}
i.e.\ to normalize by the constant period in the polynomial approximation.
In the basis \eqref{eq:genperiods}, this corresponds to
\begin{equation}\label{eq:ellLCS}
    \ell_\mathrm{LCS} = \left(1,0, \dots, 0 \right) \,.
\end{equation}
This normalization ensures that $\widehat{\Pi}^I_{\ell_\mathrm{LCS}}$ and $\Pi^I$ have the same polynomial approximation near infinity.

The equivariant transformation behavior \eqref{eq:equivariance} of the period map under monodromy transformations induces the projective transformation law
\begin{equation}\label{eq:projectiveequiv}
    \mathcal{P}(\gamma \cdot z) = \rho(\gamma) \cdot \mathcal{P}(z) \equiv \left[ \rho(\gamma)^I{}_J \Pi^J(z) \right] \,, \qquad \gamma \in \Gamma_\mathrm{F} \,.
\end{equation}
Similarly, the normalized periods transform non-linearly under a monodromy transformation,
\begin{equation}\label{eq:normperiodsequiv}
    \widehat{\Pi}_\ell^I(\gamma \cdot z) = \frac{\rho(\gamma)^I{}_J\Pi^J(z)}{\ell_K \, \rho(\gamma)^K{}_L\Pi^L(z)} \,.
\end{equation}
As in \eqref{eq:tentativeboundarylimit}, for an $\infty$-equivalent point $x \in \mathbb{R}$, we may attempt to compute the boundary value $\widehat{\Pi}_\ell(x)$ by taking the limit
\begin{equation}
    \lim_{\Im z \rightarrow \infty} \widehat{\Pi}_\ell(\gamma_x \cdot z) \,.
\end{equation}
As we discuss in more detail below, for certain Calabi-Yau fourfolds, these limits exist:
whenever the numerator in \eqref{eq:normperiodsequiv} grows no faster than the denominator, $\widehat{\Pi}_{\ell_\mathrm{LCS}}(x)$ remains finite at $\infty$-equivalent points on the real line.

We note en passant that the behavior of the holomorphic period map on the upper half-plane is reminiscent of strong quantum modularity \cite{MR2757599}.
Upon suitable normalization,%
\footnote{The normalization \eqref{eq:normalizedperiods} implies the non-linear transformation law \eqref{eq:normperiodsequiv}.
Alternatively, one could divide by a weight-four Eisenstein series $E_{4,\Gamma_\mathrm{F}}$, which gives a vector-valued modular transformation law of weight $-4$ with vanishing defect.
Either normalization may introduce poles in the interior of 
$\mathbb{H}$.
}
the nilpotent orbit expansion \eqref{eq:nilpotent} defines a formal power series at each $\infty$-equivalent point on the real axis,
which agrees at all orders with the asymptotic expansion of the periods when approaching these points vertically.
Moreover, equivariance provides the required vector-valued modular transformation law.

\subsection{Continuity of the periods on the boundary}
\label{sec:continuity}

Finiteness of the normalized period vector $\widehat \Pi(z)$ on the boundary $\partial \mathbb{H}$ of the covering space does not by itself guarantee that it extends continuously to $\overline{\mathbb{H}} = \mathbb{H} \cup \partial \mathbb{H}$.
Continuity at a boundary point $x\in\partial\mathbb H$ requires that, for every sequence $z_n \in \mathbb{H}$ converging to $x$,
\begin{equation}
    \lim_{n\to \infty} \widehat \Pi(z_n) = \widehat \Pi(x) \,.
\end{equation}
In particular, the limit must be independent of how $x$ is approached.

As we will see, the normalized periods $\widehat \Pi^I$ and also the projective periods $\mathcal{P}$ are not guaranteed to admit such a continuous extension.

\subsubsection{Relatively Anosov representations}

To formulate conditions under which such an extension exists, we first introduce the notion of a relatively Anosov representation, following \cite{Zhu2022RelativelyAR, AIF_2021__71_5_2169_0}.

Let $\Gamma_\mathrm{F} \subset \mathrm{PSL}(2, \mathbb{R})$ be a Fuchsian group of finite covolume.%
\footnote{A Fuchsian group $\Gamma_\mathrm{F}$ is said to have finite covolume if the hyperbolic area of $\mathbb{H}/ \Gamma_\mathrm{F}$ is finite.
This implies that $\Gamma_\mathrm{F}$ is of the first kind.
The converse is not true in general.}
A representation $\rho : \Gamma_\mathrm{F} \to \mathrm{GL}(d, \mathbb{C})$ is called \emph{$P_1$-relatively Anosov} if there exist continuous boundary maps
\begin{equation}\label{eq:xidefinition}
    \xi : \partial \mathbb{H} \,\to\, \mathrm{Gr}(1,\mathbb{C}^d) \simeq \mathbb{CP}^{d-1} \,,\qquad
    \xi^* : \partial \mathbb{H} \,\to\, \mathrm{Gr}(d-1, \mathbb{C}^d) \simeq (\mathbb{CP}^{d-1})^*
\end{equation}
that satisfy the following three properties:
\begin{itemize}
\item \textbf{$\rho (\Gamma_\mathrm{F})$-equivariance:} For any $\gamma \in \Gamma_\mathrm{F}$,
\begin{equation}
    \xi (\gamma \cdot x) = \rho(\gamma) \cdot \xi (x) \,,
\end{equation}
with the projective action of $\rho(\gamma)$ as in \eqref{eq:projectiveequiv},
and similarly for $\xi^*$.

\item \textbf{Transversality:} For any $x, y \in \partial \mathbb{H}$ with $x \neq y$,
\begin{equation}
    \label{eq:transversality}
    \xi (x) \oplus \xi^* (y) = \mathbb{C}^d\,,
\end{equation}
equivalently, $\xi (x) \notin \xi^*(y)$, viewing $\xi^*(y)$ as a projective hyperplane in $\mathbb{CP}^{d-1}$.

\item \textbf{Strong dynamics preservation:} For any $z_0 \in \mathbb{H}$ and $x, y \in \partial \mathbb{H}$, if $(\gamma_n) \in \Gamma_\mathrm{F}$ is a sequence such that $\lim \gamma_n \cdot z_0 = x$ and $\lim \gamma_n^{-1} \cdot z_0 = y$, then for any compact subset
$K \subset \mathbb{CP}^{d-1} \setminus \xi^* (y)$,
\begin{equation}
    \label{eq:strongdynpre}
    \lim_{n \to \infty} \rho(\gamma_n) \cdot K = \xi (x)
\end{equation}
uniformly.
This notion of convergence is independent of the choice of $z_0$.
\end{itemize}
An explicit construction of the boundary maps $\xi$ and $\xi^\ast$ in terms of the representation $\rho$ is summarized in Appendix~\ref{app:boundarymaps}.

A simple example of a relatively Anosov representation is given by the defining action of $\Gamma_\mathrm{F}$ on $\mathbb{H}$.
Identifying $\overline{\mathbb{H}}$ with a subset of $\mathbb{CP}^1$ via
\begin{equation}
    z \mapsto [z:1] \,, \qquad \infty \mapsto [1:0] \,,
\end{equation}
M\"obius transformations \eqref{eq:moebius} act linearly on homogeneous coordinates. Under the canonical identification $\mathrm{Gr}(1, \mathbb{C}^2) \simeq \mathbb{CP}^{1} \simeq  (\mathbb{CP}^{1})^*$, both boundary maps $\xi$ and $\xi^*$ are given by the identity.
While $\rho$-equivariance and transversality are immediate, strong dynamics preservation reduces to a standard convergence property of Fuchsian groups:
if $\gamma_n\cdot z_0\to x \in \partial \mathbb{H}$ and $\gamma_n^{-1}\cdot z_0\to y \in \partial \mathbb{H}$, then $\gamma_n$ converges uniformly to $x$ on compact subsets of $\mathbb{CP}^1\setminus\{y\}$.

\subsubsection{Sufficient conditions for a continuous boundary extension}
\label{sec:continuityconditions}

With this preparation, we can now give sufficient conditions under which a $\rho$-equivariant map from the upper half-plane to the projective period space extends continuously to $\overline{\mathbb{H}}$.

We consider a continuous, not necessarily holomorphic map%
\footnote{In physics notation, a non-holomorphic map is often denoted by $F = F(z, \bar z)$.
Here, to simplify notation, we sometimes also write $F = F(z)$ even if it is not holomorphic.}
\begin{equation}
    \label{eq:def_F_fun}
    F : \mathbb{H} \rightarrow \mathbb{CP}^{b_H(X_4)-1} \,,
\end{equation}
and denote a homogeneous representative of $F(z)$ by the vector $F^I(z)$, so that $F(z) = [F^I(z)]$.
We assume that $F$ is equivariant with respect to the period representation $\rho$, i.e.
\begin{equation}
    F(\gamma \cdot z ) = \left[\rho(\gamma)^I{}_J F^J(z) \right] \,, \qquad \gamma \in \Gamma_\mathrm{F} \,.
\end{equation}
In Appendix~\ref{app:anosov}, we show that $F$ extends continuously to $\overline{\mathbb{H}}$ if the following three conditions hold:
\begin{enumerate}[(i)]
    \item \label{item:anosov}
    The monodromy representation $\rho$ is relatively Anosov.
    This implies the existence of a continuous boundary map $\xi$
as in \eqref{eq:xidefinition}.

    \item \label{item:limitcondition}
    For each puncture $p_i \in \partial \mathcal{M}_{X_4}$, choose a lift $x_i \in \partial\mathbb{H}$ and $\sigma_i \in \mathrm{PSL}(2, \mathbb{R})$ such that $\sigma_i \cdot \infty = x_i$.
    We require that
    \begin{equation}\label{eq:limitcondition}
        \lim_{\Im z \rightarrow \infty} F(\sigma_i \cdot z) = \xi(x_i) \,,
    \end{equation}
    without any restriction on $\Re z$.
    By the $\rho$-equivariance of $F$ and $\xi$, this condition is independent of the choice of the lift $x_i$.
    
    \item \label{item:nullvectorcondition}
    For every nonzero vector $g \in \mathbb{R}^{b_H(X_4)}$ that is null with respect to the intersection form \eqref{eq:eta}, i.e.\ $g_I \eta^{IJ} g_J = 0$,
    one has
    \begin{equation}\label{eq:nullvectorcondition}
        g_I F^I(z) \neq 0 
    \end{equation}
    for all $z \in \mathbb{H}$.
\end{enumerate}
Then the unique continuous extension of $F$ is the map $\overline{F} : \overline{\mathbb{H}} \to \mathbb{CP}^{b_H(X_4)-1}$ defined by
\begin{equation}\label{eq:overlineF}
    \overline F(z)=
    \begin{cases}
        F(z) \,, & z\in\mathbb{H} \,,\\[0.2mm]
        \xi(z) \,, & z\in\partial\mathbb{H} \,.
    \end{cases}
\end{equation}
Since the period representation is real, $\xi$ takes values in $\mathbb{RP}^{b_H(X_4)-1}$, as follows from the explicit formula \eqref{eq:xiandxistar}.

In general, even if $F$ is smooth on the interior, the boundary map $\xi$ need not be differentiable.
However, if $F$ is holomorphic, $\xi$ is necessarily real analytic and hence $\xi \in C^\infty$.
See Appendix~\ref{app:dxicontinuity} for more details.

\subsubsection{The boundary map}

Assuming that the period representation $\rho$ is relatively Anosov, we discuss the boundary map $\xi$ in more detail.
Here we only summarize its most important properties, and refer to Appendix~\ref{app:boundarymaps} for a more detailed discussion and all derivations.

First, we note that
\begin{equation}\label{eq:nulltransversality}
    \xi^I(x) \eta_{IJ} \xi^J(y) = 0 \qquad \Longleftrightarrow \qquad x = y \,.
\end{equation}
In particular, at every point $x \in \partial \mathbb{H}$, $\xi^I(x)$ is null with respect to the intersection form $\eta_{IJ}$.
This property follows from transversality and Lemma~\ref{lemma:etaorthogonality} for $x \neq y$, and from Corollary~\ref{cor:xinull} for $x = y$.

Now, consider a puncture $p_i \in \partial \mathcal{M}_{X_4}$ and choose a lift $x_i \in \partial \mathbb{H}$.
Let $\gamma_i \in \Gamma_\mathrm{F}$ be a generator of the stabilizer of $x_i$.
As in \eqref{eq:Ni}, we define the nilpotent matrix
\begin{equation}
    N_i = \log \rho(\gamma_i)^{k_i} \,,
\end{equation}
with $k_i$ such that $\rho(\gamma_i)^{k_i}$ is unipotent.
We denote the degree of nilpotency of $N_i$ by $r_i$, as defined in \eqref{eq:degreeofnilpotency}.

According to Proposition~\ref{prop:boundary}, if $\rho$ is relatively Anosov, then
$\operatorname{rank} N_i^{r_i} = 1$,
and
viewing the boundary maps as Grassmannian-valued, 
\begin{equation}\label{eq:nilpotentboundaryvalue}
    \xi(x_i) = \operatorname{Im} N_i^{r_i} \,, \qquad \xi^*(x_i) = \ker N_i^{r_i} \,.
\end{equation}
Equivalently, choosing a vector $v_i \in \mathbb{R}^{b_H(X_4)}$ spanning the image, $\operatorname{Im} N_i^{r_i} = \left< v_i \right>$, we have
\begin{equation}
     \qquad \xi(x_i) = [v_i] \,,
\end{equation}
where $\xi(x_i)$ is viewed as a point in projective
space.

With a suitable normalization of the real homogeneous representative $\xi^I(x_i)$, Corollary~\ref{cor:Nxixi} gives
\begin{equation}\label{eq:Nxixi}
    \left( N_i^{r_i} \right)^I{}_J = \pm \, \xi^I(x_i) \xi_J(x_i) \,.
\end{equation}

\subsubsection{Relatively Anosov monodromy representations of Calabi-Yau fourfolds}
\label{sec:relanosovfourfolds}

Not every Calabi-Yau manifold exhibits a monodromy representation that is relatively Anosov.
In particular, the requirement 
\begin{equation}
    \label{eq:rk_requirement}
    \operatorname{rank} N_i^{r_i} = 1 \,,
\end{equation}
is not always satisfied at every puncture $p_i$ and
strongly constrains the types of infinite-order singularities that can occur in a moduli space with a relatively Anosov monodromy representation $\rho$.

Boundary loci of complex structure moduli spaces can be classified in terms of their limiting mixed Hodge structures.
For threefolds, such a classification was given in \cite{kerr2019polarizedrelationshorizontalsl2s, Grimm:2018cpv}, and for fourfolds in \cite{Grimm:2019ixq}.%
\footnote{The one-modulus case for fourfolds was also discussed in more detail in \cite{Grimm:2025cpq}.}
Using Table 3.1 of \cite{Grimm:2019ixq}, we find that, for an arbitrary number of complex structure moduli, only singularities of type $\mathrm{I}_{1, a'}$, $\mathrm{II}_{1, b'}$, and $\mathrm{V}_{e, e'}$ satisfy \eqref{eq:rk_requirement}.
Requiring $h^{3,1}=1$ further restricts the possibilities to two types:
\begin{itemize}
    \item $\mathrm{V}_{1, 1}$, corresponding to an LCS singularity,
    \item $\mathrm{I}_{1, 1}$, a finite-distance, infinite-order singularity, similar to a conifold singularity of a threefold.
\end{itemize}
Nevertheless, there exist single-modulus Calabi-Yau fourfolds with other types of punctures.
Examples can be found in \cite{vandeHeisteeg:2024lsa}, where some moduli spaces contain a $\mathrm{II}_{0, 0}$ puncture, referred to as a CY3-point.
The corresponding monodromy representations are therefore not relatively Anosov.%
\footnote{For these examples, we find numerical evidence that the putative boundary map $\xi$ is discontinuous, in agreement with $\rho$ not being relatively Anosov.}

The requirement that all infinite-order punctures be of type $\mathrm{V}_{1,1}$ or $\mathrm{I}_{1,1}$ is necessary but not sufficient for the monodromy representation to be relatively Anosov.
For hypergeometric Calabi-Yau three- and fourfolds, sufficient conditions were obtained in \cite{filip2022uniformizationweight3variations, MR4615758}.
These conditions are satisfied for the mirror Fermat sextic, our main example in Section~\ref{sec:sextic}, and for some other fourfolds in the list of \cite{vandeHeisteeg:2024lsa}.%
\footnote{\label{note:filip}The application to the mirror sextic relies on an updated unpublished version of \cite{filip2022uniformizationweight3variations}.
We would like to thank Simion Filip for sharing this version with us and for helpful exchanges.}

\subsubsection{Continuity of the holomorphic periods}

Given a Calabi-Yau fourfold whose period representation $\rho$ is relatively Anosov, we can ask if its holomorphic periods, or more specifically the projective periods $\mathcal{P}$ defined in \eqref{eq:projectiveperiods}, extend continuously to the boundary of the upper half-plane.
To establish continuity on $\overline{\mathbb{H}}$, it remains to verify whether $\mathcal{P}$ satisfies the second and third condition stated in Section~\ref{sec:continuityconditions}.

For a given puncture, condition~(\ref{item:limitcondition}) can be tested using the nilpotent orbit expansion \eqref{eq:nilpotent} and comparing the resulting limits with expression \eqref{eq:nilpotentboundaryvalue} for the corresponding value of the boundary map $\xi$.
Concretely, from the expansion \eqref{eq:nilpotent} and \eqref{eq:nilpotentexpansion} one has
\begin{equation}\label{eq:generalconditioniilimit}
        \lim_{\Im z \rightarrow \infty} \mathcal{P}(\sigma \cdot z) = 
        \lim_{\Im z \rightarrow \infty} \left[e^{z N /T} \sum_{n = 0}^\infty a_n e^{2 \pi i n z /T} \right] \,.
\end{equation}
The existence and value of this limit depend on the leading coefficients $a_n$ and therefore on the type of puncture in question.
The relevant information from asymptotic Hodge theory is established in Appendix~\ref{app:asymptotichodge}.

At an LCS (type $\mathrm{V}_{1,1}$) singularity, we have $N^5=0$ and $N^4 a_0 \neq 0$.
Therefore, \eqref{eq:generalconditioniilimit} yields
\begin{equation}
        \lim_{\Im z \rightarrow \infty} \mathcal{P}(\sigma \cdot z) = \left[N^4 a_0 \right] = \xi(x_i) \,,
\end{equation}
where, in the last step, we used \eqref{eq:nilpotentboundaryvalue}.
On the other hand, at a type $\mathrm{I}_{1,1}$ singularity, we have
$N a_{0}=0$ and $a_{0} \notin\operatorname{im}N$.
Hence,
\begin{equation}
        \lim_{z \rightarrow i \infty} \mathcal{P}(\sigma \cdot z) = \left[a_0 \right] \neq \xi(x_i) \,.
\end{equation}
Moreover, denoting the first coefficient that is not proportional to $a_0$ by $a_M$, we have $N^2 a_M \neq 0$, and therefore the limit \eqref{eq:generalconditioniilimit} with unrestricted $\Re z$ does not exist.
We conclude that condition~(\ref{item:limitcondition}) is only satisfied for LCS singularities, but not for $\mathrm{I}_{1,1}$ singularities.

Since we assume the existence of at least one LCS singularity, any continuous boundary extension must agree with $\xi$ at any lift of this puncture, i.e.\ on a dense subset of the boundary $\partial \mathbb{H}$.
Continuity therefore implies that $\xi$ is the only possible boundary map.
Condition~(\ref{item:limitcondition}) is hence also necessary for the existence of a continuous boundary extension.
Consequently, in the presence of an additional $\mathrm{I}_{1,1}$ singularity, the projective period map $\mathcal{P}$ does not admit a continuous extension to $\overline{\mathbb{H}}$.

We now turn our attention to condition~(\ref{item:nullvectorcondition}).
This condition is not satisfied by the holomorphic periods, independently of the types of singularities present in the moduli space.
Indeed, by the discussion below \eqref{eq:eta}, the restriction of the intersection form $\eta$ to the orthogonal complement
\begin{equation}
    \Pi(z)^\perp = \left\{ g \in \mathbb{R}^{b_H(X_4)} \,\middle|\, g^I \eta_{IJ} \Pi^J(z) = 0 \right\} \subseteq H^{3,1} \oplus  H^{2,2}_H \oplus H^{1,3}
\end{equation}
has signature $(h^{2,2}_H, 2)$.
Therefore, at any point $z \in \mathbb{H}$, $\Pi(z)^\perp$ contains a non-zero real vector that is null with respect to $\eta$.
It is hence not possible to establish the existence of a continuous boundary extension of $\mathcal{P}$ using the sufficient conditions given in Section~\ref{sec:continuityconditions}.

However, failure of condition~(\ref{item:nullvectorcondition}) alone does not imply the absence of a continuous boundary extension.
As mentioned at the end of Section~\ref{sec:continuityconditions} and discussed in more detail in Appendix~\ref{app:dxicontinuity}, holomorphicity of $\mathcal{P}$ implies that any continuous boundary extension requires the boundary map $\xi$ to be smooth.
Moreover, in Appendix~\ref{app:C1test} we develop a necessary condition for $\xi$ to be $C^1$, expressed in terms of the eigenvalues of $\rho(\gamma)$ for a hyperbolic element $\gamma \in \Gamma_\mathrm{F}$.
In Section~\ref{sec:sexticC1} we apply this test to a concrete example, the mirror Fermat sextic, and show that it fails.
This implies that, for this example, no continuous boundary extension of $\mathcal{P}$ exists.

We finally note that the indefinite signature of $\Pi(z)^\perp$ neither establishes nor excludes the existence of a normalized period vector $\widehat{\Pi}^I_\ell$, as introduced in \eqref{eq:normalizedperiods}, that is non-singular on the entire upper half-plane.

\subsection{The (3,1)-periods}
\label{sec:31per}

To compute the winding number of $DW$ along the boundary $\partial \mathbb{H}$ as an index for flux vacua, as proposed in Section~\ref{sec:preliminaries}, $DW$ must extend continuously to $\partial \mathbb{H}$ and be non-singular on the interior $\mathbb{H}$.
As discussed in the previous subsection, neither of these two requirements is in general satisfied by the normalized holomorphic periods, and neither is guaranteed for the superpotential $W$.

While this is discouraging, these conclusions do not directly carry over to the actual object of interest, the K\"ahler-covariant derivative $DW$ of the superpotential, defined in \eqref{eq:fterm}.
$DW$ is a linear combination of the periods of the $(3,1)$-form $D\Omega$.
These periods are non-holomorphic and given by the K\"ahler-covariant derivatives $D\Pi^I$ of the holomorphic periods $\Pi^I$, see \eqref{eq:development}.
We therefore have to repeat our previous analysis for the $(3,1)$-periods $D\Pi^I$.

As for the holomorphic periods in \eqref{eq:projectiveperiods}, we introduce the projective period map
\begin{equation}\label{eq:projective31periods}
    F : \mathbb{H} \rightarrow \mathbb{CP}^{b_H(X_4)-1} \,, \qquad F(z) = \left[D\Pi(z) \right] \,.
\end{equation}
This map is independent of the choice of local holomorphic coordinate on $\mathbb{H}$ and invariant under K\"ahler transformations, since either multiplies all components of $D\Pi$ by the same nonzero factor.

\subsubsection{Nilpotent orbit expansion}

The nilpotent orbit expansion \eqref{eq:nilpotent}, \eqref{eq:nilpotentcovering} of the holomorphic $(4,0)$ periods implies a similar expansion of the projective $(3,1)$-period map,
\begin{equation}\label{eq:nilpotent31}
    F = \left[ e^{z N / T} B(z, \bar z) \right] \,,
\end{equation}
where the vector-valued function $B(z,\bar z)$  admits the expansion 
\begin{equation}\label{eq:Bexpansion}
    B(z, \bar z) = \sum_{n,m \geq 0} b_{nm}(y) q^n \bar q^m \,, \qquad q = e^{2 \pi i z / T} \,.
\end{equation}
In contrast to the holomorphic case, the coefficients $b_{nm}(y)$ are in general polynomials in $y = \Im z$ rather than constants.
They can be computed explicitly from \eqref{eq:fterm} and \eqref{eq:linearWK}.
We first introduce
\begin{equation}
    V(z, \bar z) = e^{-K(z, \bar z)} = \Pi(z)^\dagger \eta \Pi(z) \,,
\end{equation}
so that
\begin{equation}\label{eq:FVPi}
    F(z, \bar z) = \left[ V \partial_z \Pi - \partial_z V \Pi \right] \,.
\end{equation}
Using the nilpotent orbit expansion of the holomorphic periods $\Pi(z)$, we obtain the expansion
\begin{equation}\label{eq:Vexpansion}
    V(z, \bar z) = \sum_{n,m \geq 0} v_{nm}(y) q^n \bar q^m \,, \qquad v_{nm}(y) = a_m^\dagger \eta e^{2iy N /T} a_n \,,
\end{equation}
where the $a_n$ are the vector-valued coefficients in the expansion \eqref{eq:nilpotentexpansion} of $\Pi(z)$.
The coefficients $v_{nm}(y)$ are scalars, and, since $N$ is nilpotent, polynomials in $y$.
Inserting this result and \eqref{eq:nilpotent}, \eqref{eq:nilpotentexpansion} back into \eqref{eq:FVPi} yields the expansion \eqref{eq:nilpotent31}, \eqref{eq:Bexpansion} with 
\begin{equation}\label{eq:bnm}
    b_{nm}(y) = \sum_{k=0}^n \left [ \frac{N + 2 (n - 2k) \pi i}{T} v_{km}(y)  - \frac{v'_{km}(y)}{2 i}\right] a_{n-k} \,.
\end{equation}
As in the case of the $(4,0)$ periods, this expansion is in general only valid at small $q$ and hence large $y$ but not at arbitrary large $|z|$.
However, as we will see below, unlike the holomorphic case, the $(3,1)$-periods allow for a related boundary expansion that is valid in every $|z| \to \infty$ limit, provided that the period representation is relatively Anosov.

\subsubsection{Boundary continuity}
\label{sec:continuity31}

To establish the existence of a continuous boundary extension of the projective $(3,1)$-periods, we test if the sufficient conditions stated in Section~\ref{sec:continuityconditions} are satisfied.
We assume that $X_4$ is a Calabi-Yau fourfold whose period representation $\rho$ is relatively Anosov, so that condition~(\ref{item:anosov}) holds.

To check condition~(\ref{item:limitcondition}), we need to evaluate the boundary limit \eqref{eq:limitcondition} of $F$ at each infinite-order puncture.
We proceed analogously to the $(4,0)$ case, and evaluate the limit using the nilpotent orbit expansion \eqref{eq:nilpotent31}, \eqref{eq:Bexpansion}, 
\begin{equation}
    \lim_{\Im z \rightarrow \infty} F(\sigma_i \cdot z) = \lim_{\Im z \rightarrow \infty} \left[e^{zN /T} \sum_{n,m\geq 0} b_{nm}(y) e^{2 \pi i (n z - m \bar z)/T}\right] \,.
\end{equation}
The existence and value of the limit depends on the leading coefficient $b_{nm}(y)$ in the expansion.
These coefficients can be computed using \eqref{eq:bnm}.
However, this computation requires additional information on the coefficients $a_n$ from asymptotic Hodge theory that is provided in Appendix~\ref{app:asymptotichodge}.

At an LCS (type $\mathrm{V}_{1,1}$) singularity, we show in Appendix~\ref{app:LCS31} that the leading coefficient $b_{00}(y)$ is non-vanishing and that $e^{z N/T} b_{00}(y) \propto y^3 z^3 \bar z N^4 a_0 + \mathcal{O}(y^2 |z|^4)$ for sufficiently large $y$.
Concretely, using \eqref{eq:app:LCSb00} we obtain
\begin{equation}
    \lim_{\Im z \to \infty} F(\sigma_i \cdot z) = \lim_{\Im z \to \infty} \left[e^{z N/T} b_{00}(y) \right] = \left[N^4 a_0 \right] = \xi(x_i) \,,
\end{equation}
where the last equality follows from \eqref{eq:nilpotentboundaryvalue}.

For type $I_{1,1}$ singularities, we denote the first coefficient in \eqref{eq:nilpotentexpansion} that is not proportional to $a_0$ by $a_M$.
Then, as we show in Appendix~\ref{app:I1131},
\begin{equation}
    b_{nm}(y) = 0 \,, \qquad n < M \,, m \geq 0 \,,
\end{equation}
and
\begin{equation}
    N^2 b_{M0}(y) \neq 0 \,,
\end{equation}
where $b_{M0}(y)$ is constant.
Therefore, $b_{M0}(y)$ is the first non-vanishing coefficient in the expansion, and, since $N^3 =0$, we find
\begin{equation}
    \lim_{\Im z \to \infty} F(\sigma_i \cdot z) = \left[N^2 b_{M0}(y) \right] = \xi(x_i) \,,
\end{equation}
using again \eqref{eq:nilpotentboundaryvalue} in the last step.
We conclude that condition~(\ref{item:limitcondition}) is satisfied for the two types of singularites that can occur for one-modulus Calabi-Yau fourfolds with a relatively Anosov period representation.

Lastly, we note that the orthogonal complement
\begin{equation}
    F(z)^\perp = \left\{ g \in \mathbb{R}^{b_H(X_4)} \,\middle|\, g^I \eta_{IJ} D\Pi^J(z) = 0 \right\} \subseteq H^{4,0} \oplus  H^{2,2}_H \oplus H^{0,4}
\end{equation}
has definite signature $(2+h^{2,2}_H,0)$.
This follows from the discussion below \eqref{eq:eta} and implies that $F(z)^\perp$ cannot contain any nonzero null vector with respect to $\eta$.
Therefore, condition~(\ref{item:nullvectorcondition}) is also satisfied, and $F$ extends continuously to $\overline{\mathbb{H}}$ with boundary values $\xi$.

Note that this does not contradict our previous observation that the boundary map $\xi$ is not $C^1$ for certain Calabi-Yau fourfolds, since $F$ is not holomorphic.

\subsubsection{Normalized representatives}
\label{sec:normalized31}

As for the holomorphic case in \eqref{eq:normalizedperiods}, we introduce the normalized representatives
\begin{equation}\label{eq:normalized31periods}
    F^I_\ell(z) = \frac{D\Pi^I(z)}{\ell_J D\Pi^J(z)} \,.
\end{equation}
They are independent of the choice of local holomorphic coordinate on $\mathbb{H}$.

Since the $(3,1)$-periods satisfy condition~(\ref{item:nullvectorcondition}), a particularly convenient choice is a nonzero real null vector $\ell$, i.e.
\begin{equation}
    \ell_I \eta^{IJ} \ell_J = 0 \,.
\end{equation}
This ensures that $\ell_J D\Pi^J(z) \neq 0$ for all $z \in \mathbb{H}$, and $F^I_\ell$ is well-defined and non-singular on the entire upper half-plane.

By \eqref{eq:nulltransversality}, the boundary map $\xi(x_0)$ is null for every $x_0 \in \partial \mathbb{H}$.
We may therefore set
\begin{equation}
    \ell_I = \eta_{IJ} \xi^J(x_0) \,,
\end{equation}
where $\xi^J(x_0)$ is a real homogeneous representative of $\xi(x_0)$.
Then, also by \eqref{eq:nulltransversality},
\begin{equation}
    \ell_I \xi^I(y) = \xi^I(x_0) \eta_{IJ} \xi^J(y) \neq 0
\end{equation}
for every $y \neq x_0$.
This implies that the boundary extension
\begin{equation}
     \xi^I_{\ell}(y) = \frac{\xi^I(y)}{\ell_J \xi^J(y)}
\end{equation}
of $F^I_{\ell}(z)$ is non-singular at every $y \neq x_0$.
Consequently, $\overline F^I_{\ell}(z)$ is non-singular on $\overline{\mathbb{H}} \setminus \{x_0\}$.
This means that the continuous extension of the projective $(3,1)$-period map \eqref{eq:projective31periods} takes values in a single affine patch when restricted to $\overline{\mathbb{H}} \setminus \{x_0\}$.

If the point at infinity corresponds to an LCS boundary, we can choose the homogeneous representative $\xi^I(\infty)$ such that, in the basis \eqref{eq:genperiods},
\begin{equation}\label{eq:LCSnormalizationvector}
    \left(\ell_{LCS}\right)_I = \eta_{IJ} \xi^J(\infty) \,,
\end{equation}
where $\ell_\mathrm{LCS}$ is the vector introduced in \eqref{eq:ellLCS}.
In the remainder of this paper, unless specified otherwise, we adopt this normalization and write
\begin{equation}\label{eq:LCSrepresentative}
    F^I(z) = F^I_{\ell_\mathrm{LCS}}(z) = F^I_{\xi(\infty)}(z) \,.
\end{equation}
Its continuous boundary extension is non-singular on $\overline{\mathbb{H}} \setminus \{\infty\}$, and therefore in particular on the entire real line.

\subsubsection{Polynomial form of the normalized representative}
\label{sec:polynomialapprox}

We restrict to the case of an LCS singularity at infinity, and consider again the normalized representative \eqref{eq:LCSrepresentative}.

As demonstrated in Appendix~\ref{app:LCS31}, for an LCS singularity the leading polynomial coefficient $b_{00}(y)$ in the expansion \eqref{eq:Bexpansion} satisfies $N^4 b_{00}(y) \neq 0$. 
By \eqref{eq:Nxixi}, this implies that $\ell_I b^I_{00}(y) \neq 0$.
We may therefore perform polynomial division and write
\begin{equation}\label{eq:b00polynonmialdevision}
    \frac{b_{00}(y)}{\ell_I b_{00}^I(y)} = p(y) + h(y) \,, 
\end{equation}
where $p(y)$ is a polynomial and $h(y)$ is properly rational, i.e.~$h(y) = \mathcal{O}(y^{-1})$.
Here $ \ell_I = \xi_I(\infty)$ as in \eqref{eq:LCSnormalizationvector}.
Contracting \eqref{eq:b00polynonmialdevision} with $\ell_I$ yields
\begin{equation}\label{eq:ellp}
    \ell_I p^I(y) = 1 \,, \qquad \ell_I h^I(y) = 0 \,.
\end{equation}
We introduce the polynomial part
\begin{equation}\label{eq:31P}
    P(z, \bar z) = e^{z N / T} p(y) = \sum_{n = 0}^4 \frac{z^n}{T^n\, n!} N^n p(y)\,,
\end{equation}
as well as the non-polynomial remainder term
\begin{equation}\label{eq:r}
    r(z, \bar z) = \frac{B(z, \bar z)}{\ell_I B^I(z, \bar z)} - p(y) \,,
\end{equation}
such that
\begin{equation}
    F_\ell(z, \bar z) = P(z, \bar z) + R(z, \bar z) \,, \qquad R(z, \bar z) = e^{z N / T} r(z, \bar z) \,.
\end{equation}
By contracting \eqref{eq:r} with $\ell_I$ and using \eqref{eq:ellp} we obtain $\ell_I r^I(z, \bar z) = 0$.
By \eqref{eq:Nxixi}, this implies that $N^4 r(z, \bar z) = 0$,
and therefore
\begin{equation}
    R(z, \bar z) = \sum_{n = 0}^3 \frac{z^n}{T^n\, n!} N^n r(z, \bar z) \,.
\end{equation}
We therefore expect that $R(z,\bar z) = \mathcal{O}(|z|^3)$ and that it is subleading with respect to the polynomial part $P(z, \bar z) = \mathcal{O}(|z|^4)$ in the large $|z|$ limit.

To formalize this observation, we note that $r(z, \bar z)$ is $T$-periodic in $x$ and satisfies, by construction,
\begin{equation}\label{eq:rlimit}
    \lim_{y \to \infty} r(x + iy, x-iy) = 0 \,,
\end{equation}
uniformly for $0 \leq x \leq T$.
Moreover, in the previous subsection we showed that $F_\ell$ extends continuously and non-singularly to $\overline{\mathbb{H}} \setminus \{\infty\}$.
Therefore, the remainder $r(z, \bar z)$ also extends continuously to this domain.
Together with \eqref{eq:rlimit} and periodicity in $x$, this implies that there exists a constant $C > 0$ such that
\begin{equation}\label{eq:rbound}
    \| r(z, \bar z) \| < C \,,
\end{equation}
for all $z \in \overline{\mathbb{H}} \setminus \{\infty\}$, and therefore indeed $R(z, \bar z) = \mathcal{O}(|z|^3)$.

To show that the remainder term is subleading with respect to the polynomial part, we need to argue that the quartic term in the latter is non-vanishing.
In Appendix~\ref{app:LCSrationalbasis} we evaluate the decomposition \eqref{eq:b00polynonmialdevision} explicitly and find
\begin{equation}
    \label{eq:py}
    p(y) = a_0 - \frac{i y}{2 T} N a_0 \,.
\end{equation}
Inserting this result into \eqref{eq:31P} we obtain
\begin{equation}
    P(z, \bar z) = \frac{z^3 \bar z}{24 T^4} N^4 a_0 + \mathcal{O}(|z|^3) \,.
\end{equation}
Therefore, since $N^4 a_0 \neq 0$, at sufficiently large $|z|$, there exists another constant $C' > 0$ such that 
\begin{equation}
    \frac{\|R(z, \bar z)\|}{\|P(z, \bar z)\|} < \frac{C'}{|z|} \rightarrow 0 \,,
\end{equation}
for $|z| \to \infty$.

Even though the polynomial part $P(z, \bar z)$ is strictly dominant over $R(z, \bar z)$ in every large $|z|$ limit, the individual subleading terms in $P(z, \bar z)$ may still be comparable or even subleading with respect to the remainder $R(z, \bar z)$.
However, slightly more can be said in the basis introduced in \eqref{eq:genperiods} where%
\footnote{This basis corresponds to a Jordan basis $e_I$ of $N$ with $N e_I = e_{I + 1}$ and $\ell \cdot e_0 = 1$.}
\begin{equation}
    P(z, \bar z) = \begin{pmatrix}
    1\\
        \frac{3 z+\overline{z} }{4T}  \\
        \frac{z (z+\overline{z})}{4 T^2}    \\
        \frac{z^2 (z+3 \overline{z})}{24 T^3}    + ic  \\
        \frac{z^3 \overline{z}}{24 T^4}   + \frac{ic (3z + \bar z)}{4T}
    \end{pmatrix} ,\qquad R(z, \bar z) = \begin{pmatrix}
        0 \\
        r^1 \\
        \frac{z}{T} r^1 + r^2 \\
        \frac{z^2}{2 T^2} r^1 + \frac{z}{T} r^2 + r^3 \\
        \frac{z^3}{6 T^3} r^1 + \frac{z^2}{2 T^2} r^2 + \frac{z}{T} r^3 + r^4
    \end{pmatrix} \,,
\end{equation}
with $r^I = r^I(z, \bar z)$.
Therefore, by the bound \eqref{eq:rbound}, $R^I(z, \bar z) = \mathcal{O}(|z|^{I-1})$.
We thus obtain the following polynomial approximation of the normalized (3,1)-periods,
\begin{equation}
    \label{eq:approxFLCS}
    F^I(z, \bar z) = \begin{pmatrix}
        1\\
        \frac{1}{4T} (3 z+\overline{z}) + \mathcal{O}(1)\\
        \frac{1}{4 T^2}  z (z+\overline{z}) + \mathcal{O}(|z|)\\
        \frac{1}{24 T^3}  z^2 (z+3 \overline{z}) + \mathcal{O}(|z|^2)\\
        \frac{1}{24 T^4}  z^3 \overline{z} + \mathcal{O}(|z|^3)
    \end{pmatrix} \,.
\end{equation}
It is valid in every $|z| \rightarrow \infty$ limit, including the real line.

\section{The index on the universal cover}
\label{sec:coveringspaceindex}

In this section we discuss the computation of the index $\ind(DW)$ introduced in Section~\ref{sec:index} for the Gukov-Vafa-Witten superpotential on the universal covering space of the complex structure moduli space.

We first summarize the main results from the previous section that are needed in the subsequent discussion.
The universal covering space of the moduli space $\mathcal{M}_{X_4}$ is given by the upper half-plane $\mathbb{H}$.
To make this explicit, one identifies the monodromy group with a subgroup $\Gamma_\mathrm{F} \subset \mathrm{PSL}(2,\mathbb{R})$, acting on $\mathbb{H}$ by M\"obius transformations \eqref{eq:moebius}, such that
\begin{equation}
    \mathcal{M}_{X_4} \simeq \mathbb{H} / \Gamma_\mathrm{F} \,.
\end{equation}
The period vector $\Pi^I(z)$ then satisfies the equivariance property
\begin{equation}\label{eq:summaryequivariance}
    \Pi^I(\gamma \cdot z) = \rho(\gamma)^I{}_J \Pi^J(z) \,,
\end{equation}
where $\gamma \cdot z$ denotes the action of $\gamma \in \Gamma_\mathrm{F}$ by M\"obius transformations and $\rho(\gamma)^I{}_J \Pi^J(z)$ denotes the action of the corresponding monodromy transformation on the periods. 

The boundary of the upper half-plane is given by
\begin{equation}
    \partial \mathbb{H} = \mathbb{R} \cup \{ \infty \} \,.
\end{equation}
We choose the point at infinity to correspond to a large complex structure (LCS) singularity.
The corresponding monodromy $\gamma_\infty \in \Gamma_\mathrm{F}$ acts on $\mathbb{H}$ by the deck transformation
\begin{equation}
    \gamma_\infty \cdot z = z + T \,,
\end{equation}
for some constant $T > 0$.
In particular, at sufficiently large $y = \Im z$, the periods $\Pi^I(z)$ on $\mathbb{H}$ admit the nilpotent orbit expansion \eqref{eq:nilpotent} with $t = z/T$ and thus a polynomial approximation in $z$.

The remaining elements of $\Gamma_\mathrm{F}$ that are not generated by $\gamma_\infty$ map the point at infinity onto the real line.
Moreover, the image
\begin{equation}
    \Gamma_\mathrm{F} \cdot \infty 
\end{equation}
is a dense subset of the boundary $\partial \mathbb{H}$.
We call the points on the real line that belong to this orbit $\infty$-equivalent and all other points on the real line $\infty$-inequivalent.

In the large-complex-structure basis \eqref{eq:genperiods}, we introduce the normalized $(3,1)$-periods
\begin{equation}\label{eq:summary31}
    F^I(z, \bar z) = \frac{D \Pi^I(z, \bar z)}{D \Pi^0(z, \bar z)} \,,
\end{equation}
where $D\Pi^I$ denotes the K\"ahler-covariant derivative of $\Pi^I$, defined in \eqref{eq:development}.
This map has a few favorable properties.
Firstly, it is independent of the choice of local holomorphic coordinate with respect to which the derivative is taken.
Secondly, it is non-singular on the entire upper half-plane, since $D \Pi^0$ is nonzero there, and each component $F^I$ vanishes if and only if $D \Pi^I$ vanishes.
Thirdly, it inherits a non-linear (or projective) equivariant transformation law from \eqref{eq:summaryequivariance}.

Lastly, if the period representation $\rho$ is relatively Anosov, $F^I$ extends continuously and finitely to $\mathbb{H} \cup \mathbb{R}$.%
\footnote{In general, the analogous normalized $(4,0)$-periods may neither be non-singular nor extend continuously to the real line.}
In this case, $F^I$ allows for a polynomial approximation of the form%
\footnote{The component-dependent proportionality factors are given in \eqref{eq:approxFLCS}.}
\begin{equation}\label{eq:summaryapprox}
    F^I(z, \bar z) \propto \left(4 - I\right) z^I + I z^{I-1}\bar z + \mathcal{O}(|z|^{I-1}) \,, \qquad I = 0, \dots, 4 \,,
\end{equation}
which is valid in any large $|z|$ limit and not only in the large $\Im z$ regime.

This polynomial approximation together with the transformation law inherited from \eqref{eq:summaryequivariance} allows us to compute $F^I(x)$ algebraically for every $\infty$-equivalent point $x \in \mathbb{R}$.
Its value at $\infty$-inequivalent points is then determined by continuity.

\subsection{Integration contour and boundary formula}

To compute the index $\ind_\mathbb{H}(DW)$ for the Gukov-Vafa-Witten superpotential $W$ defined in \eqref{eq:linearWK} on the upper half-plane $\mathbb{H}$,
we introduce 
\begin{equation}
    f_g(z, \bar z)  = g_I F^I(z, \bar z) = \frac{DW(z,\bar z)}{D\Pi^0(z,\bar z)} \,,
\end{equation}
where $F^I$ are the normalized $(3,1)$-periods \eqref{eq:summary31}.
Since our period basis is not necessarily integral, we perform the analysis for arbitrary real flux coefficients $g_I = \eta_{IJ} g^J$.
Imposing the correct quantization condition later does not alter our results.%
\footnote{We illustrate how to translate the basis \eqref{eq:genperiods} to an integral basis by means of an example in Section~\ref{sec:sextic}.}

Since the denominator $D\Pi^0$ is nonzero and finite on $\mathbb{H}$, as also discussed in detail in Section~\ref{sec:normalized31}, this normalization preserves the zeros and the local index.
We therefore have
\begin{equation}
   \ind_\mathbb{H} (DW) = \ind_\mathbb{H} (f_g) \,.
\end{equation}
As discussed in Section~\ref{sec:index}, the computation of the index amounts to the evaluation of a contour integral along the boundary $\partial \mathbb{H} = \mathbb{R} \cup \{\infty\}$ of the upper half-plane.
Assuming that the period representation $\rho$ is relatively Anosov, $f_g$ extends continuously and finitely to the real line, unlike $DW$.
It is hence a suitable choice for evaluating this boundary integral.

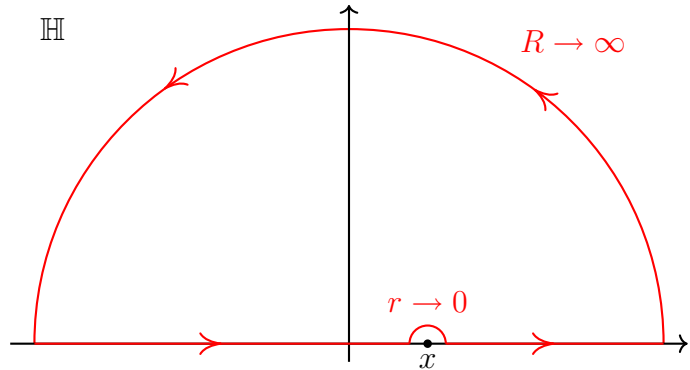
\begin{figure}[htb]
    \tikzset{
      arrow at/.style={
        postaction={decorate, decoration={markings,
          mark=at position #1 with {\arrow{To[length=3mm, width=3mm]}}}}
      },
      small arrow at/.style={
        postaction={decorate, decoration={markings,
          mark=at position #1 with {\arrow{To[length=3mm, width=3mm]}}}}
      }
    }
    
    \centering
    \begin{tikzpicture}[scale=0.8]

    \newcommand{\arrowL}{
        \tikz \draw[very thick, latex-] (-0.1,0) -- (0.1,0);
    }
    \newcommand{\arrowR}{
        \tikz \draw[very thick, -latex] (-0.1,0) -- (0.1,0);
    }
    \begin{scope}
        \clip (-5.5,0) rectangle (5.5,6);
        \draw [red, ,thick, arrow at=0.15, arrow at=0.35] (0,0) circle(5.2);
    \end{scope}
    \draw[black, thick, ->] (-5.6,0) -- (5.6,0);
    \draw[black, thick, ->] (0,-0.3) -- (0,5.6);
    \draw[red, thick, arrow at=0.5] (-5.2,0) -- (1,0);

    \draw[red, thick]
    (1,0) arc[start angle=180, end angle=0, radius=0.3];

    \fill[black] (1.3,0) circle (2pt);

    \draw[red, thick, arrow at=0.5] (1.6,0) -- (5.2,0);
    \node[below] at (1.3,0) {$x$};
    \node at (3.7,5) {\color{red}$R \to \infty$};
    \node at (1.3,0.7) {\color{red}$r \to 0$};
    \node at (-4.9,5.2) {\color{black}$\mathbb{H}$};
    \end{tikzpicture}%
    \caption{Integration contour (red) in the upper half-plane $\mathbb{H}$, illustrated for a single zero $f_g(x)=0$ on the real axis.
    It follows the real axis, bypassing the zero along a small semicircle of radius $r\to0$, and closes along a large semicircle of radius $R\to\infty$.
    This generalizes straightforwardly to finitely many zeros on the real axis.
    }
    \label{fig:intcontour}
\end{figure}

According to \eqref{eq:winding_number}, the index is computed by the winding number integral
\begin{equation}\label{eq:section4windingnumber}
    \ind_\mathbb{H} (DW) = \frac{1}{2 \pi i} \oint_{\partial \mathbb{H}} d \log f_g \,,
\end{equation}
on a positively oriented contour avoiding the zeros of $f_g$.
It tracks the change in the argument (or complex phase) of $f_g$.
Here, our integration contour follows the real axis from left to right and closes through a large upper semicircle traversed counter-clockwise, 
as illustrated in Figure~\ref{fig:intcontour}.

We observe two potential issues:
the function $f_g$ is not necessarily differentiable on the real axis and may have zeros there.
Nevertheless, since $f_g$ is purely real on the real axis, its argument is constant on every interval where it is nonzero.
Such intervals therefore make no contribution to the winding number, even if $f_g$ is not differentiable there.%
\footnote{Although $f_g$ is not required to be differentiable on the real line, its continuous extension there is essential and is ensured if the period representation $\rho$ is relatively Anosov.}
The remaining contributions come from infinity and the zeros on the real axis. 
The contribution of an isolated zero can be computed by deforming the contour to include an infinitesimal semicircle in $\mathbb{H}$ centered at that zero.
We thus arrive at
\begin{equation}\label{eq:indexsum}
    \ind_\mathbb{H} (f_g) = \ind_\infty (f_g) + \sum_{\left\{x \in \mathbb{R} \,\middle|\, f_g(x) = 0 \right\}} \ind_x (f_g) \,,
\end{equation}
where $\ind_\infty (f_g)$ denotes the contribution from the counter-clockwise large semicircle at infinity, and $\ind_x (f_g)$ the contribution from a clockwise infinitesimal semicircle centered at a zero $x$ of $f_g$.
We discuss their calculation in turn.

We note that the presence of infinitely many zeros or non-isolated zeros on the real axis cannot be excluded, as illustrated in the last example in Section~\ref{sec:examples}.
Both situations are in fact equivalent
and non-isolated zeros can only occur at $\infty$-inequivalent points. 
Moreover, an isolated zero on the real axis at an $\infty$-inequivalent point may not be isolated in $\overline{\mathbb{H}}$, but instead be the accumulation point of infinitely many interior zeros.
In either case, the convergence of \eqref{eq:indexsum} is not guaranteed and our method may not be directly applicable.
We will revisit these issues in future work.

\subsection{Contribution from infinity}

To evaluate the contribution from infinity, we consider as integration contour the counter-clockwise upper semicircle
\begin{equation}\label{eq:largecircle}
    \mathcal{C}_R : \; z = R \, e^{i \theta} \,, \qquad 0 \leq \theta \leq \pi 
\end{equation}
in the infinite-$R$ limit.
The contribution to the index is then given by
\begin{equation}\label{eq:largecircleintegral}
    \ind_\infty(f_g) = \frac{1}{2 \pi} \lim_{R\to \infty}  \Im \int_{\mathcal{C}_R} d\log f_g \,.
\end{equation}

Since the polynomial approximation \eqref{eq:summaryapprox}  of $F^I(z, \bar z)$ holds uniformly as $|z| \to \infty$ , it is valid along the entire contour \eqref{eq:largecircle} and may be used to compute \eqref{eq:largecircleintegral}, provided that the leading polynomial term of $f_g$ has no zeros along the contour.
Importantly, an approximation valid only in the conventional large complex structure limit $\Im z\to\infty$ would not suffice, since $\Im z$ becomes small near the endpoints of the contour.

Each leading term in the approximation \eqref{eq:summaryapprox} contains at most two monomials, so we first evaluate the contributions of such expressions. 
For a monomial in $z$ and $\bar z$, the contour integral \eqref{eq:largecircleintegral} along $\mathcal{C}_R$ is readily computed,
\begin{equation}\label{eq:monomialindex}
    \ind_\infty (z^p \bar z^q) = \frac{1}{2 \pi} \lim_{R\to \infty}  \Im \int_{\mathcal{C}_R} d\log \left(R^{p+q} e^{i (p-q) \theta } \right) = \frac{1}{2 \pi} \int_0^\pi (p-q) d \theta = \frac{p-q}{2} \,.
\end{equation}
Similarly, for a sum of two monomials of equal total degree, the contribution is determined by the monomial with the coefficient of larger modulus,
\begin{equation}\label{eq:twomonomialindex}
    \ind_\infty \left( \alpha z^p \bar z^q + \beta z^r \bar z^s\right) = \frac{p-q}{2} \,, \qquad p+q = r+s \,, \quad | \alpha | > | \beta | \,,
\end{equation}
since the second term can be continuously tuned to zero without creating a zero of the sum anywhere on the contour.
This argument does not apply when $|\alpha|=|\beta|$, which requires a separate analysis.

We now apply these results to $f_g(z, \bar z) = g_I F^I(z, \bar z)$.
For a given flux $g_I$, its leading term can be read off from \eqref{eq:summaryapprox}.
For example, if $g_4 \neq 0$, we have at leading order
\begin{equation}
    f_g(z, \bar z) \propto z^3 \bar z \,,
\end{equation}
and hence, by using \eqref{eq:monomialindex}, we find
    $\ind_\infty (f_g) = 1$.
Analogously, if $g_4 = 0$ but $g_3 \neq 0$,
\begin{equation}
    f_g(z, \bar z) \propto z^3 + 3 z^2 \bar z \,,
\end{equation}
so \eqref{eq:twomonomialindex} gives
$\ind_\infty (f_g) = \frac12$.
The same reasoning may now be applied to all remaining cases unless
\begin{equation}\label{eq:exceptionalflux}
    g_4 = g_3 = 0 \,, \qquad g_2 \neq 0 \,,
\end{equation}
in which case
\begin{equation}\label{eq:g2leading}
    f_g(z, \bar z) \propto z^2 + z \bar z \,.
\end{equation}
The two coefficients have equal absolute value, so \eqref{eq:twomonomialindex} cannot be applied.
Indeed, the leading polynomial term \eqref{eq:g2leading} vanishes on the contour at $z=- \bar z$.
The contribution to the index therefore depends also on subleading terms.
The details of this analysis can be found in Appendix~\ref{app:exceptionalflux}.

We summarize the results for all cases, including the exceptional case \eqref{eq:exceptionalflux}, in Table~\ref{table:result_infty_gen}.
In particular, we observe that the contribution from infinity is bounded by
\begin{equation}\label{eq:infinitybound}
    0 \leq \ind_\infty(f_g) \leq 1 \,,
\end{equation}
for every nonzero flux.

\begin{table}[htbp]
\centering
\renewcommand{\arraystretch}{1.4}
\begin{tabular}{|c | c | c | c | c | c | c | c |}
\hline
$g_4$ & $g_3$ & $g_2$ & $g_1$ & $g_0$ & $\Delta_g$ & $c$ & $\ind_\infty(f_g)$ \\
\hline\hline

$\neq0$ & --- & --- & --- & --- & --- & --- & $1$ \\
\hline
\multirow{8}{*}{$0$} & $\neq0$ & --- & --- & --- & --- & --- & $\frac12$ \\
\cline{2-8}

& \multirow{7}{*}{$0$} & \multirow{5}{*}{$\neq0$} & \multirow{5}{*}{---} & \multirow{5}{*}{---} & $<0$ & --- & $1$ \\
\cline{6-8}
& & & & & $>0$ & --- & $0$ \\
\cline{6-8}
& & & & & \multirow{3}{*}{$0$} & $>0$ & $1$ \\
\cline{7-8}
& & & & & & $<0$ & $0$ \\
\cline{7-8}
& & & & & & $0$ & Not determined \\

\cline{3-8}
& & \multirow{2}{*}{$0$} & $\neq0$ & --- & --- & --- & $\frac12$ \\
\cline{4-8}
& & & $0$ & $\neq0$ & --- & --- & $0$ \\
\hline
\end{tabular}
\captionsetup{singlelinecheck=false}
\caption[Index at infinity for nonzero real fluxes]{
Contribution to the index at infinity for nonzero real fluxes in the basis \eqref{eq:genperiods}.
A dash indicates no separate restriction on the corresponding component.
For $g_2\neq0$, we define
\begin{equation*}
    \Delta_g= \frac{g_0}{g_2} - \frac{g_1^2}{2g_2^2} \,,
\end{equation*}
and $c$ denotes the model-specific constant in \eqref{eq:genperiods}.
The undetermined case depends on exponentially suppressed corrections, see Appendix~\ref{app:exceptionalflux}.
}
\label{table:result_infty_gen}
\end{table}

\subsection{Contributions from zeros on the real axis}

We now turn to the evaluation of $\ind_x (f_g)$ for $x$ a zero of $f_g$ on the real axis.
We distinguish between $\infty$-equivalent and inequivalent zeros, and discuss each case separately.

\subsubsection{$\infty$-equivalent zeros}

If $x \in \mathbb{R}$ is $\infty$-equivalent, then, by definition, there exists a monodromy transformation $\gamma_x \in \Gamma_\mathrm{F}$ such that
\begin{equation}\label{eq:gammax}
    x = \gamma_x \cdot \infty \,.
\end{equation}
It follows from equivariance \eqref{eq:summaryequivariance} that 
\begin{equation}
    f_g(\gamma_x \cdot z)
    = \frac{g_I \rho(\gamma_x)^I{}_J D\Pi^J(z, \bar z)}{\rho(\gamma_x)^0{}_K D\Pi^K(z, \bar z)}
    = \frac{f_{g_x} (z)}{f_{\tilde g} (z)} \,,
\end{equation}
where we have introduced
\begin{equation}\label{eq:gprime}
    \left(g_x\right)_I = g_J \, \rho(\gamma_x)^J{}_I \,,\qquad \tilde g_I = \rho(\gamma_x)^0{}_I \,.
\end{equation}
Since $\gamma_x$ maps the integration contour around infinity of the previous subsection onto a clockwise infinitesimal semicircle around $x$, we obtain
\begin{equation}
    \ind_x (f_g) = \ind_\infty (f_g \circ \gamma_x) =  \ind_\infty \left( \frac{f_{g_x}}{f_{\tilde g}} \right) = \ind_\infty (f_{g_x}) - \ind_\infty (f_{\tilde g}) \,.
\end{equation}
To evaluate the index of $f_{\tilde g}$ around infinity, we note that in the basis \eqref{eq:genperiods}
\begin{equation}\label{eq:projectiveboundarymap}
    F^I(x) = \lim_{y \to \infty} \frac{\rho(\gamma_x)^I{}_J D\Pi^J(i y, -iy )}{\rho(\gamma_x)^0{}_K D\Pi^K(i y, -i y)} = \frac{\rho(\gamma_x)^I{}_4}{\rho(\gamma_x)^0{}_4} \,.
\end{equation}
Since the last column of $\rho(\gamma_x)$ is nonzero and $F^I$ is non-singular on the real axis, we have%
\footnote{
The argument can also be formulated using the machinery of relatively Anosov representations and their boundary maps.
In the basis \eqref{eq:genperiods}, the  projective boundary map $\xi$ introduced in Section~\ref{sec:continuity} satisfies
$\xi(\infty) = [0,0,0,0,1]$.
This follows by inserting the periods \eqref{eq:genperiods} into the boundary limit \eqref{eq:limitcondition}.
For this representative, equivariance and transversality in the form \eqref{eq:nulltransversality} give
\begin{equation*}
    0 \neq \xi^I(\infty) \eta_{IJ} \xi^J(x) \propto \xi^I(\infty) \eta_{IJ} \rho(\gamma_x)^J{}_K \xi^K(\infty) = \rho(\gamma_x)^0{}_4 \,,
\end{equation*}
using the intersection form \eqref{eq:def_eta}.
}
\begin{equation}
    \tilde g_4 = \rho(\gamma_x)^0{}_4 \neq 0 \,.
\end{equation}
Using the results from the previous subsection, this implies that
\begin{equation}
    \ind_\infty (f_{\tilde g}) = 1 \,,
\end{equation}
and consequently
\begin{equation}
    \ind_x (f_g) = \ind_\infty (f_{g_x}) - 1 \,.
\end{equation}
Thus, the contribution of each $\infty$-equivalent zero to the index can be computed by mapping it to infinity and applying Table~\ref{table:result_infty_gen} to the transformed flux \eqref{eq:gprime}.%
\footnote{
Note that the choice of $\gamma_x$, and hence of the transformed flux $(g_x)_I$, is not unique.
Replacing $\gamma_x$ by $\gamma_x \gamma_\infty^k$, for $k\in \mathbb{Z}$, in \eqref{eq:gammax} leaves $x$ invariant but transforms $(g_x)_I \to (g_x)_J \rho(\gamma_\infty^k)^J{}_I$ .
However, this does not change $\ind_\infty(f_{g_x})$, as can be verified using the form of $\rho(\gamma_\infty)$ in \eqref{eq:LCSmonodromy} and Table~\ref{table:result_infty_gen}.
}
In particular, \eqref{eq:infinitybound} implies that
\begin{equation}\label{eq:equivalentbound}
    \ind_x (f_g) \leq 0 \,,
\end{equation}
for every $\infty$-equivalent zero.

\subsubsection{$\infty$-inequivalent zeros}

An $\infty$-inequivalent zero cannot be mapped to infinity by an element of $\Gamma_\mathrm{F}$, so the preceding method does not apply.

We have not found a general method for computing these contributions.
However, our numerical results lead us to conjecture that
\begin{equation}\label{eq:inequivalentbound}
    \ind_x(f_g) \leq 0
\end{equation}
for every $\infty$-inequivalent zero $x$ of $f_g$.
Thus, we expect the bound \eqref{eq:equivalentbound} established for $\infty$-equivalent zeros to hold also for $\infty$-inequivalent zeros.

This conjecture is consistent with our numerical tests for the mirror Fermat sextic and another fourfold with triangle monodromy group, summarized in Section~\ref{sec:scan}. 
Although most of the results in this section do not rely on any assumption about other punctures in $\mathcal{M}_{X_4}$ besides the LCS singularity, we have not tested the bound \eqref{eq:inequivalentbound} for Calabi-Yau fourfolds with type $\mathrm{I}_{1,1}$ singularities (called C-points in \cite{vandeHeisteeg:2024lsa}).
A more general discussion is left to future work.

\subsection{Sufficient condition for the existence of flux vacua}
\label{sec:sufficientindex}

Assuming that the period representation $\rho$ is relatively Anosov, the preceding results provide a sufficient condition for the existence of flux vacua using boundary data.

If all real zeros of $f_g$ are $\infty$-equivalent, the index can be computed exactly:
\begin{equation}
\ind_\mathbb{H}(DW) = \ind_\infty(f_g) +\sum_{f_g(x) = 0} \left(\ind_\infty(f_{g_x})-1\right) \,,
\end{equation}
where $g_x$ denotes the transformed flux defined in \eqref{eq:gprime}.
Each contribution at infinity is determined by Table~\ref{table:result_infty_gen}.
A nonzero index guarantees the existence of at least one flux vacuum in $\mathbb{H}$, whereas a vanishing index is inconclusive.
This result does not require the conjecture \eqref{eq:inequivalentbound}.

If $\infty$-inequivalent zeros are present, we assume the conjectured bound \eqref{eq:inequivalentbound}, implying that 
all zeros on the real axis contribute non-positively to the index.
Since $f_g$ is real on the real axis, a zero $x$ at which its sign changes contributes a half-integer and hence $\ind_x (f_g) \leq -1/2$.
Denoting the number of such zeros by $n_{\mathrm{half}}$, we obtain
\begin{equation}
    \ind_\mathbb{H}(DW) \leq \ind_\infty(f_g)-\frac{n_{\mathrm{half}}}{2} \,.
\end{equation}
Consequently,
\begin{equation}
    \ind_\infty(f_g) < \frac{n_{\mathrm{half}}}{2}
\end{equation}
is a sufficient condition for the existence of a flux vacuum in $\mathbb{H}$.
For this purpose, it suffices to identify enough sign-changing zeros to obtain a negative index.
It is not necessary to determine all zeros on the real axis or their exact contribution to the index.

We describe some practical strategies to find such zeros of $f_g$ in Section~\ref{sec:findingboundaryzeros} for the example of the mirror sextic.

\section{Example: the mirror Fermat sextic}
\label{sec:sextic}

We illustrate our results for the mirror Fermat sextic, a Calabi-Yau fourfold with a single complex structure modulus.
It is obtained by applying the Greene-Plesser construction \cite{Greene:1990ud, Greene:1993vm} to the Dwork family of hypersurfaces
\begin{equation}
     \sum_{n=1}^6 x_n^6 - 6 \psi \prod_{n=1}^6 x_n = 0
\end{equation}
in $\mathbb{CP}^5$.
The complex structure of this family is parametrized by $\psi^6$.

The holomorphic periods of the mirror sextic satisfy a fifth-order Picard-Fuchs equation \cite{Klemm:2007in, Grimm:2009ef}; see also \cite{vandeHeisteeg:2024lsa, Lust:2024aeg} for more recent discussions.
This equation exhibits three singular points: a large complex structure (LCS) point at $\psi^6 = \infty$, a conifold point at $\psi^6=1$, and a Landau-Ginzburg (LG) point at $\psi^6 = 0$.

\subsection{Periods and integral basis}

Since its Picard-Fuchs equation is of order five, the mirror sextic has five linearly independent periods, i.e.\ $b_H =5$ and $h^{2,2}_H = 1$.
Near the LCS point, the logarithmic coordinate is given by
\begin{equation}\label{eq:sextict}
    t = \frac{\log \bigl((-6 \psi)^{-6}\bigr)}{2 \pi i} \,,
\end{equation}
and the polynomial LCS period expansion \eqref{eq:genperiods} applies with
\begin{equation}\label{eq:sextickappac}
    \kappa = 6 \,, \qquad c = -\frac{35\zeta(3)}{4\pi^3} \,.
\end{equation}
We recall that the basis \eqref{eq:genperiods} is not integral.
Following \cite{vandeHeisteeg:2024lsa, Lust:2024aeg}, we may introduce an integral basis
\begin{equation}
    \Pi_\mathrm{int} = \Lambda \Pi \,,
\end{equation}
where the transition matrix $\Lambda$ is given by
\begin{equation}\label{eq:integerbasisLambda}
    \Lambda = \begin{pmatrix}
        1 & 0 & 0 & 0 & 0\\
        0 & 1 & 0 & 0 & 0\\
        \frac{5}{8} & \frac{1}{2} & - 1 & 0 & 0\\
        -\frac{15}{8} & - \frac{3}{4} & 3 & -6 & 0\\
        -\frac{75}{64} & 0 & \frac{15}{4} & 0 & 6
    \end{pmatrix} \,.
\end{equation}
Unless stated otherwise, we use this integral basis throughout the remainder of this section, as it simplifies the monodromy matrices and allows us to impose flux quantization.

\subsection{Monodromy and covering space}

We denote the monodromy matrices around the three singular points by $M_\mathrm{LCS}$, $M_\mathrm{C}$, and $M_\mathrm{LG}$.
In the integral basis, they are given by \cite{Lust:2024aeg, vandeHeisteeg:2024lsa}
\begin{equation}\label{eq:sexticmonodromy}
    M_\mathrm{LCS} =  \!\begin{pmatrix} 
        \!1 & 0 & 0 & 0 & 0 \\
        \!1 & 1 & 0 & 0 & 0 \\
        \!0 & \!\!-1 & 1 & 0 & 0 \\
        \!\!-4 & \!\!-3 & 6 & 1 & 0 \\
        \!4 & 7 & \!\!-6 & \!\!-1 & 1 \\
    \end{pmatrix} \!,\,
    M_\mathrm{C} = \!\begin{pmatrix}
        0 & \!0 & 0 & 0 & \!\!-1 \\
        0 & \!1 & 0 & 0 & 0 \\
         0 & \!0 & 1 & 0 & 0 \\
        0 & \!0 & 0 & 1 & 0 \\
        \!-1 & \!0 & 0 & 0 & 0 \\
    \end{pmatrix} \!,\,
    M_\mathrm{LG} = \!\begin{pmatrix}
        \!-4 & 4 & 0 & \!\!-1 & \!\!-1 \\
        \!-1 & 1 & 0 & 0 & 0 \\
        \!-1 & 1 & 1 & 0 & 0 \\
        7 & \!\!-3 & \!\!-6 & 1 & 0 \\
        \!-1 & 0 & 0 & 0 & 0 \\
    \end{pmatrix} \!.
\end{equation}
While $M_\mathrm{LCS}$ is maximally unipotent and hence of infinite order, $M_\mathrm{C}$ and $M_\mathrm{LG}$ have orders $2$ and $6$.
The generators of the monodromy group thus satisfy
\begin{equation}\label{eq:sexticmonodromyrelations}
    M_\mathrm{C}^2 = M^6_\mathrm{LG} = M_\mathrm{LCS} M_\mathrm{C} M_\mathrm{LG} = \mathbf{1} \,.
\end{equation}
The last relation follows because a contour around one singular point can be deformed into a contour encircling the other two points in the opposite direction.

\subsubsection{The Fuchsian group}

To identify the upper half-plane $\mathbb{H}$ with the universal covering space of the moduli space, we realize the monodromy group as a Fuchsian subgroup $\Gamma_{\mathrm{F}} \subset \mathrm{PSL}(2, \mathbb{R})$, with a parabolic generator $\gamma_{\mathrm{LCS}}$ and two elliptic generators $\gamma_\mathrm{C}$ and $\gamma_{\mathrm{LG}}$ of orders two and six, respectively.
The group $\Gamma_{\mathrm{F}}$ then acts on $\mathbb{H}$ by M\"obius transformations \eqref{eq:moebius}, so that the moduli space can be identified with the quotient $\mathbb{H}/\Gamma_\mathrm{F}$ as in \eqref{eq:defM}.

To find suitable generators, we place the fixed point of $\gLCS$ at $z=\infty$.\footnote{A similar construction for the Fermat quintic can be found in \cite{Candelas:1990rm}.}
Since $\gLCS$ is parabolic, it takes the form
\begin{equation}
    \gLCS
    = \begin{pmatrix} 1&T\\ 0&1 \end{pmatrix} \,,
\end{equation}
for $T>0$.
One also easily verifies that the candidate
\begin{equation}
    \gLG = \begin{pmatrix}
        \cos\frac{\pi}{6} & \sin\frac{\pi}{6}\\
        -\sin\frac{\pi}{6} & \cos\frac{\pi}{6}
      \end{pmatrix}
\end{equation}
is of order six in  $\mathrm{PSL}(2,\mathbb R)$.
The constant $T$ is then determined by demanding that
\begin{equation}
    \gamma_\mathrm{C}^2 = \left(\gamma_\mathrm{LCS}^{-1}\gamma_\mathrm{LG}^{-1}\right)^2 = \pm \mathbf{1} \,,
\end{equation}
so that it is of order two in $\mathrm{PSL}(2,\mathbb R)$.
It follows that $T = 2 \sqrt{3}$.
In summary, the generators are
\begin{equation}\label{eq:sexticfuchsian}
    \gLCS = \begin{pmatrix}
            1 & 2 \sqrt{3}\\
            0 & 1
        \end{pmatrix} \,, \qquad
    \gLG = \frac12 \begin{pmatrix}
        \sqrt{3} & 1\\
        -1 & \sqrt{3}
        \end{pmatrix} \,, \qquad
    \gC = \frac12\begin{pmatrix}
            -\sqrt{3} & -7\\
        1 & \sqrt{3}
        \end{pmatrix} \,.
\end{equation}
They satisfy the relations \eqref{eq:sexticmonodromyrelations} in $\mathrm{PSL}(2,\mathbb R)$ by construction.

Using that hyperbolic metric \eqref{eq:hyperbolicmetric} has constant negative curvature, the Gauss-Bonnet formula \eqref{eq:orbifoldeuler} implies that the quotient $\mathbb H/\Gamma_{\mathrm F}$ has finite hyperbolic area.
Consequently, $\Gamma_{\mathrm F}$ has finite covolume and is a Fuchsian group of the first kind.

\subsubsection{Fundamental domain and holomorphic coordinate}

By construction, the fixed point of the parabolic generator $\gLCS$ is located on the boundary of the upper half-plane at $z= \infty$.
On the other hand, the fixed points of the two elliptic generators $\gLG$ and $\gC$ are located in its interior at $z=i$ and $z=-\sqrt{3}+2i$, respectively.

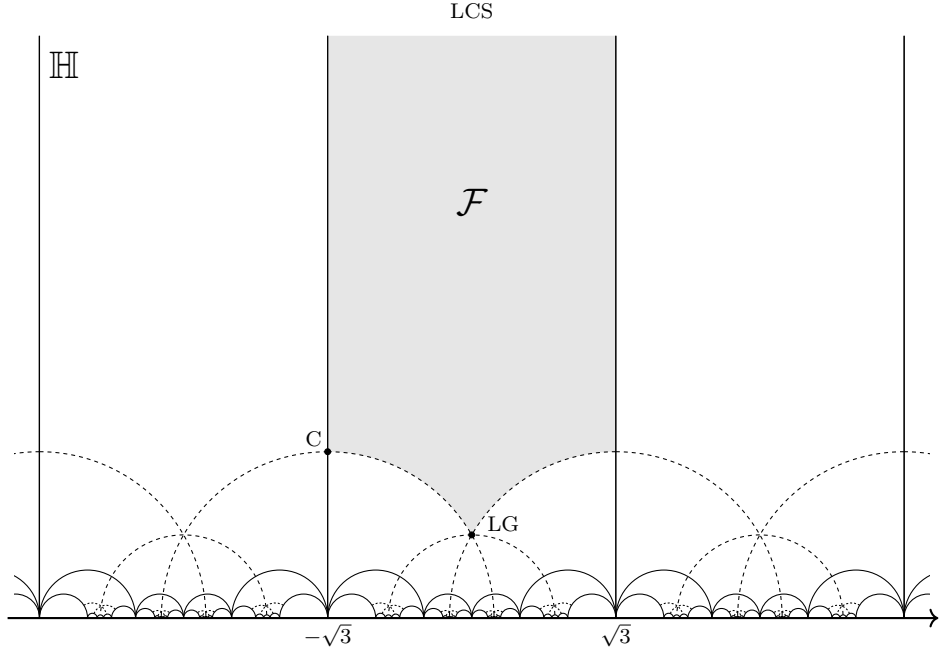
\begin{figure}[htb]
    \centering
    \begin{tikzpicture}[scale=1.1]

    \begin{scope}
        \clip (-5.5,0) rectangle (5.5,7);
            \fill[color=gray!20, fill opacity=1]
            (-1.732,7)-- (-1.732,7)--  plot [domain=90:30] ({-1.732 + 2*cos(\x)}, {2*sin(\x)})-- plot [domain=150:90] ({1.732 + 2*cos(\x)}, {2*sin(\x)})-- (1.732,7)-- cycle;
            \draw[black] (-1.732,0) -- (-1.732,7);
            \draw[black] (-5.2,0) -- (-5.2,7);
            \draw[black] (5.2,0) -- (5.2,7);
            \draw[black] (1.732,0) -- (1.732,7);
            
            \foreach \k in {-2,-1,0,1} {
                \draw[black]({(2*\k+1)*sqrt(3)},0) -- ({(2*\k+1)*sqrt(3)},7);
            }

            \foreach \k in {-2,-1,0,1,2} {
                \foreach \s in {-1,1} {

                    \draw[black]({sqrt(3)*(2*\k+\s*2/3)},0) circle[radius={sqrt(3)/3}];
                    \draw[black]({sqrt(3)*(2*\k+\s/6)},0) circle[radius={sqrt(3)/6}];

                    \foreach \a/\b in {1/18, 8/63, 13/84, 11/60, 4/15, 5/12, 19/36, 26/45, 19/30, 5/6} {
                        \draw[black]({sqrt(3)*(2*\k+\s*\a/\b)},0) circle[radius={sqrt(3)/\b}];
                    }
                }
            }
            
            \foreach \k in {-2,-1,0,1,2} {
                \begin{scope}[shift={({2*\k*sqrt(3)},0)}]

                \draw[dash pattern=on 1.5pt off 1.5pt]({cos(2.2042275)},{sin(2.2042275)}) arc[start angle=2.2042275, end angle=177.7957725, radius=1];

                \foreach \s in {-1,1} {
                    \begin{scope}[xscale=\s]

                    \draw[dash pattern=on 1.5pt off 1.5pt]({sqrt(3)},2) arc[start angle=90, end angle=179.4093114, radius=2];

                    \foreach \a/\b/\r/\from/\to in {
                    7/11/2/57.7957725/165.1781789,
                    7/13/2/20.5699921/136.8264489,
                    2/11/1/51.7867893/163.1735511,
                    3/23/2/18.3649487/132.1034489
                    } {
                        \draw[dash pattern=on 0.75pt off 0.75pt]
                            ({sqrt(3)*\a/\b + (\r/\b)*cos(\from)}, {(\r/\b)*sin(\from)})
                            arc[start angle=\from, end angle=\to, radius={\r/\b}];
                    }
                    \end{scope}
                }
                \end{scope}
            }
            
            \filldraw (0,1) circle (1pt);
            \node at (0.38,1.13) {\scriptsize LG};
            \filldraw (-1.732,2) circle (1pt);
            \node at (-1.9,2.15) {\scriptsize C};

    \end{scope}
    \draw[black, thick, ->] (-5.58,0) -- (5.61,0);
    \node at (0,7.3) {\scriptsize LCS};
    
    \node at (-4.9,6.65) {\large\color{black}$\mathbb{H}$};
    \node at (0,5) {\large\color{black}$\mathcal{F}$};
    
    \node at (-1.732,-0.2) {\scriptsize $-\sqrt{3}$};
    \node at (1.732,-0.2) {\scriptsize $\sqrt{3}$};
    
    \end{tikzpicture}
    \caption{The upper half-plane $\mathbb{H}$ as the universal covering space of the complex-structure moduli space of the mirror Fermat sextic.
    The shaded area corresponds to the fundamental domain $\mathcal{F}$ described in the text.
    The fixed points of the three generators of $\Gamma_\mathrm{F}$ are denoted by LCS (at infinity), LG, and C.
    Solid lines are geodesics connecting the images of C and LCS under $\Gamma_\mathrm{F}$, while dashed lines are geodesics connecting the images of C and LG.
    }
    \label{fig:fundz}
\end{figure}

We may therefore choose a fundamental domain $\mathcal{F}$ bounded by two vertical rays extending from the conifold point and its image under $\gamma_\mathrm{LCS}$ at $z=\pm\sqrt{3}+2i$ to the LCS point at infinity, and two circular arcs connecting these points to the LG point at $z=i$.
The images of $\mathcal{F}$ under $\Gamma_\mathrm{F}$ tessellate the upper half-plane, as illustrated in Figure~\ref{fig:fundz}.

The fundamental domain $\mathcal{F}$ represents the complex structure moduli space $\mathcal{M}$, with holomorphic coordinate $\psi^6$, in $\mathbb{H}$.
As discussed in Section~\ref{sec:nilpotentcovering}, in the vicinity of the LCS point, for sufficiently large $\Im z$, the coordinate $z$ on $\mathbb{H}$ is related to the logarithmic coordinate $t$ defined in \eqref{eq:sextict} by
\begin{equation}\label{eq:sexticaffinecoordinate}
    z = 2 \sqrt{3} t + z_0 + o(1) \,,
\end{equation}
for some constant $z_0 \in \mathbb{C}$.
This identification fixes the polynomial expansion of the periods on $\mathbb{H}$ at large $\Im z$, and therefore, as discussed below, the entire boundary map.%
\footnote{A constant shift changes lower-degree polynomial terms but leaves the highest-degree term of each component unchanged.
In particular, the leading polynomial approximation \eqref{eq:summaryapprox} of the normalized $(3,1)$-periods at large $|z|$ remains valid.
See also footnote~\ref{note:tshift}.}

However, if we wished to evaluate the periods also in the interior of $\mathbb{H}$, we would need an explicit coordinate map $z(\psi^6)$.
Such a map can be obtained using a Schwarz triangle map \cite{Candelas:1990rm,Bateman:100233}.
Applying this procedure to a triangle with vertices $(i, \sqrt{3} + 2 i, \infty)$ (corresponding to LG, conifold, and LCS points, respectively) and angles $(\pi/6, \pi/2, 0)$ yields the corresponding coordinate map on the Poincaré disk,
\begin{equation}
    \phi (\psi^6) = \sqrt{3} \, \psi \,   \frac{\Gamma(\frac{2}{6})^2 \Gamma(\frac{5}{6})}{\Gamma(\frac{7}{6}) \Gamma(\frac{1}{6})^2}  \frac{ {}_2F_1(\frac{2}{6}, \frac{2}{6}, \frac{7}{6}; \psi^6)}{{}_2F_1 (\frac{1}{6}, \frac{1}{6}, \frac{5}{6}; \psi^6)} \,,
\end{equation}
where ${}_2F_1$ denotes the Gauss hypergeometric function and the constant prefactor is required to normalize the disk to unit radius.
This is mapped back to the upper half-plane by
\begin{equation}\label{eq:schwarztriangle}
    z(\psi^6) = i\,\frac{e^{i\pi/6}+\phi(\psi^6)}{e^{i\pi/6}-\phi(\psi^6)} \,,
\end{equation}
and taking the hypergeometric functions on their principal branches with $0<\arg\psi<\pi/3$ gives the full fundamental domain $\mathcal{F}$ by analytic continuation.

Expanding \eqref{eq:schwarztriangle} at large $|\psi|$ confirms the affine relation \eqref{eq:sexticaffinecoordinate} with
\begin{equation}
    z_0=\sqrt{3}-\frac{i\sqrt{3}}{\pi}\log432 
\end{equation}
near the LCS point.

\subsection{Boundary maps}

\subsubsection{Relatively Anosov monodromy representation}

As established in Section~\ref{sec:coverinspace} and Appendix~\ref{app:anosov}, the normalized $(3,1)$-periods $F^I$ defined in \eqref{eq:summary31} extend continuously and finitely to the real line if the period representation $\rho$ is relatively Anosov.
For the mirror Fermat sextic, this representation is defined by
\begin{equation}
    \rho(\gamma_\mathrm{LCS}) = M_\mathrm{LCS} \,, \qquad  \rho(\gamma_\mathrm{LG}) = M_\mathrm{LG} \,, \qquad  \rho(\gamma_\mathrm{C}) = M_\mathrm{C} \,.
\end{equation}
In Section~\ref{sec:relanosovfourfolds}, we identified the absence of infinite-order singularities other than LCS or type-$\mathrm{I}_{1,1}$ points as a necessary condition for $\rho$ to be relatively Anosov.
The mirror sextic satisfies this condition, since its only infinite-order singularity is the LCS point.
Moreover, it was shown by S.~Filip that $\rho$ is indeed relatively Anosov (refs.~\cite{filip2022uniformizationweight3variations, MR4615758} and private communication; see footnote~\ref{note:filip}).

\subsubsection{The boundary map}
\label{sec:sexticboundarymap}

For an $\infty$-equivalent point $x$ on the real axis (i.e.\ $x \in \Gamma_\mathrm{F} \cdot \infty \cap \mathbb{R}$), we can compute the boundary value $F^I(x)$ of the normalized $(3,1)$-periods using the equivariance property \eqref{eq:summaryequivariance} and the explicit matrix representations given in \eqref{eq:sexticmonodromy} and \eqref{eq:sexticfuchsian}.

Denoting the element of $\Gamma_\mathrm{F}$ that sends the LCS point at infinity to an $\infty$-equivalent point $x \in \mathbb{R}$ by $\gamma_x$,
\begin{equation}
    \gamma_x \cdot \infty = x \,,
\end{equation}
we obtain, using \eqref{eq:projectiveboundarymap},%
\footnote{
This formula remains valid in the integral basis.
The projective boundary map in the rational basis \eqref{eq:genperiods} satisfies $\xi(\infty) = [0, 0, 0, 0, 1]$.
Therefore $\xi_\mathrm{int}(\infty) = [ \Lambda^I{}_J \xi^J(\infty)]  = \xi(\infty)$ for the transition matrix $\Lambda$ in \eqref{eq:integerbasisLambda}.
}
\begin{equation}\label{eq:sexticboundaryfunction}
    F^I(x) = \frac{\rho(\gamma_x)^I{}_4}{\rho(\gamma_x)^0{}_4} \,.
\end{equation}
For example, $\gamma_\mathrm{LG} \cdot \infty = -\sqrt{3}$ and therefore,
\begin{equation}
    F^I(-\sqrt{3}) = \frac{\left(M_\mathrm{LG}\right)^I{}_4}{\left(M_\mathrm{LG}\right)^0{}_4} = \left(1,0,0,0,0\right)^T \,. 
\end{equation}
By considering longer words in $\gamma_\mathrm{LG}$ and $\gamma_\mathrm{LCS}$, we can similarly compute boundary values at further $\infty$-equivalent points.

\begin{figure}[htb]
\centering
\includegraphics[width = 0.49\textwidth]{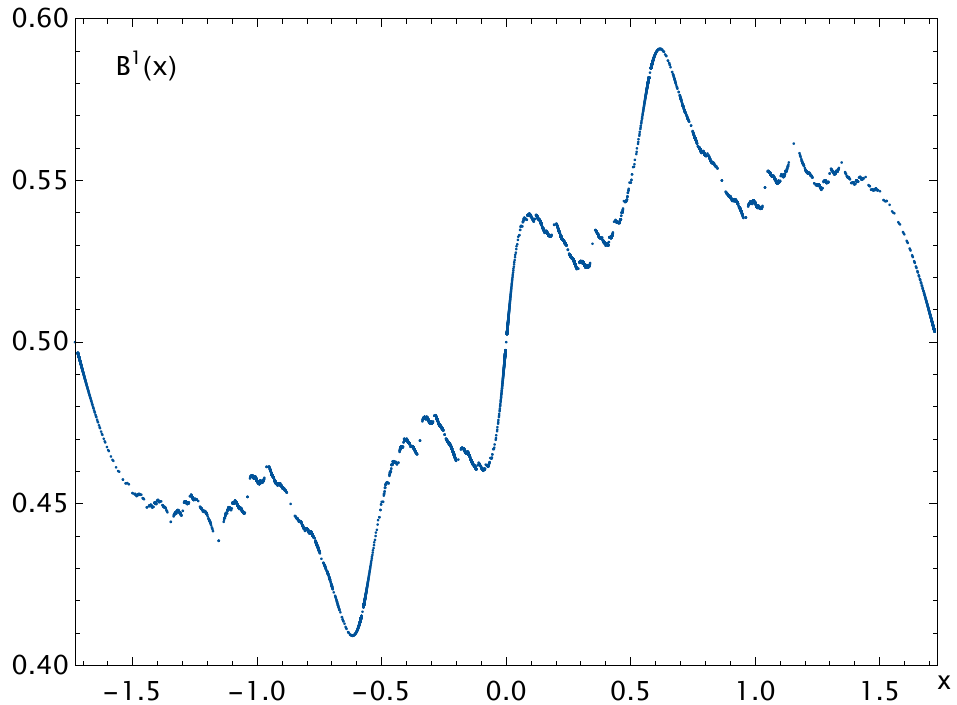}
\hfill
\includegraphics[width = 0.49\textwidth]{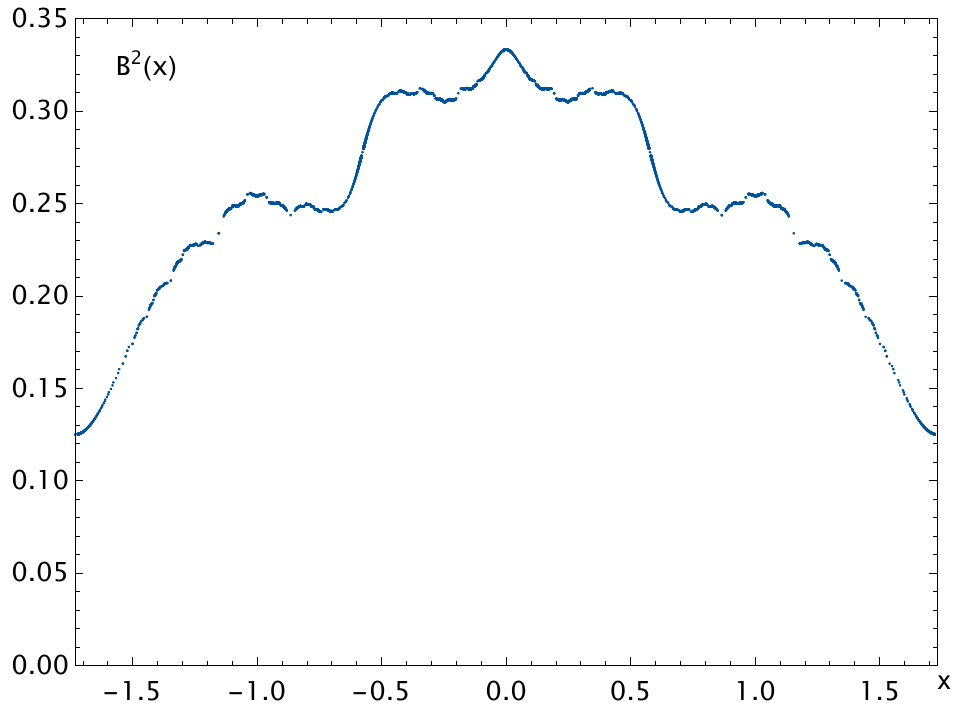}\\
\vspace{0.2cm}
\includegraphics[width = 0.49\textwidth]{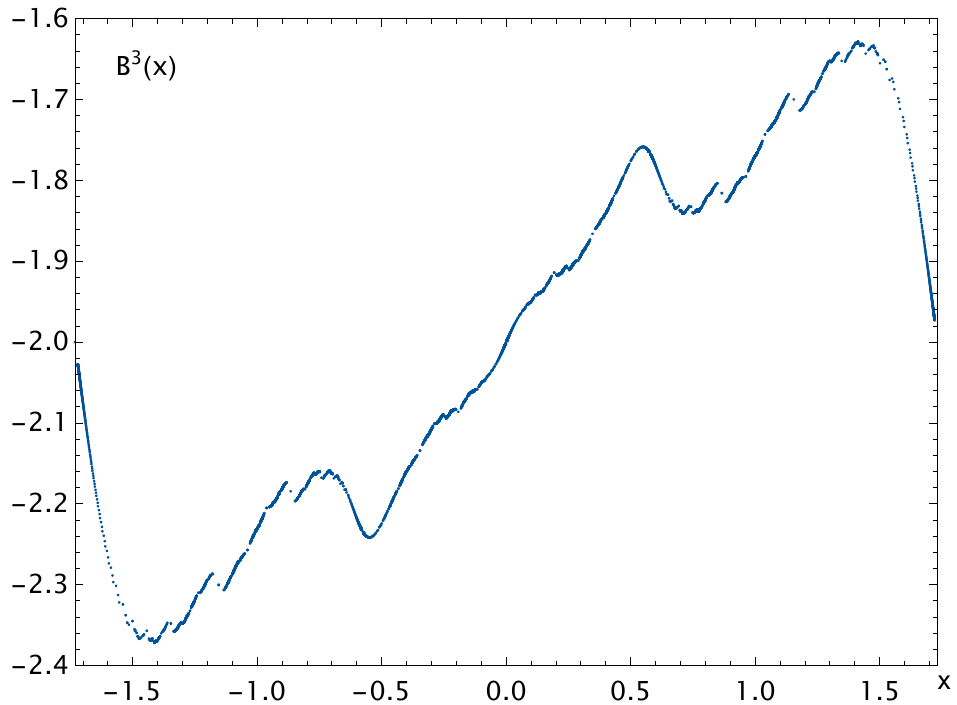}
\hfill
\includegraphics[width = 0.49\textwidth]{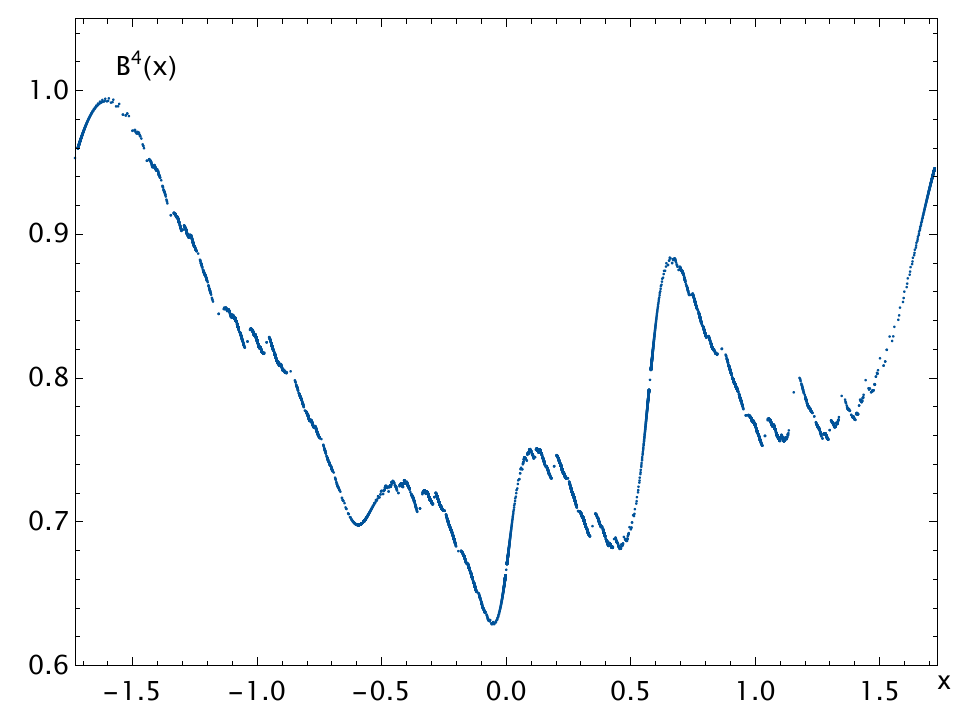}
\caption{The $T$-periodic boundary functions $B^I(x)$, defined in \eqref{eq:sexticB}, on the interval $[T/2, T/2] = [-\sqrt{3}, \sqrt{3}]$.
The first function satisfies $B^0(x) = 1$ and is not displayed. 
}
\label{fig:sexticboundaryfunctions}
\end{figure}

To describe $F^I(x)$ on the entire real axis, we write, analogously to \eqref{eq:nilpotent31}, 
\begin{equation}\label{eq:sexticB}
    F(x) = e^{x N / T} B(x) \,,
\end{equation}
where $N = \log M_\mathrm{LCS}$.
It then follows from $\rho$-equivariance that $B(x)$ is $T$-periodic,
\begin{equation}\label{eq:sexticperiodicity}
    B^I(x + T) = B^I(x) \,.
\end{equation}
It therefore suffices to compute $F^I(x)$ at $\infty$-equivalent points in an interval of width $T=2\sqrt{3}$ using \eqref{eq:sexticboundaryfunction}.
These points are dense, so continuity determines $F^I(x)$ throughout the interval.
Eqs.~\eqref{eq:sexticB} and \eqref{eq:sexticperiodicity} then determine $F^I(x)$ on the entire real axis.
We illustrate the resulting functions $B^I(x)$ in Figure~\ref{fig:sexticboundaryfunctions}.

\subsubsection{Differentiability of the boundary map}
\label{sec:sexticC1}

The boundary functions displayed in Figure~\ref{fig:sexticboundaryfunctions} appear to exhibit a fractal or self-similar behavior.
Even though they are continuous, it is not clear whether they are smooth or even differentiable.
In Appendix~\ref{app:C1test} we formulate a simple necessary condition for the boundary map to be continuously differentiable.

To this end, we consider the element
\begin{equation}
    \gamma = \gamma_{LCS}^{-1} \cdot \gamma_{LG}= \frac 12 \begin{pmatrix}
        3 \sqrt{3} & -5 \\
        -1 & \sqrt{3} \\
    \end{pmatrix}  \in \Gamma_\mathrm{F} \,. 
\end{equation}
Being hyperbolic, it has two fixed points on the boundary of $\mathbb{H}$.
The attracting one is located at 
\begin{equation}
    x_+ = -2 \sqrt{2}-\sqrt{3} \,.
\end{equation}
Writing $\gamma = \left(\begin{smallmatrix}a & b \\ c & d \end{smallmatrix}\right)$ and using \eqref{eq:app:kappagamma2}, we compute
\begin{equation}
    \kappa_\gamma = \gamma'(x_+) = \frac{1}{(c x_+ + d)^2} = 5 - 2 \sqrt{6} \,.
\end{equation}
On the other hand, using the matrix representation \eqref{eq:sexticmonodromy}, we obtain the eigenvalues of $\rho(\gamma)$,
\begin{equation}
    \operatorname{Spec}\left(\rho(\gamma)\right) = \left\{ - 7 - 4 \sqrt{3},\, -1,\, - e^{2 \pi i / 6},\, e^{2 \pi i / 3},\, -7 + 4 \sqrt{3} \right\} \,,
\end{equation}
sorted by non-increasing absolute value.
Normalizing the remaining eigenvalues by the first yields the eigenvalues of the differential of the corresponding projective map at the attracting fixed point.
We find that
\begin{equation}
    \kappa_\gamma \notin\left\{7-4 \sqrt{3}, \left(7-4 \sqrt{3}\right) e^{\pm \pi i / 3},97-56 \sqrt{3} \right\}\,.
\end{equation}
Therefore, the projective boundary map $\xi(x) = [F^I(x)]$ and its normalized representative are not $C^1$.
In particular, as explained in Appendix~\ref{app:schwarzreflection}, this implies that the projective period map \eqref{eq:projectiveperiods} of the mirror Fermat sextic does not admit a continuous extension to $\overline{\mathbb{H}}$.
Consequently, neither do the normalized holomorphic $(4,0)$-periods \eqref{eq:normalizedperiods}.

\subsection{Computation of the index and numerical examples}

Having established that the period representation of the mirror Fermat sextic is relatively Anosov, we can apply our results from Section~\ref{sec:coveringspaceindex}.
There we found that the index of $DW$, for $W$ the flux superpotential \eqref{eq:GVW}, can be computed as a contour integral along the boundary of the upper half-plane.
This integral is given by the sum \eqref{eq:indexsum} of the contributions from infinity and from each zero of $f_g = g_I F^I$ on the real axis.
While the contribution from infinity is determined by Table~\ref{table:result_infty_gen}, the contribution of an $\infty$-equivalent zero is obtained by first transforming the flux vector according to \eqref{eq:gprime}, then applying Table~\ref{table:result_infty_gen} to the transformed flux vector and subtracting one.

\subsubsection{Integral fluxes}

Table~\ref{table:result_infty_gen} applies to the flux coefficients $g_I$ in the rational basis \eqref{eq:genperiods}.
To impose flux quantization,%
\footnote{For a discussion of further subtleties regarding flux quantization on the mirror sextic, see, e.g., \cite{Lust:2024aeg}.}
we express these coefficients in terms of the fluxes in the integral basis using the transition matrix \eqref{eq:integerbasisLambda},
\begin{equation}
    g_I^{\mathrm{rat}} = (\eta_{\mathrm{rat}})_{IJ}  (\Lambda^{-1})^J{}_K \, g_{\mathrm{int}}^K \,.
\end{equation}
This gives, denoting the fluxes in the integral basis by $g^I$,
\begin{equation}
\begin{aligned}
    \bigl(g_I^{\mathrm{rat}}\bigr)
    ={}& \left(
    g^4+\tfrac{15}{4}g^2-\tfrac{15}{8}g^1-\tfrac{75}{64}g^0,\,
    g^3+3g^2-\tfrac34 g^1,\,
    -6g^2+3g^1+\tfrac{15}{4}g^0,\,
    -6g^1,\,
    6g^0
    \right) \,.
\end{aligned}
\end{equation}
The resulting conditions on $g^I \in \mathbb{Z}$ in the integral basis are summarized in Table~\ref{table:result_infty_sextic}.
The rest of Section~\ref{sec:coveringspaceindex} applies unchanged.

\begin{table}[htb]
\centering
\renewcommand{\arraystretch}{1.4}
\begin{tabular}{|c | c | c | c | c | c | c |}
\hline
$g^0$ & $g^1$ & $g^2$ & $g^3$ & $g^4$ & $\Delta_g$  & $\ind_\infty(f_g)$ \\
\hline\hline

$\neq0$ & --- & --- & --- & ---  & --- & $1$ \\
\hline
\multirow{5}{*}{$0$} & $\neq0$ & --- & --- & ---  & --- & $\frac12$ \\
\cline{2-7}

& \multirow{4}{*}{$0$} & \multirow{2}{*}{$\neq0$} & \multirow{2}{*}{---} & \multirow{2}{*}{---} & $<0$  & $1$ \\
\cline{6-7}
& & & & & $\geq0$ & $0$ \\
\cline{6-7}

\cline{3-7}
& & \multirow{2}{*}{$0$} & $\neq0$ & ---  & --- & $\frac12$ \\
\cline{4-7}
& & & $0$ & $\neq0$  & --- & $0$ \\
\hline
\end{tabular}
\captionsetup{singlelinecheck=false}
\caption[Index at infinity for nonzero real fluxes]{
Contribution to the index at infinity for nonzero real fluxes in the integral period basis of the mirror Fermat sextic,
where, for $g^2 \neq 0$,
$
    \Delta_g = -\frac{(g^3+3g^2)^2+3g^2(4g^4+15g^2)}{72(g^2)^2} 
$.
}
\label{table:result_infty_sextic}
\end{table}

\subsubsection{Boundary zeros}
\label{sec:findingboundaryzeros}

Once an $\infty$-equivalent zero on the real axis is known, determining its contribution to the index is relatively straightforward.
Locating zeros reliably, or excluding their existence on a given interval, is, however, more challenging.
In principle, one may compute $f_g$ at a large number of $\infty$-equivalent points on the real axis using the method outlined in Section~\ref{sec:sexticboundarymap} and search for zeros.
However, only finitely many points can be evaluated in practice, so this procedure does not guarantee that all zeros on the real axis are found.
An additional difficulty arises from the possibility of zeros at $\infty$-inequivalent points.

Nonetheless, precise statements can often be made using the properties of the boundary map.
For example, if two points are found where $f_g$ has opposite signs, continuity guarantees the existence of a zero between them.
This motivates the use of a bisection-type root-finding algorithm.

Moreover, the functions $B^I(x)$ in \eqref{eq:sexticB} are bounded by continuity and periodicity, which can be used to exclude zeros in certain intervals.
For example, at sufficiently large $|x|$, the leading polynomial term of $f_g$ dominates, so no zeros can occur.
By $\rho$-equivariance, the same argument excludes zeros in a sufficiently small punctured interval around any $\infty$-equivalent point.
The size of this interval depends on the point, the flux, and the bounds on $B^I(x)$.

Finally, assuming the conjectured bound \eqref{eq:inequivalentbound}, all zeros on the real axis contribute non-positively, so it is often unnecessary to determine all of them.
Instead, as outlined in Section~\ref{sec:sufficientindex}, it suffices to find sufficiently many sign changes of $f_g$ on the real axis to conclude that the index must be negative.

\subsubsection{Numerical examples}
\label{sec:examples}

In the following, we illustrate some of the above considerations by means of a few examples.

\paragraph{a) No zeros on the boundary.}

\begin{figure}[p]
  \centering

  \subfloat[$g^I = (1, 1, 1, 1, 10)^T$\label{fig:example1}]{%
    \includegraphics[width=0.45\textwidth]{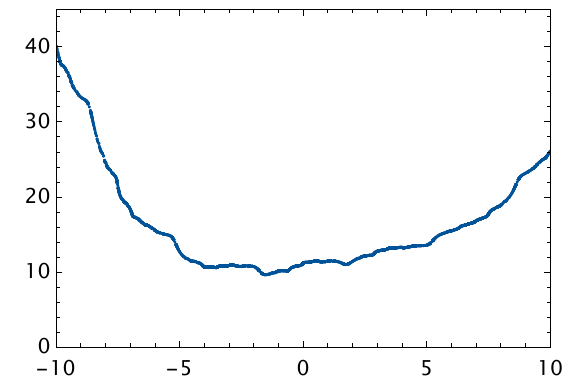}\hspace{0.05\textwidth}%
    \includegraphics[width=0.45\textwidth]{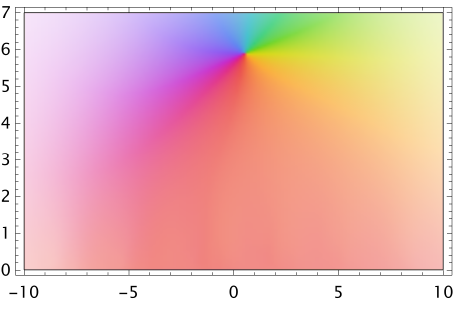}}

  \subfloat[$g^I = (0, 1, 0, -1, 0)^T$\label{fig:example2}]{%
    \includegraphics[width=0.45\textwidth]{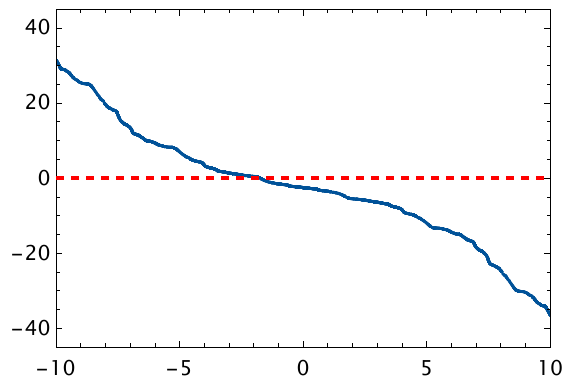}\hspace{0.05\textwidth}%
    \includegraphics[width=0.45\textwidth]{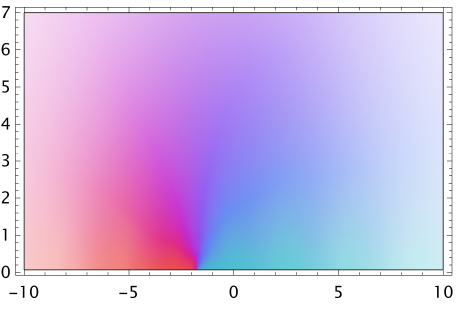}}

  \subfloat[$g^I = (1, 1, 2, 1, -1)^T$\label{fig:example3}]{%
    \includegraphics[width=0.45\textwidth]{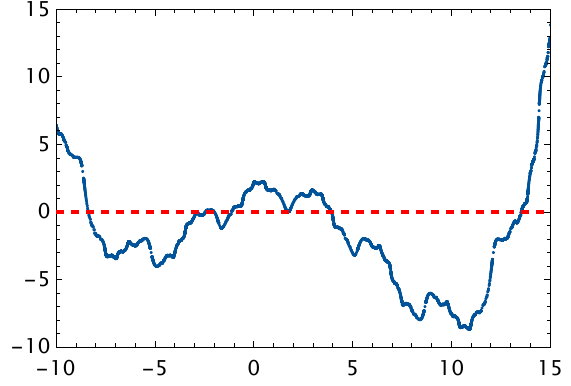}\hspace{0.05\textwidth}%
    \includegraphics[width=0.45\textwidth]{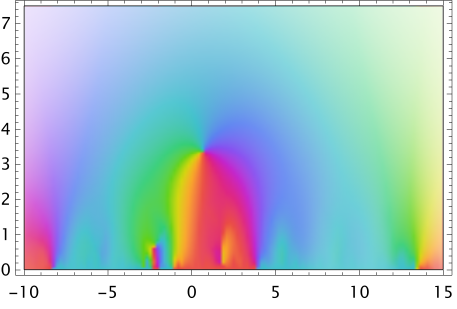}}

  \subfloat[$g^I = (0, 0, 1, 0, 0)^T$\label{fig:example4}]{%
    \includegraphics[width=0.45\textwidth]{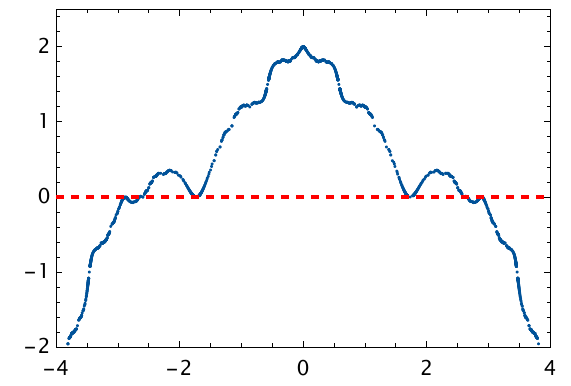}\hspace{0.05\textwidth}%
    \includegraphics[width=0.45\textwidth]{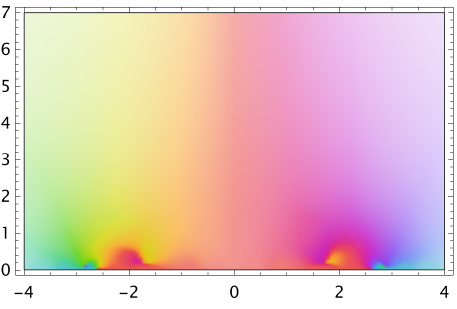}}

    \caption{Boundary functions $f_g (x)$ (on the left) and phase profiles of $f_g$ (on the right) for the examples considered in this section. The red line corresponds to $f(x)=0$.}
    \label{fig:examples}
\end{figure}

We first consider the integral flux
\begin{equation}
    g^I =(1,1,1,1,10) \,.
\end{equation}
Numerical evaluation of $f_g$ on the real axis using the method of Section~\ref{sec:sexticboundarymap} indicates that it has no boundary zeros, see Figure~\ref{fig:example1}.
Therefore, using the sum formula \eqref{eq:indexsum}, the index is
\begin{equation}
    \ind_{\mathbb{H}}(DW) = \ind_\infty(f_g) = 1 \,,
\end{equation}
which implies the existence of a point $z \in \mathbb{H}$ with $f_g(z) = 0$ and hence a flux vacuum in the interior of the upper half-plane.
This is consistent with the numerical plot of $f_g$ on $\mathbb{H}$ in Figure~\ref{fig:example1}, which indicates a single zero of index $1$.

More generally, consider arbitrary fluxes with $g^0\neq0$.
The polynomial asymptotics \eqref{eq:summaryapprox} imply that $f_g(x)/g^0$ tends to $+\infty$ as $x\to\pm\infty$, so continuity ensures that it has a finite global minimum.
Since changing $g^4$ adds a constant to $f_g$, taking $g^4/g^0$ sufficiently large makes $f_g/g^0$ strictly positive on the entire real axis.
The index is then $1$, independently of the conjecture \eqref{eq:inequivalentbound}.

This construction is not specific to the sextic but applies to all one-parameter Calabi--Yau fourfolds in the class considered here.
However, at fixed $g^0,\dots,g^3$, tadpole cancellation places an upper bound on the ratio $g^4/g^0$, limiting the number of flux vacua that can be constructed in this way.

\paragraph{b) An $\infty$-equivalent boundary zero.}

We next consider the flux
\begin{equation}
    g^I=(0,1,0,-1,0) \,.
\end{equation}
Its self-intersection is $g^I\eta_{IJ}g^J=-2$.
By the self-duality properties of the Hodge decomposition \eqref{eq:hodgedecomposition}, this flux class admits no self-dual representative at any point in moduli space.
This excludes flux vacua in $\mathbb{H}$ and hence implies $\ind_{\mathbb{H}}(DW)=0$.

The boundary calculation reproduces this result.
Since $g^0=0$ and $g^1\neq0$, Table~\ref{table:result_infty_sextic} gives
\begin{equation}
    \ind_\infty(f_g)=\frac12 \,.
\end{equation}
At the $\infty$-equivalent zero $x=-\sqrt3=\gamma_\mathrm{LG}\cdot\infty$, applying \eqref{eq:gprime} yields
\begin{equation}
    g_x=(0,1,-1,-4,8) \,,
\end{equation}
so its contribution is
\begin{equation}
    \ind_{-\sqrt3}(f_g) =\ind_\infty(f_{g_x})-1 =-\frac12 \,.
\end{equation}
The two contributions therefore cancel.
Numerical evaluation reveals no additional boundary zeros.
This is consistent with the vanishing index and the absence of interior zeros in Figure~\ref{fig:example2}.

\paragraph{c) A negative index from sign changes.}

We now consider the flux
\begin{equation}
    g^I=(1,1,2,1,-1) \,.
\end{equation}
Since $g^0\neq0$, the contribution from infinity is $\ind_\infty(f_g)=1$.
Moreover, evaluating $f_g$ at $\infty$-equivalent points on the real axis reveals at least six sign changes, see Figure~\ref{fig:example3}.
Assuming the conjectured bound \eqref{eq:inequivalentbound}, each sign-changing zero contributes at most $-1/2$, while any additional zeros contribute non-positively.
We therefore obtain
\begin{equation}
    \ind_{\mathbb{H}}(DW)  \leq 1-\frac62=-2 \,.
\end{equation}
This guarantees the existence of a flux vacuum without requiring a complete determination of the boundary zeros.

The numerical phase plot in Figure~\ref{fig:example3} illustrates the presence of several zeros of $f_g$ in $\mathbb{H}$, consistent with the existence of flux vacua established by the boundary index.

\paragraph{d) Non-isolated and $\infty$-inequivalent boundary zeros.}

Finally, we consider the  flux
\begin{equation}
    g^I=(0,0,1,0,0) \,,
\end{equation}
for which $f_g$ has infinitely many zeros on the real axis.
This flux is invariant under the subgroup $H=\langle\gamma\rangle$ generated by the hyperbolic element
\begin{equation}
    \gamma=\gLCS\gLG^2\gLCS \,.
\end{equation}
The points $x_n=\gamma^n\cdot\infty$, for $0 \neq n\in\mathbb{Z}$, are $\infty$-equivalent zeros.
Since, by \eqref{eq:gprime},
\begin{equation}
    g_{x_n} = \rho(\gamma)^{-n}g = g \,,
\end{equation}
and $\ind_\infty(f_g)=1$, each contributes
\begin{equation}
    \ind_{x_n}(f_g)=\ind_\infty(f_g)-1=0 \,.
\end{equation}
These zeros accumulate at the two fixed points
\begin{equation}
    x_\pm=\lim_{n\to\infty}\gamma^{\pm n}\cdot\infty
    =\pm\sqrt7 \,,
\end{equation}
which are non-isolated, $\infty$-inequivalent zeros.
Numerically, we observe that $f_g$ changes sign at both points.

Assuming that the boundary sum formula and the conjectured bound \eqref{eq:inequivalentbound} extend to this situation, the sign changes imply
\begin{equation}
    \ind_{x_\pm}(f_g)\leq-\frac12 \,, 
\end{equation}
and therefore
\begin{equation}
    \ind_{\mathbb{H}}(DW) \leq 1-\frac12-\frac12 = 0 \,.
\end{equation}
This bound is consistent with a numerical analysis of $f_g$ in the upper half-plane, which indicates the absence of interior zeros; see also Figure~\ref{fig:example4}.

This result can be confirmed by a more rigorous analysis on the quotient $\mathbb{H}/H$, identifying zeros related by the action of $H$.
We leave this analysis to future work.

\subsubsection{A systematic flux scan}
\label{sec:scan}

Beyond individual examples, the sufficient condition of Section~\ref{sec:sufficientindex} can be used to search systematically for fluxes admitting vacua.
Such a search may also be used to provide heuristic evidence for our conjectured bound \eqref{eq:inequivalentbound}.

To carry out the search, we implemented a systematic scan of integral fluxes.
For each flux, our algorithm computes the contribution from infinity and searches for sign changes on the real axis to obtain a bound on the index.
To restrict the region where zeros can appear, it uses the $\gLCS$-equivariant expansion \eqref{eq:sexticB} and combines it with numerical bounds on the functions $B^I(x)$, as in Section~\ref{sec:findingboundaryzeros}.
If no zeros are found, it assumes the index to be equal to the contribution from infinity.

Moreover, to simplify and accelerate the evaluation of $f_g(x)$ on the real axis, it approximates the periodic functions $B^I(x)$ by truncated Fourier series with modes $|n|\leq45$.
Further numerical safeguards are implemented to reduce the risk of spurious sign changes or missed zeros caused by approximation errors.
For details of the implementation, we refer to the accompanying Mathematica notebook.

We first applied this algorithm to all nonzero integral fluxes with components $|g^I| \leq 4$ and non-positive self-intersection.
Such fluxes cannot lead to a zero of $f_g$ in the interior of $\mathbb{H}$ (see, for example, the argument at the end of Section~\ref{sec:continuity31}), and therefore must have vanishing index.
Indeed, for all of the $10\,288$ fluxes in this class, our algorithm obtained a bound which is compatible with $\ind_\mathbb{H}(f_g) = 0$.
This provides a consistency check of our method and the conjectured bound \eqref{eq:inequivalentbound}.%
\footnote{We performed an analogous consistency test for the seventh model in Table~3.2 of \cite{vandeHeisteeg:2024lsa}.
By the arguments of \cite{filip2022uniformizationweight3variations}, it has a relatively Anosov period representation and hence belongs to the class of models to which our methods can be applied.
For all tested nonzero fluxes with $|g^I|\leq4$ and non-positive self-intersection, the numerical results were compatible with a vanishing index.}
However, we could not reliably establish the existence of $\infty$-inequivalent zeros in the sample set.

We then applied the algorithm to all integral fluxes with components $|g^I| \leq 2$.
This gives a total number of $3\,125$ fluxes, of which $585$ have non-positive self-intersection and are subsequently excluded from the scan.
Among the remaining $2\,540$ fluxes, the algorithm yielded a nonzero index in $492$ cases, corresponding to approximately $19\%$.
For the other $2\,048$ cases, our bounds did not establish a nonzero index and therefore remained inconclusive regarding the existence of flux vacua.

\section{Discussion}
\label{sec:discussion}

In this paper we demonstrated that classical flux vacua of M-theory compactified on Calabi-Yau fourfolds with a single complex structure modulus can be detected from just asymptotic period data and monodromy data without requiring any information on the periods in the interior of the complex structure moduli space.

Working on the covering space instead of a single fundamental domain of the monodromy group allowed us to overcome a main obstacle in the index construction of \cite{Lust:2024aeg}:
By construction, period integrals are single-valued on the covering space, so the index can be computed along a contour that remains at the asymptotic boundary of the covering space and does not probe its interior.

As in \cite{Lust:2024aeg}, we applied our results to the mirror Fermat sextic, including explicit examples and a systematic flux scan.
However, most of the discussion remains general and applicable to other one-parameter Calabi-Yau fourfolds, as classified recently in \cite{vandeHeisteeg:2024lsa}.
It remains to establish rigorously the conjectured bound \eqref{eq:inequivalentbound} for $\infty$-inequivalent points, which we have so far tested only heuristically.

\medskip

A main technical requirement in our construction is that the period maps extend continuously to the boundary of the covering space, which we identified with the complex upper half-plane.
Due to the intricate and distinct action of the monodromy group on the periods and on the upper half-plane, it is a priori not evident that  this property holds.
Indeed, upon visual inspection (see, e.g., Figure~\ref{fig:sexticboundaryfunctions}), the restriction of the period maps to the boundary of the upper half-plane appears to take a rather complicated and non-smooth form.

For Calabi-Yau fourfolds with a single complex structure modulus, we could prove that the projective non-holomorphic $(3,1)$-periods extend continuously to the boundary of the covering space if the corresponding monodromy representation is relatively Anosov. 
These are the relevant periods for our computation as they enter the F-terms of the GVW flux superpotential.
Interestingly, the same does, in general, not hold for the projective holomorphic $(4,0)$-periods, as we could show for concrete examples.
It was previously established in the mathematics literature \cite{filip2022uniformizationweight3variations} that certain one-parameter Calabi-Yau fourfolds indeed have relatively Anosov period representations, ensuring that our continuity result applies to these models.

Relatively Anosov representations are an active topic of mathematical research, with connections to differential geometry, geometric group theory, and dynamical systems (see, e.g., \cite{Labourie:2006, Kapovich:2018, AIF_2021__71_5_2169_0, filip2022uniformizationweight3variations, Zhu2022RelativelyAR, ZhuZimmer2, Canary:2022, Weisman:2023, canary2025patterson, Weisman:2026}).
Their role in physics is far less established, and their physical interpretation in the context of Calabi-Yau compactifications of string theory or M-theory remains unclear. 
Importantly, not all Calabi-Yau period representations satisfy this property; see, e.g., Section~\ref{sec:relanosovfourfolds} above, as well as \cite{filip2022uniformizationweight3variations}.
It would thus be interesting to understand what distinguishes compactifications with relatively Anosov monodromy representations, and whether this property can be linked to a physical consistency condition, possibly in the context of the Swampland program where certain constraints on monodromy and duality groups have already been identified \cite{Delgado:2024skw}.

\medskip

It is natural to ask whether our results extend to other classes of Calabi-Yau manifolds.
For elliptic curves and certain one-parameter families of K3-surfaces, the projective holomorphic period maps can be written as $[z,1]$ or $[z^2, z, 1]$, respectively, in suitable rational bases, with holomorphic coordinate $z$ on the upper half-plane (see, e.g., \cite{Nagura:1993kd, Dolgachev:1996xw, hartmann2013period}).
They therefore extend continuously to its boundary.

On the other hand, relatively Anosov period representations have also been identified for certain Calabi-Yau threefolds with a single complex structure modulus \cite{filip2022uniformizationweight3variations}.
However, none of their period maps satisfy our transversality condition \eqref{eq:nullvectorcondition}.
While the differentiability criterion developed in Appendix~\ref{app:dxicontinuity} may still rule out a continuous boundary extension of the holomorphic $(3,0)$-period map, our methods do not allow to establish whether the non-holomorphic $(2,1)$-period map admits such an extension.
Analogous statements apply to Calabi-Yau manifolds of complex dimension five or higher.

\medskip

Perhaps even more interesting physically would be an extension to Calabi-Yau fourfolds with more than one complex-structure modulus.%
\footnote{This would also include Type IIB orientifold compactifications on threefolds, via their duality with F-theory on $(CY_3 \times T^2)/\mathbb{Z}_2$ in special cases, or, more generally, by the Sen limit of F-theory on elliptically fibred fourfolds \cite{Sen:1996vd,Sen:1997gv}.}
This would potentially open up applications to landscape questions of flux compactifications, such as the tadpole conjecture \cite{Bena:2020xrh}.

When applying our technique to statistical analyses of the flux landscape, one should be aware of potential over-counting issues related to the transition from a single fundamental domain to the covering space.
Vacua that are related by a simultaneous monodromy transformation of both the fluxes and the moduli should be regarded as physically equivalent but are not identified on the covering space.
A statistical count of physically inequivalent vacua therefore requires identifying configurations related by the monodromy group.

A special case arises for fluxes that are stabilized by an infinite subgroup of the full monodromy group, as in our last example in Section~\ref{sec:examples}.
Here, for a fixed flux, each vacuum is accompanied by infinitely many equivalent copies on the covering space.
Treating such cases consistently would require quotienting the universal covering space by the respective stabilizing subgroup and performing the index computation on the resulting intermediate covering space.
We intend to revisit this construction in a future publication \cite{quotientpaper}.

\medskip

Even though the boundary period maps can be obtained in a simple algebraic way from the leading-order asymptotic expansion of the periods at a given infinite-distance singularity combined with monodromy information, the resulting functions are not simple but rather intricate, possibly reflecting aspects of the complexity of the flux landscape (see, e.g., \cite{Denef:2006ad, Halverson:2018cio, Grimm:2023lrf, Grimm:2026haa}).
Nonetheless, assuming the conjectured bound \eqref{eq:inequivalentbound}, our sufficient condition in Section~\ref{sec:sufficientindex} allows us in certain cases to establish the existence of a flux vacuum from evaluations of the boundary functions at a small number of boundary points.

\section*{Acknowledgements}

We thank Sergei Alexandrov, Daniele Atzori, Markus Dierigl, Simion Filip, Th\'eodore Fischer, Thomas Grimm, Damian van de Heisteeg, Dieter L\"ust, Fernando Marchesano, Miguel Montero, Eliott Morgensztern, Jakob Moritz, Boris Pioline, David Prieto, Sanjay Raman, Andr\'es Sambarino, Cumrun Vafa, Max Wiesner, and Kai Xu for fruitful discussions and exchanges.
We would also like to thank Simion Filip for sharing an unpublished updated version of his paper \cite{filip2022uniformizationweight3variations}.
The work of PB was partially supported by an STSM Grant from COST Action CA22113: Fundamental challenges in theoretical physics, funded by COST (European Cooperation in Science and Technology).
During the preparation of this work, generative AI software was used for language editing and proofreading.

\appendix

\section{Relatively Anosov period representations}
\label{app:anosov}

In this appendix we collect technical results on relatively Anosov period representations used in the main text and prove the boundary extension criterion stated in Section~\ref{sec:continuityconditions}.

In Appendix~\ref{app:boundarymaps}, we summarize and derive properties of the boundary maps $\xi$ and $\xi^*$.
In Appendix~\ref{app:continuityproof}, we show that the map $\overline{F} : \overline{\mathbb{H}} \to \mathbb{CP}^{b_H(X_4)-1}$ defined in \eqref{eq:overlineF} is continuous provided that  conditions~(\ref{item:anosov})--(\ref{item:nullvectorcondition}) are satisfied.
In Appendix~\ref{app:dxicontinuity}, we discuss the differentiability of the boundary map.

\subsection{The boundary maps \texorpdfstring{$\xi$}{xi} and \texorpdfstring{$\xi^\ast$}{xi*}}
\label{app:boundarymaps}

We begin by collecting some properties of the boundary maps $\xi : \partial \mathbb{H} \rightarrow \mathbb{CP}^{d-1}$ and $\xi^\ast : \partial \mathbb{H} \rightarrow (\mathbb{CP}^{d-1})^*$ that will be needed below and in the main text.

\subsubsection{Singular-value characterization}
\label{app:singularvalue}

For a $P_1$-relatively Anosov representation $\rho$, strong dynamics preservation and the singular value decomposition imply that the boundary maps can be recovered as limits of its leading singular subspaces \cite{Kapovich:2018,Zhu2022RelativelyAR, AIF_2021__71_5_2169_0, Canary:2022}.%
\footnote{Some of these references formulate the result for the Bowditch boundary. For a Fuchsian group of finite covolume, this boundary is canonically identified with the limit set $\Lambda (\Gamma_\mathrm{F}) = \partial\mathbb{H}$.}

We recall that any matrix $M \in M_d (\mathbb{C})$ admits a singular value decomposition
\begin{equation}
    M = U D V^\dagger \,, \qquad D = \operatorname{diag}(\mu_1, \dots, \mu_d) \,,
\end{equation}
where $U$ and $V$ are unitary and the singular values $\mu_i$ are ordered non-increasingly, i.e.\ $\mu_i \geq \mu_{i+1}$.
For $i = 1, \dots, d-1$ and $\mu_i > \mu_{i+1}$, we define
\begin{equation}\label{eq:Uidef}
    U_i(M) = \operatorname{span}\{u_1, \dots, u_i\} \in \mathrm{Gr} (i, \mathbb{C}^d) \,,
\end{equation}
where $u_1, \dots, u_i$ are the first $i$ columns of $U$.
Equivalently, $U_i(M)$ is spanned by the $i$ leading singular directions of $M$.
The condition $\mu_i>\mu_{i+1}$ ensures that $U_i(M)$ is well-defined.

Fix an arbitrary point $z_0 \in \mathbb H$.
For any sequence $(\gamma_n) \subset \Gamma_\mathrm{F}$ such that
\begin{equation}
    \lim_{n \to \infty} \gamma_n \cdot z_0 = x \in \partial \mathbb{H} \,,
\end{equation}
relative Anosovness implies the singular gaps\cite{Zhu2022RelativelyAR, AIF_2021__71_5_2169_0, Canary:2022}
\begin{equation}\label{eq:singularvaluegap}
    \lim_{n\to \infty} \frac{\mu_2\!\left(\rho(\gamma_n)\right)}{\mu_1\!\left(\rho(\gamma_n)\right)} = 
    \lim_{n\to \infty} \frac{\mu_d\!\left(\rho(\gamma_n)\right)}{\mu_{d-1}\!\left(\rho(\gamma_n)\right)} = 0 \,.
\end{equation}
This implies that $U_1(\rho(\gamma_n))$ and $U_{d-1}(\rho(\gamma_n))$ are well-defined for all sufficiently large $n$,
and the boundary maps are given by \cite{AIF_2021__71_5_2169_0}
\begin{equation}\label{eq:xiandxistar}
    \xi(x) =  \lim_{n \to \infty} U_1 (\rho(\gamma_n))  \qquad \xi^*(x) = \lim_{n \to \infty} U_{d-1} (\rho(\gamma_n)) \,.
\end{equation}
These limits are independent of the sequence $(\gamma_n)$ and the base point
$z_0$.

\subsubsection{Boundary maps and the intersection form}
\label{app:xiproperties}

We now consider $d = b_H (X_4) $ and $\rho$ to be the period representation on the horizontal middle cohomology of $X_4$.
By \eqref{eq:etamonodromy}, $\rho$ preserves the intersection form $\eta$,
\begin{equation}\label{eq:rhoeta}
    \rho^T(\gamma) \eta \rho(\gamma) = \eta \,,
\end{equation}
for any $\gamma \in \Gamma_\mathrm{F}$.
We collect some consequences for the boundary maps.

\begin{lemma}\label{lemma:etaorthogonality}
For every $x \in \partial \mathbb{H}$, the boundary maps satisfy
\begin{equation}\label{eq:etaorthogonality}
\xi^*(x) = \xi(x)^{\perp_\eta} \equiv \left\{v \,\middle|\, \xi^I(x) \eta_{IJ} v^J = 0 \right\} \,,
\end{equation}
where $\xi^I(x)$ is any homogeneous representative of $\xi(x)$.
\end{lemma}
\begin{proof}
The singular vectors of $\rho(\gamma_n)$ are given by the eigenvectors of
\begin{equation}
    A_n = \rho(\gamma_n) \rho(\gamma_n)^T \,,
\end{equation}
while the eigenvalues of $A_n$ are the squares of the singular values of $\rho(\gamma_n)$.

In a real basis where $\eta^2 \propto \mathbf{1}$,%
\footnote{A suitable change of basis always exists and does not affect the boundary maps, since they do not depend on the inner product used in the singular value decomposition.}
it follows from \eqref{eq:rhoeta} that
\begin{equation}
    A_n \eta = \eta A_n^{-1} \,.
\end{equation}
Therefore, if $v$ is an eigenvector of $A_n$ with eigenvalue $\lambda$, $\eta v$ is an eigenvector of $A_n$ with eigenvalue $\lambda^{-1}$.
In particular, $\eta$ maps the leading singular direction of $\rho(\gamma_n)$ to its smallest singular direction.
Since the eigenspaces of the symmetric matrix $A_n$ are mutually orthogonal with respect to the canonical inner product, 
\begin{equation}
    U_{d-1}(\rho(\gamma_n)) = U_1(\rho(\gamma_n))^{\perp_\eta} \,.
\end{equation}
Taking the limit in \eqref{eq:xiandxistar} yields \eqref{eq:etaorthogonality}.
\end{proof}

Viewing $\xi^*(x)$ as a point of the dual projective space, this lemma motivates the notation
\begin{equation}
    \xi^*(x) = \left[\xi_I(x) \right] \,,
\end{equation}
where $\xi_I(x) = \eta_{IJ} \xi^J(x)$.

\begin{corollary}\label{cor:xinull}
For every $x \in \partial \mathbb{H}$,
\begin{equation}\label{eq:xinull}
    \xi^I(x) \eta_{IJ} \xi^J(x) = 0 \,,
\end{equation}
i.e.\ $\xi(x)$ is null with respect to the intersection form $\eta$.
\end{corollary}
\begin{proof}
By definition \eqref{eq:Uidef}, $U_1(M) \subset U_{d-1}(M)$.
Taking the limit in \eqref{eq:xiandxistar} therefore gives
\begin{equation}
    \xi(x) \subset \xi^*(x) \,.
\end{equation}
Together with Lemma~\ref{lemma:etaorthogonality} this implies \eqref{eq:xinull}.
\end{proof}

\begin{corollary}\label{cor:Ftransversality}
Suppose that $F$ satisfies \eqref{eq:nullvectorcondition}.
Then, for every $x\in\partial\mathbb{H}$ and
$z\in\overline{\mathbb{H}}\setminus\{x\}$,
\begin{equation}\label{eq:Ftransversality}
    \overline F(z)\notin\xi^*(x) \,.
\end{equation}
\end{corollary}
\begin{proof}
For $z\in\partial\mathbb{H}\setminus\{x\}$, \eqref{eq:Ftransversality} follows from
the transversality \eqref{eq:transversality} of $\xi$ and $\xi^\ast$.

For $z\in\mathbb{H}$, Corollary~\ref{cor:xinull} and
\eqref{eq:nullvectorcondition} imply
\begin{equation}
    \xi^I(x)\eta_{IJ}F^J(z)\neq 0 \,.
\end{equation}
The result then follows from Lemma~\ref{lemma:etaorthogonality}.
\end{proof}

\subsubsection{Boundary values at punctures}

Consider now a puncture $p \in \partial \mathcal{M}_{X_4}$ and choose a lift $x \in \partial \mathbb{H}$.
Let $\gamma \in \Gamma_\mathrm{F}$ be a generator of the stabilizer of $x$, and denote the corresponding local monodromy by
\begin{equation}
    M = \rho(\gamma) \,.
\end{equation}
We recall from \eqref{eq:Ni} that there exists a nilpotent matrix $N$ such that
\begin{equation}\label{eq:Niapp}
    M^{k} = e^{N}
\end{equation}
for some $k \in \mathbb{N}$.
We denote the degree of nilpotency of $N$ by $r$, as defined in \eqref{eq:degreeofnilpotency}.

\begin{proposition}
\label{prop:boundary}
Suppose that the period representation $\rho$ is relatively Anosov.
Then 
\begin{equation}\label{eq:rankcondition}
    \operatorname{rank} N^r = 1 \,,
\end{equation}
and the boundary maps at $x$ are given by
\begin{equation}\label{eq:xiatpunctures}
    \xi(x) = \operatorname{im} N^r \,, \qquad \xi^*(x) = \ker N^r \,.
\end{equation}
\end{proposition}
\begin{proof}
Since $\gamma$ is parabolic,
\begin{equation}
    \lim_{n \to \infty} \gamma^{k n} \cdot z_0 = x \,,
\end{equation}
for every $z_0 \in \mathbb{H}$.
It follows from \eqref{eq:xiandxistar} that
\begin{equation}\label{eq:xipuncture}
    \xi(x)  = \lim_{n \to \infty} U_1 \!\left(M^{k n} \right) \,,\qquad
    \xi^*(x)  = \lim_{n \to \infty} U_{d-1} \!\left(M^{k n}\right) \,.
\end{equation}

Using $M^k=e^N$ and the nilpotency of $N$, we obtain
\begin{equation}\label{eq:Mexpansion}
    \lim_{n\rightarrow \infty} \frac{M^{k n}}{n^r} = \frac{N^r}{r!}  \,.
\end{equation}
Therefore, it follows from the singular value gap \eqref{eq:singularvaluegap} that
\begin{equation}
    0 = \lim_{n \to \infty} \frac{\mu_2(M^{k n})}{\mu_1(M^{k n})} =  \frac{\mu_2(N^r)}{\mu_1(N^r)} \,.
\end{equation}
Since $N^r \neq 0$, $\mu_1(N^r) \neq 0$ and hence $\mu_2(N^r) = 0$.
This implies 
\begin{equation}
    \operatorname{rank} N^r = 1 \,,
\end{equation}
and $N^r$ has a unique leading singular direction, given by its image.
Thus, \eqref{eq:xipuncture} and \eqref{eq:Mexpansion} give
\begin{equation}
    \xi(x) = \operatorname{im} N^r \,.
\end{equation}

It follows from \eqref{eq:etamonodromy} that
\begin{equation}
    \label{eq:app:etaN}
    \eta N^r = (-1)^r (N^r)^T \eta \,.
\end{equation}
Therefore, using Lemma~\ref{lemma:etaorthogonality}
\begin{equation}\begin{aligned}
    \xi^*(x) = \left(\operatorname{im} N^r\right)^{\perp_\eta} &= \left\{v \,\middle |\, v^T \eta N^r u = 0 \,,\; u \in \mathbb{R}^d \right\} \\
    &= \left\{v \,\middle |\, (N^r v)^T \eta  u = 0 \,,\; u \in \mathbb{R}^d \right\} = \ker N^r \,,
\end{aligned}\end{equation}
where in the last step we used that $\eta$ is non-degenerate.
\end{proof}

\begin{corollary}\label{cor:Nxixi}
    There exists a complex homogeneous representative of $\xi(x)$ such that
    \begin{equation}\label{eq:app:Nxixi}
        \left(N^r\right)^I{}_J = \xi^I(x) \xi_J(x) \,,
    \end{equation}
    where $\xi_J(x) = \eta_{JK} \xi^K(x)$. 
\end{corollary}
\begin{proof}
    It follows immediately from Proposition~\ref{prop:boundary} in combination with Lemma~\ref{lemma:etaorthogonality}.
\end{proof}

\subsection{Proof of continuity}
\label{app:continuityproof}

We show that $\overline F$ is continuous on $\overline{\mathbb{H}}$ under the conditions stated in Section~\ref{sec:continuityconditions}.
Since $F = \overline{F}|_\mathbb{H}$ and $\xi = \overline{F}|_{\partial\mathbb{H}}$ are continuous,
it suffices to show continuity at boundary points $x\in\partial\mathbb{H}$ along
sequences $(z_n)\subset\mathbb H$ with $z_n\to x$.

Since $\mathbb{CP}^{d-1}$ is compact, it suffices to show that every convergent subsequence of \(F(z_n)\) has limit $\xi(x)$.
After passing to an arbitrary such subsequence and relabeling, we may
assume that $F(z_n)$ converges.
We will show that
\begin{equation}
    \lim_{n\to\infty} F(z_n)=\xi(x) \,.
\end{equation}
Since $F(z_n)$ converges, we may pass to further subsequences below without changing its limit.

As in the main text, we choose a fundamental domain $\mathcal F \subset \mathbb H$ for the action of the monodromy group $\Gamma_\mathrm{F}$.
For each $n$, we write
\begin{equation}
    z_n = \gamma_n \cdot w_n\,,
\end{equation}
with $ \gamma_n \in \Gamma_\mathrm{F}$ and $w_n \in \mathcal{F}$.

We distinguish two cases.

\paragraph{Case I:}
There exists a compact subset $K \subset \mathcal{F}$ such that $w_n \in K$.

Fix an arbitrary point $z_0 \in \mathbb H$.
The sequences $(\gamma_n \cdot z_0)$ and $(\gamma_n^{-1} \cdot z_0)$ take values in the compact space $\overline{\mathbb{H}}$.
Therefore, after passing to a further subsequence of $(z_n)$ and relabeling, we may assume that both $\gamma_n \cdot z_0$ and $\gamma_n^{-1} \cdot z_0$ converge.
We define
\begin{equation}
    z_* = \lim_{n \to \infty} \gamma_n \cdot z_0 \in \partial \mathbb{H} \,, \qquad
    z_\mathrm{opp} = \lim_{n \to \infty} \gamma_n^{-1} \cdot z_0 \in \partial \mathbb{H} \,.
\end{equation}
Indeed, if either $z_*$ or $z_\mathrm{opp}$ belonged to $\mathbb{H}$, proper discontinuity would imply, after passing to a subsequence, that $\gamma_n$ is constant.
This contradicts $\gamma_n\cdot w_n \to x \in \partial \mathbb{H}$, since $(w_n)$  lies in the compact set $K$.

Since $F$ is continuous on the interior $\mathbb{H}$, $F(K)$ is compact in $\mathbb{CP}^{d-1}$.
Moreover, by Corollary~\ref{cor:Ftransversality}, it is disjoint from $\xi^\ast(z_{\mathrm{opp}})$.
Therefore, by strong dynamics preservation \eqref{eq:strongdynpre}
\begin{equation}
    \lim_{n \to \infty} F(z_n) = \lim_{n \to \infty} \rho(\gamma_n) \cdot F(w_n) = \xi(z_*) \,.
\end{equation}

It remains to show that $z_* = x$.
We use that the defining action of $\Gamma_\mathrm{F} \subset \mathrm{PSL}(2, \mathbb{R})$ on $\overline{\mathbb{H}}$ is itself strongly dynamics-preserving, with both boundary maps given by the identity.
Since $K$ is compact and does not include $z_\mathrm{opp}$, this yields
\begin{equation}
    x = \lim_{n\to\infty} \gamma_n \cdot w_n = \lim_{n\to\infty} \gamma_n \cdot z_0 = z_* \,.
\end{equation}

\paragraph{Case II:}
The sequence $(w_n)$ is not contained in any compact subset of
$\mathcal F$.

By the compactness of $\overline{\mathbb H}$, we may pass to a subsequence such that $w_n \to y$ for some $y\in\partial\mathbb H$.
After an isometric change of coordinates on $\mathbb{H}$, we may assume $y = \infty$.
In particular, this implies
\begin{equation}\label{eq:Imlimit}
    \lim_{n \to \infty} \Im w_n = \infty \,.
\end{equation}

Fix an arbitrary point $z_0 \in \mathbb H$.
We denote by $\gamma_{\infty} \in \Gamma_\mathrm{F}$ a generator of the stabilizer of the cusp at infinity, chosen such that it acts on $\mathbb{H}$ by
\begin{equation}\label{eq:app:gammaLCS}
    \gamma_{\infty} \cdot z = z + T \,, 
\end{equation}
for some $T > 0$.
It follows that there exists $m_n \in \mathbb{Z}$ such that, with 
\begin{equation}
    \tilde \gamma_n = \gamma_n \gamma_\mathrm{\infty}^{-m_n} \,,\qquad \tilde w_n = \gamma_\mathrm{\infty}^{m_n} \cdot w_n \,,
\end{equation}
the bound
\begin{equation}\label{eq:app:gammanbound}
    0 \leq \Re \left(\tilde \gamma_n^{-1} \cdot z_0\right) < T 
\end{equation}
holds.
By construction, $z_n = \tilde \gamma_n \cdot \tilde w_n$.

By \eqref{eq:app:gammaLCS}, $\Im \tilde w_n = \Im w_n$, and therefore condition \eqref{eq:limitcondition} in combination with \eqref{eq:Imlimit} implies that
\begin{equation}\label{eq:Ftildean}
    \lim_{n \to \infty} F(\tilde w_n) = \xi(\infty) \,.
\end{equation}

As in case I, after passing to a further subsequence of $(z_n)$ and relabeling, we may assume that both $\tilde \gamma_n \cdot z_0$ and $\tilde \gamma_n^{-1} \cdot z_0$ converge and define
\begin{equation}
    z_* = \lim_{n \to \infty} \tilde \gamma_n \cdot z_0 \,, \qquad
    z_\mathrm{opp} = \lim_{n \to \infty} \tilde \gamma_n^{-1} \cdot z_0 \,.
\end{equation}

We first assume that $z_* \in \mathbb{H}$.
Since the action of $\Gamma_\mathrm{F}$ is properly discontinuous on $\mathbb{H}$, the sequence $\tilde \gamma_n \cdot z_0$ becomes eventually constant and, after passing to a subsequence, $\tilde \gamma_n = \gamma_* \in \Gamma_\mathrm{F}$ for all sufficiently large $n$.
Moreover $\tilde \gamma_n \cdot \tilde w_n \rightarrow x \in \partial \mathbb{H}$ and $\tilde w_n \rightarrow \infty$ imply that $x = \gamma_* \cdot \infty$.
Therefore, using \eqref{eq:Ftildean}, we obtain
\begin{equation}
    \lim_{n \rightarrow \infty} F(z_n) = \rho (\gamma_*) \cdot \lim_{n \rightarrow \infty} F(\tilde w_n) = \rho (\gamma_*) \cdot \xi(\infty) = \xi(x) \,.
\end{equation}

We now consider $z_* \in \partial \mathbb{H}$ and therefore also $z_\mathrm{opp} \in \partial \mathbb{H}$.
Write
\begin{equation}
    \tilde\gamma_n^{-1} = \begin{pmatrix}
        a_n & b_n \\
        c_n & d_n
    \end{pmatrix} \,.
\end{equation}
If $c_n=0$, then $\tilde\gamma_n^{-1}$ fixes $\infty$ and hence belongs to $\left\langle \gamma_{\infty} \right\rangle$.
Therefore, using \eqref{eq:app:gammaLCS},
\begin{equation}
    \Im(\tilde\gamma_n^{-1}\cdot z_0)=\Im(z_0) \,.
\end{equation}
On the other hand, if $c_n \neq 0$, $\gamma_\infty$ and $\tilde\gamma_n^{-1}$ generate a non-elementary Fuchsian group,
so J{\o}rgensen's inequality \cite{Jorgensen:1976} gives
\begin{equation}
    \left| \mathrm{Tr} (\gamma_\infty)^2 - 4 \right| + \left| \mathrm{Tr} \left(\gamma_\infty \tilde\gamma_n^{-1} \gamma_\infty^{-1} \tilde\gamma_n\right) - 2 \right| \geq 1 \,.
\end{equation}
With \eqref{eq:app:gammaLCS} this reduces to
\begin{equation}
    c_n^2 \geq \frac{1}{T^2} \,.
\end{equation}
It follows that
\begin{equation}
    \Im (\tilde\gamma_n^{-1} \cdot z_0) = \frac{\Im (z_0)}{|c_n z_0 + d_n|^2} \leq \frac{1}{ c_n^2 \Im (z_0)} \leq \frac{T^2}{\Im (z_0)} \,.
\end{equation}
Therefore, $\Im(\tilde\gamma_n^{-1}\cdot z_0)$ is uniformly bounded, and so is $\Im(z_{\mathrm{opp}})$.
By \eqref{eq:app:gammanbound}, $\Re(z_{\mathrm{opp}})$ is also bounded, and hence
\begin{equation}\label{eq:zoppnotinfty}
    z_{\mathrm{opp}} \neq \infty\,.
\end{equation}

It follows from \eqref{eq:Ftildean} that the set
\begin{equation}
    \left\{F(\widetilde w_n) \,\middle|\, n \in \mathbb{N}\right\}
    \cup \left\{\xi(\infty)\right\} 
\end{equation}
is compact in $\mathbb{CP}^{d-1}$.
Moreover, by \eqref{eq:zoppnotinfty} and Corollary~\ref{cor:Ftransversality}, it is disjoint from $\xi^*(z_\mathrm{opp})$.
Therefore, by strong dynamics preservation
\begin{equation}
    \lim_{n \to \infty} F(z_n) = \lim_{n \to \infty} \rho(\tilde \gamma_n) \cdot F(\widetilde w_n) = \xi(z_*) \,,
\end{equation}
and $z_* = x$, as in case I.

\subsection{Differentiability of the boundary map}
\label{app:dxicontinuity}

The proof in the previous subsection makes no assumption on the differentiability of $F$ or the boundary map $\xi$.
In particular, even if $F$ is smooth on the interior $\mathbb{H}$, the boundary map $\xi$ need not be differentiable.

However, if $F$ is holomorphic, then, as we show in Section~\ref{app:schwarzreflection}, $\xi$ is necessarily real analytic and therefore in $C^\infty$.
Moreover, in Section~\ref{app:C1test} we give a simple necessary condition for $\xi$ to be $C^1$.
Together, these observations provide a necessary condition for a holomorphic map $F$ to admit a continuous boundary extension with boundary values $\xi$.

\subsubsection{Schwarz Reflection}
\label{app:schwarzreflection}

Consider a holomorphic map $F : \mathbb{H} \to \mathbb{CP}^{d-1}$ that admits a continuous extension $\overline F : \overline{\mathbb{H}} \to \mathbb{CP}^{d-1}$ with real boundary values $\xi = \overline F|_{\partial\mathbb{H}}$.
For the case of interest in this paper, reality of the boundary map follows from the reality of the period representation and \eqref{eq:xiandxistar}.

Fix $x \in \mathbb{R}$ and choose a constant vector $\ell \in \mathbb{R}^d$ such that $\ell_I \xi^I(x) \neq 0$, where,
as above, $\xi^I$ and $\overline F^I$ denote homogeneous representatives of $\xi$ and $\overline F$, respectively.
By continuity, there exists a neighborhood $U \subset \overline{\mathbb{H}}$ of $x$ such that $\ell_I \overline F^I(z) \neq 0$ for all $z \in U$.
The normalized components%
\footnote{Cf.\ the definition of the normalized periods in Section~\ref{sec:projectiveperiods}.}
\begin{equation}
     F_\ell^I(z) = \frac{\overline F^I(z)}{\ell_J \overline F^J(z)} \,,
    \qquad z \in U \,,
\end{equation}
are continuous on $U$, holomorphic on $U \cap \mathbb{H}$, and real-valued on $U \cap \mathbb{R}$.
By the Schwarz reflection principle, each $F_\ell^I$ extends holomorphically 
to the reflected neighborhood in the lower half-plane, with $F_\ell^I(\bar z) = \overline{F_\ell^I(z)}$.
This implies that the normalized boundary components $F_\ell^I|_{U \cap \mathbb{R}}$ are real analytic.

For $x = \infty$, the same argument applies in the coordinate $w = - 1/z$.
Consequently, the boundary map $\xi$ is real analytic on $\partial\mathbb{H}$ and hence $C^\infty$.

\subsubsection{Necessary condition for continuous differentiability}
\label{app:C1test}

Consider a relatively Anosov representation $\rho : \Gamma_\mathrm{F} \to \mathrm{GL}(d, \mathbb{C})$ with boundary map $\xi : \partial \mathbb{H} \to \mathbb{CP}^{d-1}$.

For a hyperbolic element $\gamma \in \Gamma_\mathrm{F}$, denote its attracting fixed point by $x_+ \in \partial \mathbb{H}$,%
\footnote{Analogous arguments apply to the repelling fixed point $x_-$ by replacing $\gamma$ with $\gamma^{-1}$, which exchanges the roles of $x_+$ and $x_-$.}
and define
\begin{equation}
    \kappa_\gamma = \gamma'(x_+) \,.
\end{equation}
If $x_+ \neq \infty$, 
\begin{equation}\label{eq:app:kappagamma2}
    \kappa_\gamma = \frac{d}{d z} \Bigl.\left(\gamma \cdot z \right)\Bigr|_{z = x_+} = \frac{1}{(c x_+ + d)^2} \,, \qquad \gamma = \begin{pmatrix} a & b \\ c & d \end{pmatrix} \,.
\end{equation}
If $x_+ = \infty$, use the local coordinate $w = - 1/z$, i.e.\ conjugate $\gamma$ by $\left(\begin{smallmatrix}0 & 1 \\ -1 & 0 \end{smallmatrix}\right)$.

Moreover, denote the eigenvalues of $\rho(\gamma)$ by $\lambda_i$, $i = 1, \dots, d$, ordered such that
\begin{equation}
    | \lambda_1 | > | \lambda_2 | \geq \dots \geq | \lambda_d | \,.
\end{equation}
Then, we have the following proposition.
\begin{proposition}
    If $\xi \in C^1$, then, for every hyperbolic element $\gamma \in \Gamma_\mathrm{F}$,
    \begin{equation}\label{eq:C1condition}
        \kappa_\gamma \in \left \{ \frac{\lambda_2}{\lambda_1}, \dots, \frac{\lambda_d}{\lambda_1} \right\} \,.
    \end{equation}
\end{proposition}

\begin{proof}
    We first note that the derivative of $\xi$ cannot vanish anywhere.
    If $\xi'(x) = 0$ at some point $x \in \partial \mathbb{H}$, then, by equivarance $\xi' = 0$ along the entire orbit $\Gamma_\mathrm{F} \cdot x$, which is dense in $\partial \mathbb{H}$.
    By continuity of the derivative, $\xi' = 0$ everywhere and consequently $\xi$ is constant.
    This contradicts transversality, since $\xi(x) = \xi(y) \subset \xi^*(y)$ for distinct boundary points $x$ and $y$.
    
    Taking the derivative of the equivariance condition at the attracting fixed point gives
    \begin{equation}\label{eq:equivderivative}
        D(\rho(\gamma))_{\xi(x_+)} \xi'(x_+) = \kappa_\gamma \, \xi'(x_+) \,,
    \end{equation}
    where $D(\rho(\gamma))$ denotes the differential of the projective action of $\rho(\gamma)$.
    Therefore, $\xi'(x_+)$ is a non-vanishing eigenvector of $D(\rho(\gamma))_{\xi(x_+)}$ with eigenvalue $\kappa_\gamma$.

    It also follows from equivariance that
    \begin{equation}
        \xi(x_+) = \left[ v_1 \right] \,,
    \end{equation}
    for $v_1$ an eigenvector of $\rho(\gamma)$.
    Strong dynamics preservation implies that $v_1$ corresponds to the eigenvalue $\lambda_1$ of largest absolute value; see, e.g., \cite[Proposition~2.6(2)]{ZhuZimmer2}.

    Complete $v_1$ to a Jordan basis $v_1, \dots, v_d$ for $\rho(\gamma)$.
    In a neighborhood of $\left[ v_1 \right] \in \mathbb{CP}^{d-1}$, we adopt affine coordinates $(z^2, \dots, z^d) \in \mathbb{C}^{d-1}$, represented by
    \begin{equation}
        \left[v_1 + \sum_{i = 2}^d v_i z^i \right] \,.
    \end{equation}
    The projective action of $\rho(\gamma)$ sends this element to
    \begin{equation}\label{eq:projectiveaction}
         \left[v_1 + \frac{1}{\lambda_1} \sum_{i,j = 2}^d  v_j M^j{}_i z^i \right] \,,
    \end{equation}
    with $M^j{}_i$ an upper triangular matrix, given by the restriction of $\rho(\gamma)$ to $\left<v_2, \dots, v_d \right>$.
    Consequently, $M^j{}_i$ has eigenvalues $\lambda_2, \dots, \lambda_d$,
    and taking the derivative of \eqref{eq:projectiveaction} with respect to the local coordinates $z^i$ shows that the eigenvalues of $D(\rho(\gamma))_{\xi(x_+)}$ are given by
    \begin{equation}
        \left \{ \frac{\lambda_2}{\lambda_1}, \dots, \frac{\lambda_d}{\lambda_1} \right\} \,.
    \end{equation}
    Therefore, \eqref{eq:equivderivative} implies \eqref{eq:C1condition}.
\end{proof}

\section{Asymptotic Hodge theory and period expansions}
\label{app:asymptotichodge}

In this appendix we provide supplementary information from asymptotic Hodge theory that is needed to evaluate the boundary limit in condition~(\ref{item:limitcondition}) of the continuity criterion in Section~\ref{sec:continuityconditions}
for the projective $(4,0)$- and $(3,1)$-period maps.
In particular, we determine the leading terms in the nilpotent orbit expansion of the periods near $\mathrm{V}_{1,1}$ and $\mathrm{I}_{1,1}$ punctures.

Throughout this appendix, we denote the logarithmic coordinate around a given puncture by $t = x + i y$, as defined in \eqref{eq:deft}, and use the nilpotent orbit expansion \eqref{eq:nilpotent} and \eqref{eq:nilpotentexpansion} with $q = e^{2 \pi i t}$.
We denote the corresponding nilpotent monodromy logarithm by $N$ and its degree of nilpotency by $r$, so that $N^r \neq 0$ and $N^{r+1} = 0$.

\subsection{Limiting mixed Hodge structures}

We first recall a few notions of asymptotic Hodge theory that will be needed in the following discussion.
Our presentation follows \cite{vandeHeisteeg:2022gsp}.%
\footnote{We would like to thank Damian van de Heisteeg for very helpful discussions and explanations.}

\subsubsection{Hodge filtration and limits}

The Hodge decomposition of the middle cohomology of the fourfold $X_4$ defines the Hodge filtration $H^{4,0}(X_4,\mathbb{C}) = F^4 \subseteq \dots \subseteq F^0 = H^4(X_4,\mathbb{C})$ via
\begin{equation}
    F^p (X_4) = \bigoplus_{i=p}^4 H^{i, 4-i} (X_4, \mathbb{C})\,.
\end{equation}
Conversely, the Hodge decomposition can be recovered from the filtration through
\begin{equation}
    H^{p, q} (X_4, \mathbb{C}) = F^p (X_4) \cap \overline{F^q (X_4)} \,.
\end{equation}
This filtration depends on the complex structure modulus.

The horizontal part of the cohomology is generated by the holomorphic four-form $\Omega$ and its derivatives with respect to the complex structure modulus.
Restricted to this subspace, the Hodge filtration therefore satisfies
\begin{equation}
     \left< \Pi, \dots, \partial^{4-p} \Pi \right> \subseteq F^p (X_4) \cap H^4_H (X_4, \mathbb{C}) \,.
\end{equation}
For Calabi-Yau fourfolds, the non-horizontal part of the middle cohomology lies entirely in $H^{2, 2} (X_4, \mathbb{C})$.
Therefore, for $p \in \{4, 3\}$, this simplifies to
\begin{equation}
    \label{eq:app:pergenerate}
    F^4(X_4) = \left< \Pi \right>\,, \qquad F^3 (X_4) = \left< \Pi, \partial \Pi \right>\,.
\end{equation}

At the puncture under consideration, the Hodge filtration determines a limiting filtration $F^p_0$, defined by
\begin{equation}
    \label{eq:app:limiting}
    F^p_0 = \lim_{t \rightarrow i \infty} e^{- t N} F^p (t) \,.
\end{equation}
The limiting filtration is thus independent of $t$.

When inserting the nilpotent orbit expansion \eqref{eq:nilpotent} into \eqref{eq:app:pergenerate}, the exponential factor in \eqref{eq:app:limiting} drops out and one obtains
\begin{equation}\label{eq:app:F40}
    F^4_0 = \left<a_0\right> \,,
\end{equation}
where the vector $a_0$ is the leading coefficient in the expansion \eqref{eq:nilpotentexpansion}, normalized such that $a_0 \neq 0$.
There exist similar relations between $F^3_0$, $N$, and the coefficients $a_n$, but their precise form depends on the type of boundary.

\subsubsection{The Deligne splitting}

The limiting filtration admits an additional structure, the Deligne splitting
\begin{equation}
    F^0_0 = \bigoplus_{p, q = 0}^4 I^{p, q}\,.
\end{equation}
We will use the following properties of this splitting.
The action of $N$ satisfies
\begin{equation}
    \label{eq:app:NandI}
    N I^{p, q} \subseteq I^{p-1, q-1}\,,
\end{equation}
while the degree of nilpotency $r$ of $N$, is given by
\begin{equation}\label{eq:app:degreeofnilpotency}
    r = \max \bigl\{ p+q - 4 \,\bigm|\, \dim I^{p,q} \neq 0 \bigr\} \,.
\end{equation}
This follows from the fact that, for $p+q \geq 4$, $N^{p+q-4}$ acts as an isomorphism from $I^{p,q}$ to $I^{4-q,4-p}$.
Moreover, for $p+q \geq 4$, every nonzero $v\in I^{p,q}$ with $N^{p+q-3} v =0$ satisfies the polarization condition
\begin{equation}\label{eq:polarization}
    i^{p-q} \, v^\dagger \eta N^{p+q-4} v > 0 \,.
\end{equation}
Finally, the limiting filtration is recovered as
\begin{equation}
    \label{eq:app:FandI}
    F^p_0 = \bigoplus_{i=p}^4 \bigoplus_{q=0}^4 I^{i, q} \,.
\end{equation}

\begin{figure}
    \centering
    \subfloat[\centering]{
    \begin{tikzpicture}[
            scale=0.9,
            x={(-0.8cm,0.8cm)},
            y={(0.8cm,0.8cm)}
        ]
        
        \foreach \i in {0,...,4}
          \draw[gray!40] (\i,0)--(\i,4);
        
        \foreach \j in {0,...,4}
          \draw[gray!40] (0,\j)--(4,\j);
        
        \fill (4,4) circle (2pt);
        \node at (4.2,4.2) {\scriptsize $\dim I^{4, 4} = 1$};
        \fill (0,0) circle (2pt);
        \node at (0.2,0.2) {\scriptsize $\dim I^{0, 0} = 1$};
        \fill (3,3) circle (2pt);
        \node at (3.2,3.2) {\scriptsize $\dim I^{3, 3} = 1$};
        \fill (2,2) circle (2pt);
        \node at (2.15,2.25) {\scriptsize $\dim I^{2, 2} = h^{2, 2}_H$};
        \fill (1,1) circle (2pt);
        \node at (1.3,1.1) {\scriptsize $\dim I^{1, 1} = 1$};
        \end{tikzpicture}%
    \label{fig:V11}}
    \hspace{0.8cm}
    \subfloat[\centering]{
    \begin{tikzpicture}[
            scale=0.9,
            x={(-0.8cm,0.8cm)},
            y={(0.8cm,0.8cm)},
        ]
        
        \foreach \i in {0,...,4}
          \draw[gray!40] (\i,0)--(\i,4);
        
        \foreach \j in {0,...,4}
          \draw[gray!40] (0,\j)--(4,\j);
        
        \fill (4,0) circle (2pt);
        \node at (4.0,0.4) {\scriptsize $\dim I^{4, 0} = 1$};
        \fill (0,4) circle (2pt);
        \node at (0.4,4.0) {\scriptsize $\dim I^{0, 4} = 1$};
        \fill (3,3) circle (2pt);
        \node at (3.3,3.1) {\scriptsize $\dim I^{3, 3} = 1$};
        \fill (2,2) circle (2pt);
        \node at (2.15,2.25) {\scriptsize $\dim I^{2, 2} = h^{2, 2}_H$};
        \fill (1,1) circle (2pt);
        \node at (1.3,1.1) {\scriptsize $\dim I^{1, 1} = 1$};
        \end{tikzpicture}%
    \label{fig:I11}}
    \caption{Hodge-Deligne diamonds of (a) a $\mathrm{V}_{1, 1}$-type boundary and (b) an $\mathrm{I}_{1, 1}$-type boundary. Unmarked sites correspond to zero-dimensional subspaces.}
    \label{fig:V11_and_I11}
\end{figure}
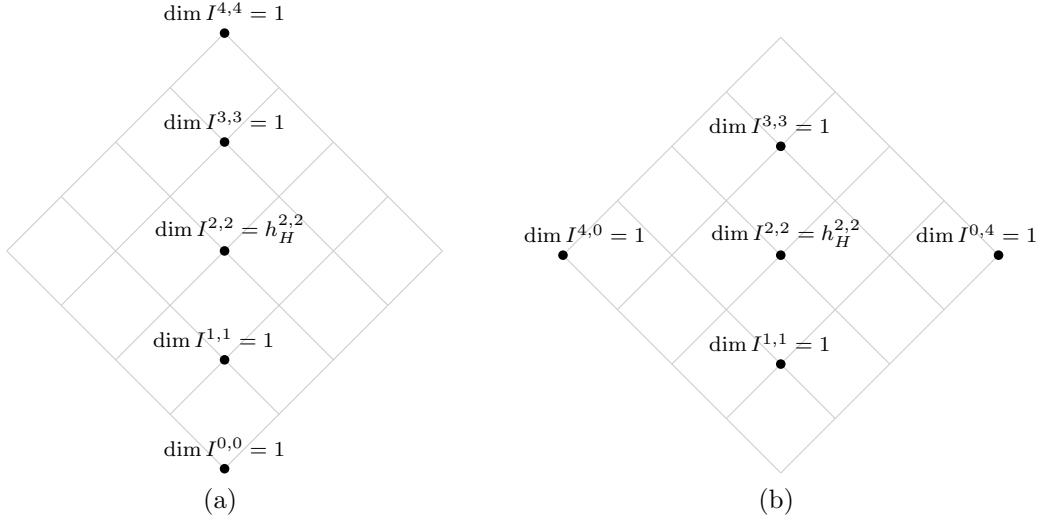

The dimensions of the Deligne subspaces $I^{p, q}$ classify the boundary types of the complex structure moduli space.
As discussed in Section~\ref{sec:relanosovfourfolds}, one-modulus Calabi-Yau fourfolds with a relatively Anosov period representation admit only two types of infinite-order punctures:
\begin{itemize}
    \item $\mathrm{V}_{1, 1}$ singularities, corresponding to LCS points;
    \item $\mathrm{I}_{1, 1}$ singularities, which lie at finite distance in the moduli space and are called C-points in \cite{vandeHeisteeg:2024lsa}.
\end{itemize}
Figure~\ref{fig:V11_and_I11} shows the corresponding Hodge--Deligne diamonds, from which the dimensions of the subspaces $I^{p, q}$ can be inferred.
Together with \eqref{eq:app:NandI} and \eqref{eq:app:FandI}, this information constrains the action of $N$ on the limiting filtration $F_0^p$.
We discuss both types of singularities in turn.

\subsection{\texorpdfstring{$\mathrm{V}_{1,1}$}{V11} boundary}

The Hodge-Deligne diamond of a $\mathrm{V}_{1, 1}$-type boundary is displayed in Figure~\ref{fig:V11}.
Using \eqref{eq:app:degreeofnilpotency}, we determine that the degree of nilpotency of $N$ is
\begin{equation}
    r = 4 \,,
\end{equation}
i.e.\ the corresponding monodromy is maximally unipotent.
Using \eqref{eq:app:FandI} we also read off
\begin{equation}
    F_0^4 = I^{4,4} \,.
\end{equation}
Together with \eqref{eq:app:F40}, this implies
\begin{equation}
    \left< a_0 \right> = I^{4,4} \,.
\end{equation}
By \eqref{eq:app:NandI}, all Deligne subspaces other than $I^{4,4}$ lie in the kernel of $N^4$, so we conclude that $N^4$ acts non-trivially on $I^{4,4}$ and hence
\begin{equation}\label{eq:app:N4a0}
    N^4 a_0 \neq 0 \,,
\end{equation}
as otherwise $N^4 = 0$, in contradiction with $r = 4$.

\subsubsection{$(3,1)$ periods}
\label{app:LCS31}

The leading coefficient in the nilpotent orbit expansion \eqref{eq:Bexpansion} of the (3,1) periods reads
\begin{equation}
    b_{00}(y) = v_{00}(y) N a_0 + \frac{i}{2} v'_{00}(y) a_0 \,,
\end{equation}
where, from \eqref{eq:Vexpansion},
\begin{equation}\label{eq:appv00}
    v_{00}(y) = a_0^\dagger \eta e^{2iy N} a_0 = \frac23 y^4 a_0^\dagger \eta N^4 a_0 + \mathcal{O}(y^3) \,.
\end{equation}
Expanding the exponential, we obtain
\begin{equation}\label{eq:app:LCSb00}
    e^{t N} b_{00}(y) =  \frac{i}{18} y^3 t^3 \bar t \left( a_0^\dagger \eta N^4 a_0 \right) N^4 a_0 + \mathcal{O}(y^2 |t|^4) \,,
\end{equation}
and note that the leading term is non-vanishing since $N^4 a_0 \neq 0$ by \eqref{eq:app:N4a0} and
$a_0^\dagger \eta N^4 a_0 > 0$ by the polarization condition \eqref{eq:polarization}.

\subsubsection{Explicit expressions in the rational basis}
\label{app:LCSrationalbasis}

Finally, we give explicit expressions in the rational basis introduced in \eqref{eq:genperiods}.
Using the LCS polynomial expansion given there, we obtain the corresponding monodromy matrix $M_\mathrm{LCS}$ and its logarithm $N$ from the shift $t \to t+1$, 
\begin{equation}\label{eq:LCSmonodromy}
    M_\mathrm{LCS} = \begin{pmatrix}
        1 & 0 & 0 & 0 & 0\\
        1 & 1 & 0 & 0 & 0\\
       \frac{1}{2} & 1 & 1 & 0 & 0\\
        \frac{1}{6} & \frac{1}{2} & 1 & 1 & 0\\
        \frac{1}{24} & \frac{1}{6} & \frac{1}{2} & 1 & 1
    \end{pmatrix} \,, \qquad N = \log M_\mathrm{LCS} =  \begin{pmatrix}
        0 & 0 & 0 & 0 & 0\\
        1 & 0 & 0 & 0 & 0\\
        0 & 1 & 0 & 0 & 0\\
        0 & 0 & 1 & 0 & 0\\
        0 & 0 & 0 & 1 & 0
    \end{pmatrix} \,.
\end{equation}
Moreover, one easily confirms that the vector
\begin{equation}\label{eq:aLCS}
   a_{0} = \left( 1, 0, 0, ic, 0\right)^T  
\end{equation}
reproduces \eqref{eq:genperiods} from the leading-order term \eqref{eq:polynomialperiodexpansion} in the nilpotent orbit expansion \eqref{eq:nilpotent}.

We also give explicit expressions for the expansion of the normalized $(3,1)$-periods used in Section~\ref{sec:polynomialapprox}.
Inserting \eqref{eq:LCSmonodromy} and \eqref{eq:aLCS} into \eqref{eq:appv00}, we obtain
\begin{equation}
    v_{00}(y) = \frac{2 \kappa}{3 T^4} y^4 - \frac{4 \kappa c}{T}y \,.
\end{equation}
Moreover, the normalization vector \eqref{eq:ellLCS} satisfies
\begin{equation}
    \ell_I a_0^I = 1 \,, \qquad \ell N = 0 \,.
\end{equation}
Therefore,
\begin{equation}
    \frac{b_{00}(y)}{\ell_I b_{00}^I(y)} = a_0 - \frac{2i}{T} \frac{v_{00}(y)}{v'_{00}(y)} N a_0 \,,
\end{equation}
and polynomial division gives
\begin{equation}
    \label{eq:pyhy}
    p(y) = a_0 - \frac{i y}{2 T} N a_0 \,, \qquad
    h(y) = \frac{9i c y T^2}{4 y^3 - 6 c T^3} N a_0 \,,
\end{equation}
with $p(y)$ and $h(y)$ as introduced in \eqref{eq:b00polynonmialdevision}.

\subsection{\texorpdfstring{$\mathrm{I}_{1,1}$}{I11} boundary}

The Hodge-Deligne diamond of a $\mathrm{I}_{1, 1}$-type boundary is displayed in Figure~\ref{fig:I11}.
Using \eqref{eq:app:degreeofnilpotency}, we read off the degree of nilpotency of $N$,
\begin{equation}
    r = 2 \,,
\end{equation}
and, using \eqref{eq:app:FandI} and \eqref{eq:app:F40},
\begin{equation}
    \left<a_0\right > = F^4_0 = I^{4,0} \,.
\end{equation}
It follows from \eqref{eq:app:NandI} that
\begin{equation}
    a_0 \in \ker N \,, \qquad a_0 \notin \operatorname{im} N \,.
\end{equation}
Since $N^2 a_0 = N a_0 = 0$, we must also inspect the exponential corrections.

Let $a_M$ be the first coefficient in the nilpotent orbit expansion \eqref{eq:nilpotentexpansion} that is not proportional to $a_0$.
Then, from \eqref{eq:app:pergenerate} and \eqref{eq:app:limiting}, we find
\begin{equation}
    \label{eq:app:genAAp}
    F_0^3 = \left < a_0, (N + 2\pi i M  ) a_M \right> \,.
\end{equation}
On the other hand, Figure~\ref{fig:I11} and \eqref{eq:app:FandI} give
\begin{equation}
    F_0^3 = I^{4, 0} \oplus I^{3, 3} \,.
\end{equation}
Similarly to the $\mathrm{V}_{1,1}$ case, all Deligne subspaces other than $I^{3,3}$ lie in the kernel of $N^2$.
Since $r=2$, $N^2$ must therefore act non-trivially on $I^{3,3}$.
Combining this with $Na_0=0$ and \eqref{eq:app:genAAp}, we find
\begin{equation}
    0 \neq N^2 (N+2\pi i M ) a_M = 2\pi i M   N^2 a_M \,,
\end{equation}
and hence
\begin{equation}\label{eq:N2aM}
    N^2 a_M \neq 0 \,.
\end{equation}

\subsubsection{$(3,1)$-periods}
\label{app:I1131}

For $n < M$, we write
\begin{equation}
    \label{eq:app:cn}
    a_n = c_n a_0 \,.
\end{equation}
Since $Na_0=0$ and $N^T\eta=-\eta N$, \eqref{eq:Vexpansion} gives
\begin{equation}
    v_{nm}(y) = \bar v_{mn}(y) = a_m^\dagger \eta a_n \,,\qquad n < M \,,
\end{equation}
and hence
\begin{equation}
    v'_{nm}(y) = v'_{mn}(y) = 0 \,, \qquad n < M \,.
\end{equation}
Inserting this into \eqref{eq:bnm} gives
\begin{equation}\label{eq:I11bnm}
    b_{nm}(y) = 2\pi i v_{0m}(y) a_0 \sum_{k=0}^n (n-2k) c_k c_{n-k} = 0 \,, \qquad n < M \,,
\end{equation}
where, in the last step, we used that the sum vanishes identically.
On the other hand,
\begin{equation}
    b_{M0}(y) =  \left( N + 2\pi  i M \right) v_{00}(y) a_M -  2  \pi i M v_{M0}(y) a_{0} \,.
\end{equation}
By the polarization condition \eqref{eq:polarization},
\begin{equation}
    v_{00}(y) = a_0^\dagger \eta a_0 > 0 \,,
\end{equation}
and therefore, with \eqref{eq:N2aM},
\begin{equation}\label{eq:N2bM0}
    N^2 b_{M0}(y) = 2 \pi i M v_{00}(y) N^2 a_M \neq 0 \,.
\end{equation}
Moreover, by \eqref{eq:I11bnm}, all other coefficients $b_{nm}(y)$ with $n+m\leq M$ vanish, and $b_{M0}(y)$ is the first non-vanishing coefficient in the expansion \eqref{eq:Bexpansion}.

\section{Index at infinity for \texorpdfstring{$g_4=g_3=0$ and $g_2\neq0$}{g4 = g3 = 0 and g2 ≠ 0}}
\label{app:exceptionalflux}

In this appendix we compute the contribution to the index at infinity for fluxes satisfying $g_4 = g_3 = 0$ and $g_2 \neq 0$.
As discussed in the main text, in this case the leading polynomial term of $f_g$ is proportional to $z(z+ \bar z)$ and therefore vanishes at $z=iR$ on the integration contour $\mathcal{C}_R$.
Its contribution cannot therefore be determined by the leading-order analysis alone, and we must retain subleading terms.

Since $\Im z$ is large near this point, the relevant subleading terms can be obtained from the nilpotent orbit expansion of the holomorphic periods.
As derived in Section~\ref{sec:polynomialapprox}, in this regime the normalized periods take the form
\begin{equation}\label{eq:app:largeyapprox}
    F (z, \bar z) \simeq e^{z N/T} \bigl( p(y) + h(y) \bigr) \,,
\end{equation}
up to exponentially suppressed corrections.
Here $p(y)$ is polynomial and $h(y)$ is properly rational in $y=\Im z$.
While the polynomial approximation $P(z,\bar z)=e^{zN/T}p(y)$ is valid in every $|z|\to\infty$ limit,
a nonzero rational correction $h(y)$ is only guaranteed to dominate the exponential corrections for sufficiently large $y$.

Computing $p(y)$ and $h(y)$ using \eqref{eq:pyhy}, for $g_4 = g_3 = 0$ we obtain 
\begin{equation}
    \label{eq:app:Fgg2}
    f_g \simeq g_2 \left( \frac{z (z+\overline{z})}{4 T^2} + \frac{i z}{T} a(y) \right) + g_1 \left( \frac{3 z+\overline{z}}{4T} + i a(y) \right)+ g_0  \,,
\end{equation}
where
\begin{equation}
    a(y) = \frac{9 c y}{4 \frac{y^3}{T^2} - 6 c T} \,.
\end{equation}

To determine the winding of $f_g$, we examine its argument along the contour.
For sufficiently large $R$, $f_g/g_2$ is real and positive at both the starting point $z=R$ and the endpoint $z=-R$.
Along the contour, it picks up a nonzero imaginary part and crosses the real axis from the upper to the lower half-plane near $z = i R$, close to the zero of the leading polynomial.
The sign of $\Re f_g/g_2$ at the crossing then determines the winding and the contribution to the index:
If the crossing occurs on the positive real axis, the total winding vanishes and $\ind_\infty(f_g)  = 0$.
On the other hand, if it occurs on the negative real axis, $f_g/g_2$ winds counter-clockwise around the origin once, and $\ind_\infty(f_g) = 1$.%
\footnote{For a more general discussion of this algorithm, see e.g.\ Appendix~A.1 in \cite{Lust:2024aeg}.}

It remains to determine the sign of $\Re f_g/g_2$ at the vanishing locus of $\Im f_g$ along the contour.
Writing $z=x+iy$ and neglecting exponentially suppressed corrections, we obtain
\begin{equation}
    \Im f_g = \left(g_1 + g_2 \frac{x}{T}\right)\left(\frac{y}{2T} + a(y) \right) \,.
\end{equation}
For sufficiently large $y$, the second factor is positive and $\Im f_g$ vanishes at
\begin{equation}
    x_\ast = - \frac{T g_1}{g_2} \,.
\end{equation}
The corresponding imaginary part satisfies $y_\ast = \sqrt{R^2 - x_\ast^2} \to \infty$ for $R \to \infty$, so the large-$y$ approximation \eqref{eq:app:largeyapprox} applies.
At this point, we have
\begin{equation}
    \Re \frac{ f_g(z_\ast, \bar z_\ast)}{g_2} = \Delta_g - \frac{y_\ast}{T}a(y_\ast) = \Delta_g - \frac{9 c T}{4 y_\ast} + \mathcal{O}(y_\ast^{-4}) \,,
\end{equation}
where
\begin{equation}
    \Delta_g = \frac{g_0}{g_2} - \frac{g_1^2}{2 g_2^2} \,.
\end{equation}
For $\Delta_g \neq 0$, we therefore find
\begin{equation}
    \ind_\infty(f_g) = \begin{cases}
        1 \,, & \Delta_g < 0 \,, \\
        0 \,, & \Delta_g > 0 \,,
    \end{cases}
\end{equation}
and, if $\Delta_g = 0$ but $c \neq 0$,
\begin{equation}
    \ind_\infty(f_g) = \begin{cases}
        1 \,, & c > 0 \,, \\
        0 \,, & c < 0 \,.
    \end{cases}
\end{equation}
If $\Delta_g = c = 0$, the index cannot be determined from \eqref{eq:app:Fgg2} alone, and the first exponentially suppressed corrections must be included.

\bibliographystyle{bibstyle}
\bibliography{sample}

\end{document}